\documentclass[twocolumn, twocolappendix]{aastex701}
\accepted{January 19, 2026}
\shorttitle{Characterizing ISM Fluctuations}
\shortauthors{Modak et al.}
\graphicspath{{./}{figures/}}

\usepackage{graphicx}
\usepackage{bm}
\usepackage{hyperref}
\usepackage{float}
\usepackage{amsmath, amssymb}
\usepackage{xcolor}
\usepackage{enumitem}

\usepackage[normalem]{ulem}

\newcommand{\Msun}{M_\odot}
\newcommand{\md}{\mathrm{d}}
\newcommand{\bx}{\mathbf{x}}
\newcommand{\bvee}{\mathbf{v}}
\newcommand{\bk}{\mathbf{k}}

\newcommand{\nn}{\nonumber}
\newcommand{\zcom}{z_\mathrm{com}}
\newcommand{\veff}{v_\mathrm{eff}}

\newcommand{\ztilde}{\tilde{z}}
\newcommand{\ktilde}{\tilde{k}}
\newcommand{\omegatilde}{\tilde{\omega}}
\newcommand{\sech}{\mathrm{sech}}

\begin{document}

\title{Characterizing Density and Gravitational Potential Fluctuations of the Interstellar Medium}

\author[0000-0002-8532-827X]{Shaunak Modak}
\affiliation{Department of Astrophysical Sciences, 4 Ivy Lane, Princeton University, Princeton, NJ 08544, USA}
\email{shaunakmodak@princeton.edu}

\author[0000-0002-0509-9113]{Eve C. Ostriker}
\affiliation{Department of Astrophysical Sciences, 4 Ivy Lane, Princeton University, Princeton, NJ 08544, USA}
\email{eostrike@princeton.edu}

\author[0000-0002-5861-5687]{Chris Hamilton}
\affiliation{School of Natural Sciences, Institute for Advanced Study, Einstein Drive, Princeton NJ 08540}
\affiliation{Department of Astrophysical Sciences, 4 Ivy Lane, Princeton University, Princeton, NJ 08544, USA}
\email{chamilton@ias.edu}

\author[0000-0002-0278-7180]{Scott Tremaine}
\affiliation{School of Natural Sciences, Institute for Advanced Study, Einstein Drive, Princeton NJ 08540}
\affiliation{Canadian Institute for Theoretical Astrophysics, University of Toronto, 60 St. George Street, Toronto, ON M5S 3H8, Canada}
\email{tremaine@ias.edu}

\begin{abstract}

Substructure in the interstellar medium (ISM) is crucial for establishing the correlation between star formation and feedback and has the capacity to significantly perturb stellar orbits, thus playing a central role in galaxy dynamics and evolution. Contemporary surveys of gas and dust emission in nearby galaxies resolve structure down to $\sim 10\,$pc scales, demanding theoretical models of ISM substructure with matching fidelity. In this work, we address this need by quantitatively characterizing the gas density in state-of-the-art MHD simulations of disk galaxies that resolve pc to kpc scales. The TIGRESS-NCR framework we employ includes sheared galactic rotation, self-consistent star formation and feedback, and nonequilibrium chemistry and cooling. We fit simple analytic models to the one-point spatial, two-point spatial, and two-point spatio-temporal statistics of the surface density fluctuation field. We find that for both solar neighborhood and inner-galaxy conditions, (i) the surface density fluctuations follow a log-normal distribution, (ii) the linear and logarithmic fluctuation power spectra are well-approximated as power laws with indices of $\approx -2.2$ and $\approx -2.8$ respectively, and (iii) lifetimes of structures at different scales are set by a combination of feedback and effective pressure terms. Additionally, we find that the vertical structure of the gas is well-modeled by a mixture of exponential and $\sech^2$ profiles, allowing us to link the surface density statistics to those of the volume density and gravitational potential. We provide convenient parameterizations for incorporating realistic ISM effects into stellar-dynamical studies and for comparison with multi-wavelength observations.

\end{abstract}

\keywords{\uat{Interstellar medium}{847} --- \uat{Molecular clouds}{1072} --- \uat{Galaxy dynamics}{591}}

%%%%%%%%%%%%%%%%%%%%%%%%%%%%%%%%%%%%%%%%%%%%%%%%%%%%%%%%%%%%
%%%%%%%%%%%%%%%%%%%%%%%%%%%%%%%%%%%%%%%%%%%%%%%%%%%%%%%%%%%%
\section{Introduction}
\label{sec:introduction}
%%%%%%%%%%%%%%%%%%%%%%%%%%%%%%%%%%%%%%%%%%%%%%%%%%%%%%%%%%%%
%%%%%%%%%%%%%%%%%%%%%%%%%%%%%%%%%%%%%%%%%%%%%%%%%%%%%%%%%%%%

The interstellar medium (ISM) plays a central role in sculpting the structure of the Milky Way and other gas-rich galaxies. Inhomogeneities in the gas (``fluctuations'') emerge as a result of the complex, nonlinear interplay between gravity, shear flow, and multi-scale magnetohydrodynamic (MHD) turbulence, with feedback processes from stars (especially expansion of \ion{H}{2} regions and supernova remnants/superbubbles) serving as key drivers of fluid motions \citep[see, e.g., reviews by][]{elmegreen2004,mckee_ostriker2007,Hacar2023,McClure2023,SchinnererLeroy2024}. Gas density fluctuations in the ISM are in turn important to many physical processes in galaxies, including initiating star formation, modulating the propagation of radiation, and creating gravitational fields that alter the orbits of stars. Thus, understanding and quantifying ISM structure is essential across many different areas of galactic astrophysics.

The ISM's gas may be classified into several different phases based on its local characteristic density, temperature, and chemical state (see, e.g., \citealt{Draine2011}). In principle it is easiest to create density fluctuations in the molecular gas, since it is the most compressible phase given its low sound speed (although in practice, gas may become molecular after compression rather than vice versa). The concentration of cold, dense molecular gas into large-scale, massive structures was already evident from CO mapping surveys of the Milky Way in the 1980s \citep{Sanders1985,Dame1986,Bally1988}, as well as studies focused on environments of individual star-forming regions \citep[e.g.,][]{Lada1978,Blitz1980,Bally1987}.

As reviewed in \cite{Blitz1993}, many early studies of ISM structure segmented the gas into individual giant molecular clouds (GMCs), and quantified the distribution of GMCs by mass and effective diameter. The resulting mass functions typically follow power laws, $\md N/ \md M \propto M^{-\gamma}$ for $\gamma\approx 1.5$ in the Milky Way \citep[e.g.,][]{Solomon1987,Heyer2015} or slightly steeper in the LMC \citep[$\gamma=1.7$,][]{Rosolowsky2005}, i.e., with most of the mass contained in the largest objects. To the extent that molecular gas is concentrated in individual clouds, the resulting mass functions may have additional consequences for the mass distributions of star clusters and for the spatial correlation of supernova feedback \citep[e.g.,][]{Williams1997,Krumholz2019,Chevance2023}.

The characterization of ISM structure as consisting of massive GMCs has the appeal of simplicity, and matches well to idealized approaches in stellar dynamics. Following the pioneering work of \cite{SpitzerSchwarzschild1951, SpitzerSchwarzschild1953}, most stellar dynamical calculations aiming to incorporate the influence of ISM fluctuations have modeled them as a collection of identical, highly idealized spherically symmetric perturbers in both analytic theory and numerical simulations (e.g., \citealt{Lacey1984, HutTremaine1985, WeinbergShapiroWasserman1987, Ida1993, ShiidsukaIda1999}). While more modern treatments have begun including more realistic aspects of ISM substructure such as perturber mass and size distributions, finite perturber lifetimes, and evolving perturber masses (e.g., \citealt{AumerBinneySchonrich2016a, AumerBinneySchonrich2016b}), the classical picture of discrete, compact (sizes of $\sim 10-100$\,pc) clouds has persisted.

However, in reality, individual ISM substructures are observed to be far from spherically symmetric, often instead taking a more extended, filamentary form (e.g., \citealt{Soler2022,Hacar2023} and references therein), and substantial fluctuations are observed in the Galaxy across a wide range of scales in atomic as well as molecular gas, and also in dust \citep[e.g.,][]{Planck2015}. As resolution has improved, surveys of external galaxies in both the Local Group and to distances of tens of Mpc have also revealed extensive structure in both the gas and the dust \citep[e.g.,][]{Bolatto2008,Rosolowsky2021,Lee2023}. Thus, empirical characterization of ISM fluctuations now often take the form of multi-scale statistics rather than GMC (or clump) mass functions (see \autoref{sec:observations}).

In parallel with the development of higher resolution observations that enable more sophisticated empirical characterization of ISM structure, the advent of numerical hydrodynamics as a tool to investigate the ISM has led to vastly more realistic theoretical models. Different kinds of simulations provide different kinds of insight to ISM structure. Global galaxy and cosmological zoom simulations are able to capture important large-scale, long-lived features such as grand design spiral arms by solving N-body and fluid dynamics equations together, and including tidal gravity from nearby companion systems (e.g., \citealt{TepperGarcia2024}). However, these simulations generally lack the spatial resolution and key ISM microphysical processes that would be needed to realistically model ISM fluctuations at scales below $\sim 100$\,pc (with $\lesssim 10$\,Myr lifetimes), since the computational cost would be prohibitive for galaxies larger than dwarfs \citep[e.g.,][]{Somerville2015, Naab2017}. 

The primary driver of fluid motions that lead to sub-kpc scale ISM fluctuations is the expansion of highly pressurized hot gas produced in supernova explosions. Properly following the spatio-temporal correlation of supernovae with each other and with the distribution of cool gas is necessary in order to achieve a realistic turbulent velocity field, and requires self-consistently modeling gravitational collapse and stellar population evolution. Additionally, the formation of over- and underdensities in the ISM is sensitive to the compressibility of the gas, which depends on its thermodynamics and magnetization. Thermal structure is controlled by the balance of heating and cooling, with rates dependent on the chemical state of the gas, and the heating rate and other microphysics are in turn sensitive to the UV radiation field, produced by young stars and attenuated by passage through the ISM. The growth and maintenance of magnetic fields depends on shear and turbulence in the ISM. To follow the relevant processes numerically, a large number of modules that accurately follow individual physical effects must be developed and integrated together, and simulations must resolve scales down to $\sim 1-10$\,pc while also extending over at least several disk scale heights laterally. To this end, several groups have developed and applied numerical codes that focus on kpc-scale ISM patches in investigations of the star-forming, multiphase ISM \citep[see][and references therein]{KimOstriker2017,Kim2023,Peters2017,Kannan2020,Brucy2020,Hu2021,Gurman2025,Rathjen2025}.

In analysis of ISM simulations, similar to other fluid dynamics simulations, structure is often characterized at a ``field'' level via one- and two-point statistics, rather than via segmentation into discrete objects \citep[e.g.,][]{Ballesteros2007,Dobbs2014}. Analysis of numerical ISM simulations makes it possible to characterize not only the spatial but also the temporal properties of ISM structure, in analogy to solar wind studies \citep[e.g.,][]{Yuen2023}. Understanding temporal properties of ISM structure may be especially important for stellar dynamical studies of galactic disks, since stellar orbital heating and migration is sensitive to the time-dependence of gravitational potential fluctuations \citep{BinneyLacey1988}. For this application and others, it is vital to develop realistic models that capture key attributes, while remaining simple enough to allow for convenient use in (semi-)analytic calculations.

In this work, we characterize ISM fluctuations that develop in the state-of-the-art TIGRESS-NCR simulations \citep{Kim2023}. These simulations incorporate the effects of galactic rotation and magnetic fields while employing self-consistent star formation and feedback (ray-tracing radiation and supernovae) as well as nonequilibrium chemistry and cooling, with a simulation resolution of $2-4$\,pc. The simulations we analyze span both solar neighborhood and inner-galaxy conditions relevant in a variety of dynamical contexts. With an eye toward extragalactic observations of face-on disk systems (as in e.g., the PHANGS project, \citealt{Leroy2021, Lee2023}), we build a model for the ISM's 2D surface density fluctuations, characterizing both their one-point and two-point statistics via probability density functions (pdfs) and power spectra. Additionally, we characterize the time dependence at each spatial frequency. For use in stellar dynamical applications, we also characterize the ISM's 3D gravitational potential fluctuations.

The rest of this paper is organized as follows. First, in \autoref{sec:simulations}, we describe in greater detail the TIGRESS-NCR simulations that we study throughout this work. In \autoref{sec:surface_density} and \autoref{sec:rho_phi} respectively, we characterize the ISM's surface density fluctuations and connect them to the structure of the ISM's volume density and gravitational potential. We discuss interpretations, implications, and caveats of our results, and connect with observations in \autoref{sec:discussion}. Finally, we summarize in \autoref{sec:summary}.

%%%%%%%%%%%%%%%%%%%%%%%%%%%%%%%%%%%%%%%%%%%%%%%%%%%%%%%%%%%%
%%%%%%%%%%%%%%%%%%%%%%%%%%%%%%%%%%%%%%%%%%%%%%%%%%%%%%%%%%%%
\section{Simulations}
\label{sec:simulations}
%%%%%%%%%%%%%%%%%%%%%%%%%%%%%%%%%%%%%%%%%%%%%%%%%%%%%%%%%%%%
%%%%%%%%%%%%%%%%%%%%%%%%%%%%%%%%%%%%%%%%%%%%%%%%%%%%%%%%%%%%

The simulations we analyze in this work are a subset of the TIGRESS-NCR simulation suite. The details of the TIGRESS-NCR numerical framework are described thoroughly in \cite{Kim2023}, but we summarize a few of the relevant features here.

The original ``TIGRESS-classic'' simulations \citep{KimOstriker2017, Kim2020} were developed to model the multiphase ISM in galactic disks with resolved, self-consistent star formation and feedback. The ``TIGRESS-NCR'' framework \citep{Kim2023} extends these simulations to include the effects of non-equilibrium cooling and radiation \citep{Kim_JG2023}. The TIGRESS-NCR simulations solve the ideal MHD equations using the \textit{Athena} code \citep{Stone2008} on a uniform cartesian mesh, while also evolving the abundances of hydrogen species, free electrons, and key atomic and molecular coolants.

As detailed in \cite{Kim_JG2023}, in TIGRESS-NCR the ionization, recombination, formation, and dissociation of hydrogen species (H, H$_2$, H$^+$) are evolved in a time-dependent fashion, based on interactions with the time-dependent Ly continuum and Ly Werner radiation fields, which are computed via adaptive ray tracing. Carbon (C, C$^+$, CO) and oxygen (O, O$^+$) species are treated as being in formation-destruction balance given the hydrogen abundances and local UV radiation field. Star formation is modeled through the use of sink particles, representing star clusters, which form when certain collapse conditions are met, and can accrete and merge. Based on a population synthesis model, the star cluster particles inject energy into the gas via a prescription for supernova feedback. The star cluster particles are also the source of EUV and FUV photons that ionize and dissociate the gas and provide heating via the photoelectric effect on small grains; cosmic ray heating and ionization proportional to the time-dependent star formation rate are also included. Heating and cooling of warm and cold neutral atomic and molecular gas is based on the abundances of hydrogen, oxygen, and carbon species following \cite{Gong2017}; cooling of photoionized gas is due to hydrogen and nebular lines; cooling of hot gas assumes collisional and ionization equilibrium. We refer readers interested in the details of the star formation and feedback prescriptions to Section 2.2 of \cite{KimOstriker2017}, and those interested in the implementations of the nonequilibrium chemistry, radiation transfer, and heating/cooling to Sections 3, 4, and 5 of \cite{Kim_JG2023}.

The TIGRESS simulations employ the ``shearing box'' approximation \citep{GoldreichLyndenBell1965, HawleyGammieBalbus1995} to model a locally Cartesian patch of the Galactic disk centered at a Galactocentric radius $R_0$, rotating at the local circular orbital frequency $\Omega$. The in-plane coordinates $x \equiv R - R_0$ and $y \equiv R_0(\varphi - \Omega t)$ represent the radial and azimuthal directions respectively, while the $z$ coordinate represents the distance above or below the midplane. The simulation domain extends a length $L_x=L_y \equiv L$ in each of the in-plane directions and $L_z$ in the vertical direction, with a cell size of $\Delta x$ in all dimensions, so that there are $N_x = N_y = L/\Delta x$ cells in each planar direction, and $N_z = L_z/\Delta x$ cells resolving the vertical direction.  The background differential rotation of the galaxy is modeled as a linear shear flow, so that a circular orbit at a radial position $x$ has a velocity $\bvee_0 =-q\Omega x \hat{y}$, where $q \equiv -(\md\ln\Omega/\md\ln R)_{R_0}$ parametrizes the shear. In the rotating frame, the tidal potential allows for the spatially varying centrifugal force, with the tidal force $2 q \Omega^2 x \hat x$ exactly balanced by the Coriolis force $-2 \Omega \hat z \times \mathbf{v}$ for circular orbits with $\mathbf{v}=\mathbf{v}_0$.  For reference, Oort's $A$ constant is related to $q$ by $A = q\Omega/2$, and the radial epicyclic frequency is given by $\kappa = \sqrt{2(2-q)}\Omega$. The simulations studied here fix $q = 1$, corresponding to a flat rotation curve.

The key advantage of the shearing box setup from a numerical standpoint is the ability to construct self-consistent ``shearing-periodic'' boundary conditions in the plane, as illustrated in, e.g., Figure 1 of \cite{HawleyGammieBalbus1995}. The $y$ direction is purely periodic with $f(x, y, z) = f(x, y+L_y, z)$ for any quantity $f$. However, upon crossing the box boundary in the $\pm x$ direction, the shear flow leads to an offset in $y$ of $\Delta y(t) = \mp q\Omega L_x (t - t_n)$, where $t_n \equiv nL_y/(q\Omega L_x)$ with $n \equiv \lfloor q\Omega tL_x/L_y \rceil$ is the ``nearest periodic time'' to $t$ and $\lfloor u \rceil$ denotes the closest integer to $u$. The $x$ boundary condition is thus $f(x,y,z)=f(x+L_x,y+\Delta y(t),z)$ for all fluid variables other than $v_y$, for which instead (accounting for the shear across the box) $\delta v_y \equiv v_y + q\Omega x$, the azimuthal velocity relative to the circular orbit at radius $R_0+x$, satisfies this condition. These boundary conditions can be more succinctly expressed with the use of ``sheared coordinates'' $(x', y')$ given by
\begin{align}
    \label{eq:sheared_x}
    x' & \equiv x, \\
    \label{eq:sheared_y}
    y' & \equiv y + q\Omega (t-t_n) x,
\end{align}
for which all variables $f$ (including $\delta v_y$) are fully periodic, i.e., they satisfy
\begin{equation}
    \label{eq:shearing_periodic}
    f(x' + L_x, y', z) = f(x', y' + L_y, z) = f(x', y', z).
\end{equation}
Because the density is periodic in the sheared coordinates, the gravitational potential of the gas may be obtained by remapping from $(x,y,z)$ to $(x',y',z)$ and applying Fourier transforms (see \citealt{Gammie2001} and below). More generally, introduction of the sheared coordinate system enables analysis of any scalar field in the box using Fourier series expansions.

Note that this change of coordinates requires a nontrivial corresponding change of differential operators, e.g., $\partial/\partial x = \partial/\partial x' + q\Omega (t-t_n) \partial/\partial y'$, which results in a time-dependent relationship between Fourier wavevectors in ordinary and sheared coordinates:
\begin{align}
    \label{eq:sheared_kx}
    k_x & \equiv k_x' + q\Omega (t-t_n) k_y', \\
    \label{eq:sheared_ky}
    k_y & \equiv k_y';
\end{align}
this transformation guarantees that $k_x x + k_y y = k_x' x' + k_y' y'$. 

For practical purposes, it is most advantageous to characterize the spectra of surface density and gravitational potential fluctuations in terms of wavevectors $k_x, k_y$ with reference to an ordinary Cartesian coordinate system; nevertheless, the sheared coordinates and sheared wavevectors are used as an intermediate step in the calculations. To be explicit, when performing a Fourier expansion of any quantity $Q(x, y, t)$, we write
\begin{align}
    \label{eq:fourier_details}
    Q(x, y, t) = \sum_{k_x} & \sum_{k_y} Q_{k_x,k_y}(t) \nn \\
    & \times e^{i[k_x'(k_x, k_y, t)x'(x) + k_y'(k_y)y'(x, y, t)]},
\end{align}
i.e., we (i) convert $(x, y)$ to sheared coordinates $(x', y')$, (ii) expand in wavevectors $(k_x', k_y')$, and (iii) re-label the Fourier components by $(k_x, k_y)$ using \autoref{eq:sheared_kx} and \autoref{eq:sheared_ky} (which we are free to do because the mapping at a fixed time is bijective)\footnote{Because our quantities of interest are periodic in sheared coordinates, note also that $(k_x', k_y') = (2\pi n_x/L_x, 2\pi n_y/L_y)$ for a pair of integers $(n_x, n_y)\in \mathbb{Z}_2$, and it is formally $(n_x, n_y)$ that we are summing over.}. In what follows, we will suppress the details of these operations, defining e.g., the power spectrum as
\begin{equation}
    \label{eq:power_def}
    P_Q(k_x, k_y) \equiv \langle |Q_{k_x, k_y}(t)|^2\rangle,
\end{equation}
where angle brackets denote an ensemble average, which we will often in practice calculate as a time average (assuming the simulation is in a statistical steady-state).

Because the fluctuations in the gravitational potential are a key focus of this paper, we devote special attention to the treatment of the potential in the simulations here. The total gravitational potential felt by the gas incorporates contributions from the following four sources. First, it includes the gas potential $\phi_\mathrm{g}$, which solves Poisson's equation $\nabla^2\phi_\mathrm{g} = 4\pi G\rho$, where $\rho$ is the gas density. Second, it includes the potential $\phi_\mathrm{sc}$ due to the star cluster particles, which solves Poisson's equation $\nabla^2\phi_\mathrm{sc} = 4\pi G\rho_\mathrm{sc}$, where $\rho_\mathrm{sc}$ is their density. In the TIGRESS simulations, the star cluster particle masses are mapped onto the grid using the triangle-shaped cloud method (see \citealt{GongOstriker2013} for the details of the implementation) in order to compute $\phi_\mathrm{sc}$. Third, it includes vertical gravity of the stellar disk and dark matter halo via a fixed, external potential,
\begin{align}
    \label{eq:external_potential}
    \phi_\mathrm{ext}(z) = 2\pi G & \Sigma_* z_* \left[\left(1 + \frac{z^2}{z_*^2}\right)^{1/2} - 1\right] \nn \\
    & + 2\pi G \rho_\mathrm{DM} R_0^2 \ln\left(1 + \frac{z^2}{R_0^2}\right),
\end{align}
where $\Sigma_*$, $z_*$, and $\rho_\mathrm{DM}$ are the stellar surface density, stellar scale height, and dark matter density respectively. Finally, the tidal potential arising from the non-inertial frame of the shearing box and centrifugal forces contributes $\phi_\mathrm{tidal}(x) = -q\Omega^2 x^2$. Although the total potential in the simulation is given by $\phi_\mathrm{tot} \equiv \phi_\mathrm{g} + \phi_\mathrm{sc} + \phi_\mathrm{ext} + \phi_\mathrm{tidal}$, here we focus on characterizing the fluctuations in the gas potential---the external and tidal potentials are fixed, while the star cluster particle potential is always subdominant. Thus, from here onward all references to the potential will be to the gas component alone, and we drop the subscript ``g'', i.e., we write $\phi_\mathrm{g} \equiv \phi$.

\begin{table*}
    \centering
    \begin{tabular}{ccccccccccccc}
    \hline
     & $R_0$ & $\Omega$ & $\overline\Sigma(t=0)$ & $\Sigma_*$ & $z_*$ & $\rho_\mathrm{DM}$ & $L_x (=L_y \equiv L)$ & $L_z$ & $\Delta x$ & $t_\mathrm{sim}$ & $\Delta t$ \\
    Sim & (kpc) & (km/s/kpc) & ($\Msun/\mathrm{pc}^2$)& ($\Msun/\mathrm{pc}^2$) & (pc)& ($\Msun/\mathrm{pc}^3$) & (pc) & (pc) & (pc) & (Myr) & (Myr) \\
    (1) & (2) & (3) & (4) & (5) & (6) & (7) & (8) & (9) & (10) & (11) & (12) \\
    \hline
    \hline
    R8 & 8 & 28 & 12 & 42 & 245 & 0.0064 & 1024 & 6144 & 4 & 200 & 1\\
    LGR4 & 4 & 30 & 50 & 50 & 500 & 0.005 & 512 & 3072 & 2 & 100 & 1\\
    \hline
    \end{tabular}
    \caption{Key simulation parameters, with columns as follows: (1) the simulation name, (2) the Galactocentric radius, (3) the angular velocity at the box center, (4) the initial mean gas surface density, (5) the stellar surface density, (6) the stellar scale height, (7) the dark matter density, (8) the box size in the planar directions, (9) the box size in the vertical direction, (10) the cell size, (11) the temporal baseline of output times analyzed, and (12) the interval between snapshots $\Delta t$.}
    \label{tab:sim_parameters}
\end{table*}

The two particular TIGRESS-NCR simulations we study here are (i) the ``R8-4\,pc'' model, with a nominal Galactocentric radius of $R_0 = 8\,$kpc and a spatial resolution of $\Delta x = 4\,$pc, and (ii) the ``LGR4-2\,pc'' model, 
with a nominal Galactocentric radius of $R_0 = 4\,$kpc and a spatial resolution of $\Delta x = 2\,$pc. \autoref{tab:sim_parameters} summarizes the simulation parameters for each model. The simulations reach a statistical steady-state by $t\approx 200\,$Myr (see, e.g., Section 3.1 of \citealt{Kim2023}). For the time interval $t\in[250\,\mathrm{Myr}, 450\,\mathrm{Myr}]$ and $t\in[250\,\mathrm{Myr}, 350\,\mathrm{Myr}]$ for the R8 and LGR4 simulations, respectively, the basic ISM properties and star formation rates (SFRs) have been analyzed in \cite{Kim2023}, and the radiation fields have been characterized in \cite{Linzer2024}. Here, we characterize their surface density and potential fields every $\Delta t = 1\,$Myr for the same intervals; we denote the temporal baseline with $t_\mathrm{sim}$ ($=200\,$Myr for R8 or $100\,$Myr for LGR4).

We choose these two simulations to span a range of gas surface densities and ISM conditions, representative of the gas structures in, e.g., nearby disk galaxies observed in the PHANGS survey \citep{Leroy2021}. Note that these simulations are not intended to model different regions within a single galaxy: the conditions in R8 are chosen to match solar neighborhood conditions and are also similar to area-weighted averages from PHANGS, while conditions in LGR4 are similar to star-formation-weighted averages from PHANGS \citep[see Table 1 of][]{Sun2022}. In principle, the analysis we present in \autoref{sec:surface_density} and \autoref{sec:rho_phi} could be straightforwardly extended to simulations incorporating a strong spiral arm potential as in \cite{KimKimOstriker2020}; we defer this generalization to future work.

\begin{figure*}
    \centering
    \includegraphics[width=0.95\textwidth]{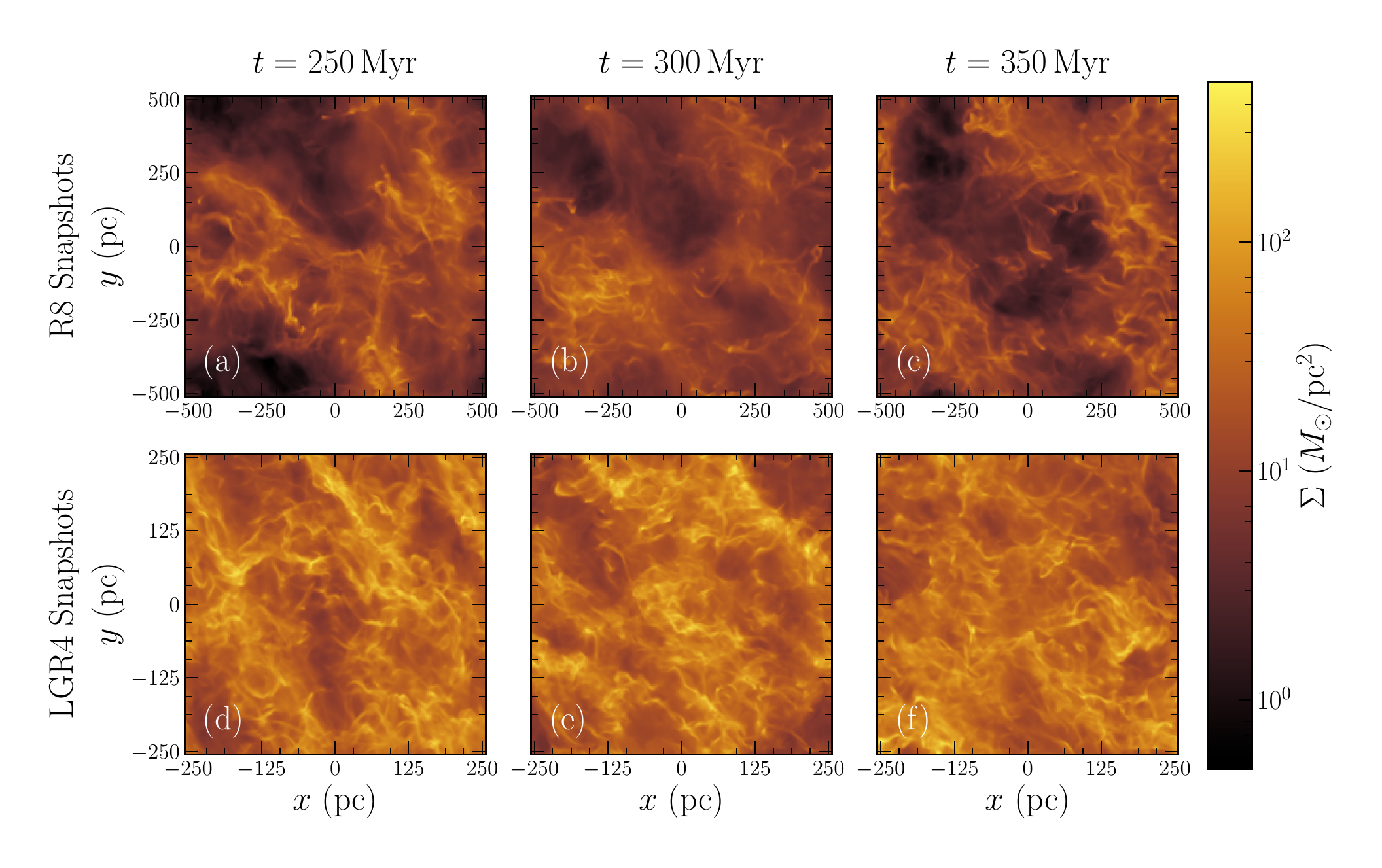}
    \caption{Snapshots of the gas surface density $\Sigma$ at different times in the R8 (top row) and LGR4 (bottom row) TIGRESS-NCR simulations.}
    \label{fig:snapshots}
\end{figure*}

To illustrate the structures present in each simulation, in \autoref{fig:snapshots} we show snapshots of the gas surface density $\Sigma$ at a few different times for the R8 model (top row) and the LGR4 model (bottom row)\footnote{See Figure 1 of \cite{Kim2023} for a detailed description of the time evolution of the global gas properties. Briefly, the mean gas surface density decreases gradually (by $\lesssim 10\%$ over the course of several hundred Myr) due to star formation and outflows, while the SFR surface density (i.e., the star formation rate per unit area, $\Sigma_\mathrm{SFR}$), gas effective velocity dispersion (which includes thermal, turbulent, and magnetic terms), and vertical scale height of the gas $H$ exhibit quasi-periodic variations of $\sim 0.2\,$dex on timescales between the vertical crossing time of the turbulent flow ($H/\sigma_\mathrm{turb}\sim 20\,$Myr) and the vertical gravitational oscillation period ($\sim 80\,$Myr).}. In both simulations, intricate nonlinear, filamentary, overdense structures develop in the gas. The lowest surface densities seen in \autoref{fig:snapshots} correspond to the projections of volumes that contain superbubbles filled with extremely tenuous, hot gas that are created by multiple correlated supernovae, while the highest surface densities correspond to gravitationally bound substructures. In what follows, we will characterize the spatio-temporal properties of these gas structures and their corresponding potential fluctuations.

%%%%%%%%%%%%%%%%%%%%%%%%%%%%%%%%%%%%%%%%%%%%%%%%%%%%%%%%%%%%
%%%%%%%%%%%%%%%%%%%%%%%%%%%%%%%%%%%%%%%%%%%%%%%%%%%%%%%%%%%%
\section{Surface Density}
\label{sec:surface_density}
%%%%%%%%%%%%%%%%%%%%%%%%%%%%%%%%%%%%%%%%%%%%%%%%%%%%%%%%%%%%
%%%%%%%%%%%%%%%%%%%%%%%%%%%%%%%%%%%%%%%%%%%%%%%%%%%%%%%%%%%%

In this section, we will quantify the fluctuations in the surface density field, calculated by integrating the volume density over the vertical extent of the box. Note that here and throughout the remainder of this paper we will use the vector $\bx$ to denote the 2D planar variables only, i.e., $\bx \equiv (x, y)$, and consider $z$-dependence explicitly when needed.

\begin{figure*}
    \centering
    \includegraphics[width=0.95\textwidth]{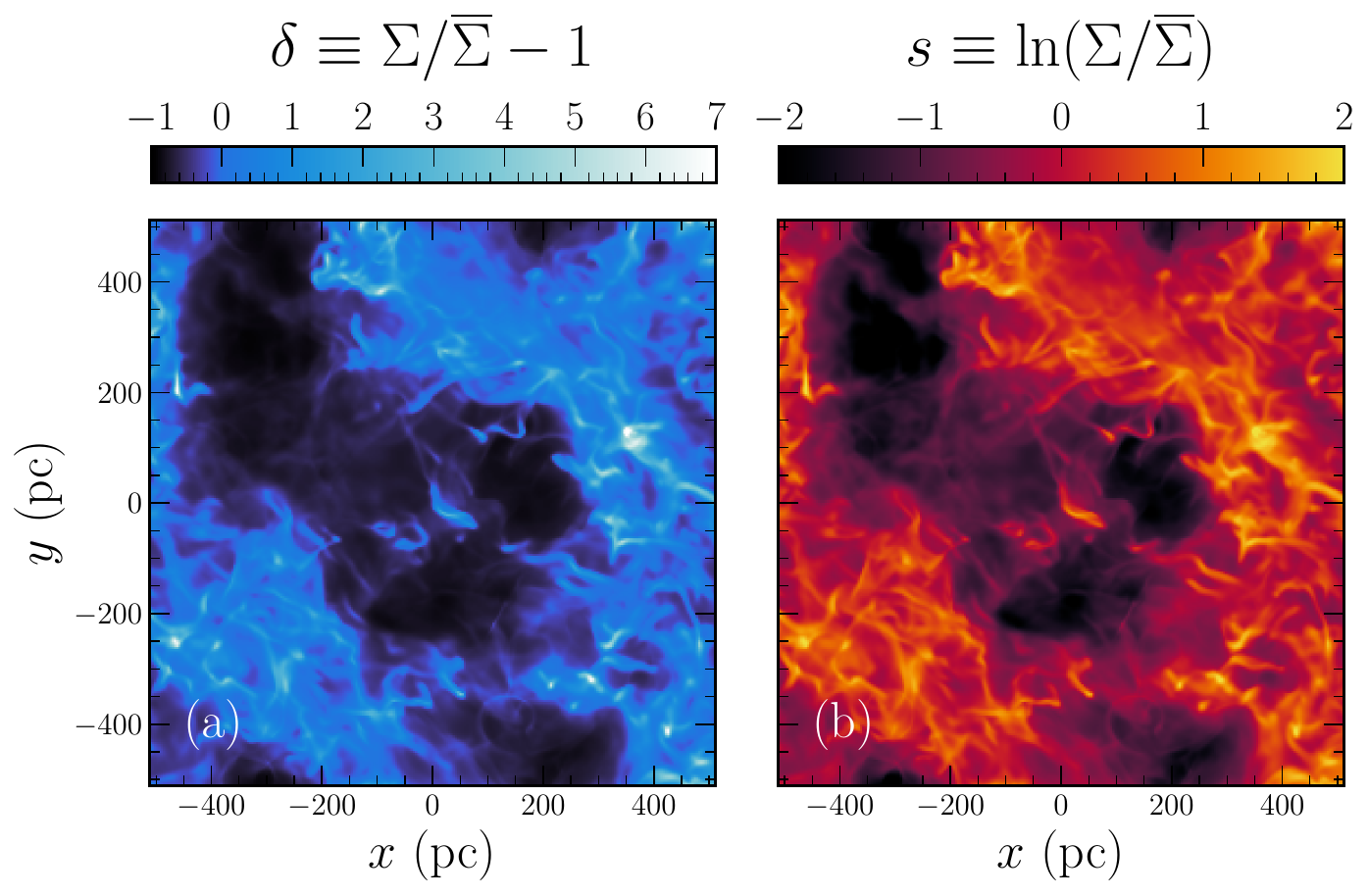}
    \caption{A snapshot of the linear surface density fluctuation $\delta$ (left) and logarithmic surface density fluctuation $s$ (right) for the R8 simulation at $t = 350\,$Myr. This is the same snapshot as in panel (c) of \autoref{fig:snapshots}.}
    \label{fig:delta_s}
\end{figure*}

Rather than studying the ``raw'' surface density field $\Sigma(\bx, t)$, we isolate the fluctuations by analyzing the linear deviation from the spatial mean, 
\begin{equation}
    \label{eq:delta_definition}
    \delta(\bx, t) \equiv \frac{\Sigma(\bx, t)}{\overline\Sigma(t)} - 1,
\end{equation}
as well as the logarithmic deviation
\begin{equation}
    \label{eq:s_definition}
    s(\bx, t) \equiv \ln\left(\frac{\Sigma(\bx, t)}{\overline\Sigma(t)}\right) = \ln(1+\delta(\bx, t)),
\end{equation}
where the overline represents a spatial average over the box at a given time. In \autoref{fig:delta_s} we illustrate the structure in the variables $\delta$ and $s$ for the R8 surface density snapshot shown in panel (c) of \autoref{fig:snapshots}. Note that for values of $|\delta| \ll 1$, we can expand to find $s \approx \delta$ to first order, but we see that many of the structures of interest are highly nonlinear with $\delta > 1$, so the properties of the two fields will differ.

Below, we will characterize the fields $\delta$ and $s$ by their spatial one-point (i.e., the pdfs of fluctuation values---see \autoref{sec:surface_density_pdfs}) and two-point (i.e., the fluctuation power spectra---see \autoref{sec:surface_density_spectra}) statistics. We also describe the two-point \textit{spatio-temporal} structure of the field in \autoref{sec:surface_density_spatiotemporal}. Although these statistics are formally dependent on the grid cell size of the simulations used to study them, in \autoref{sec:convergence} we demonstrate that our key results are converged with respect to spatial resolution by comparing measurements from our fiducial R8 simulation and a lower-resolution simulation with identical initial conditions. Note also that for the strongly nonlinear fluctuations (e.g., where $\delta \gtrsim 1$), the phases of different Fourier components of the field are correlated and information is therefore contained in higher order (beyond two-point) statistics; we discuss this caveat in \autoref{sec:nongaussianity_anisotropy}.

Throughout the rest of this paper, we will distinguish between models that we fit and quantities directly measured in the simulations by adding a subscript ``m'' to the former.

%%%%%%%%%%%%%%%%%%%%%%%%%%%%%%%%%%%%%%%%%%%%%%%%%%%%%%%%%%%%
\subsection{One-Point Spatial Surface Density Statistics}
\label{sec:surface_density_pdfs}
%%%%%%%%%%%%%%%%%%%%%%%%%%%%%%%%%%%%%%%%%%%%%%%%%%%%%%%%%%%%

\begin{figure*}
    \centering
    \includegraphics[width=0.95\textwidth]{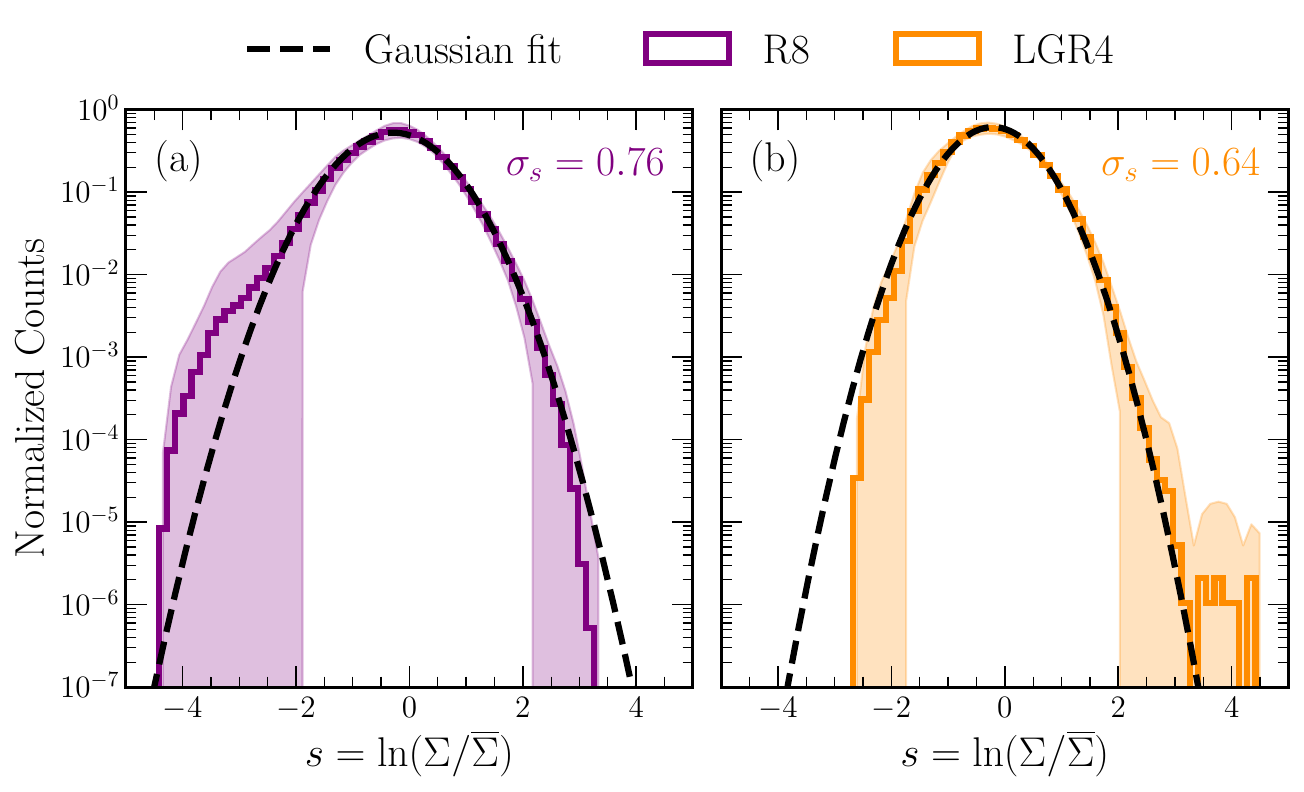}
    \caption{Histograms of the values of the logarithmic density fluctuation $s(\bx,t)$ measured at the simulation grid scale for the R8 (left, purple) and LGR4 (right, orange) simulations. The solid curves show the time-averaged histograms, while the shaded regions indicate $\pm 1\sigma$ variations about the values in each bin. A Gaussian with standard deviation $\sigma_s$ (corresponding to the measured standard deviation in the time-averaged values) and mean $\mu_s = -\sigma_s^2 / 2$ is overplotted as a black dashed curve.}
    \label{fig:s_pdfs}
\end{figure*}

In \autoref{fig:s_pdfs} we show the area-weighted pdf $p(s)$ of the logarithmic surface density fluctuation $s(\bx, t)$ measured at the simulation grid scale for the R8 (left, purple) and LGR4 (right, orange) simulations. The dark curves in each panel show the time-averaged histogram across the entire temporal baseline we consider for each simulation, while the shaded region shows the $\pm 1\sigma$ variation about the time-averaged histogram. As anticipated from numerical simulations of isothermal, supersonic turbulence (e.g., \citealt{Vazquez-Semadeni1994, OstrikerStoneGammie2001}), we find that the histograms near the peak are well-fit by Gaussian distributions\footnote{So, the distribution of the surface density itself is \textit{log}-normal.},
\begin{equation}
    \label{eq:gaussian_s}
    p_\mathrm{m}(s)\md s = \frac{1}{\sqrt{2\pi \sigma_s^2}}\exp{\left[-\frac{(s - \mu_s)^2}{2\sigma_s^2}\right]} \md s,
\end{equation}
although some deviations are present at both very small and very large $s$ values, due to the effects of feedback, shocks, and self-gravity. In particular, the extreme low-density regions result from bubbles produced through energy and momentum injection from supernovae, while extreme high-density regions occur when strong shocks are resolved and the gas has not yet reached the threshold for collapse into a star cluster particle. However, these deviations are only present for a negligibly small fraction of the total area and mass in the simulation (note the logarithmic vertical axis in the Figure), and so our Gaussian distribution model $p_\mathrm{m}$ is appropriate for the vast majority of the gas.

Using this model, and the fact that in statistical steady-state $\int \md s \, p_\mathrm{m}(s) e^s = \overline{e^{s(\bx, t)}} = \overline{1 + \delta(\bx, t)} = 1$ (see \autoref{eq:s_definition}), we can relate the mean and variance of the $s$ distribution by
\begin{equation}
    \label{eq:mean_s}
    (\mu_s)_\mathrm{m} = -\frac{(\sigma_s)_\mathrm{m}^2}{2},
\end{equation}
at every time. Thus, the model distributions of $s$ are characterized entirely by their standard deviation, which we can measure from the time-averaged histograms shown in \autoref{fig:s_pdfs}. The black dashed curves in the Figure illustrate these fits for each simulation.

\begin{figure}
    \centering
    \includegraphics[width=0.48\textwidth]{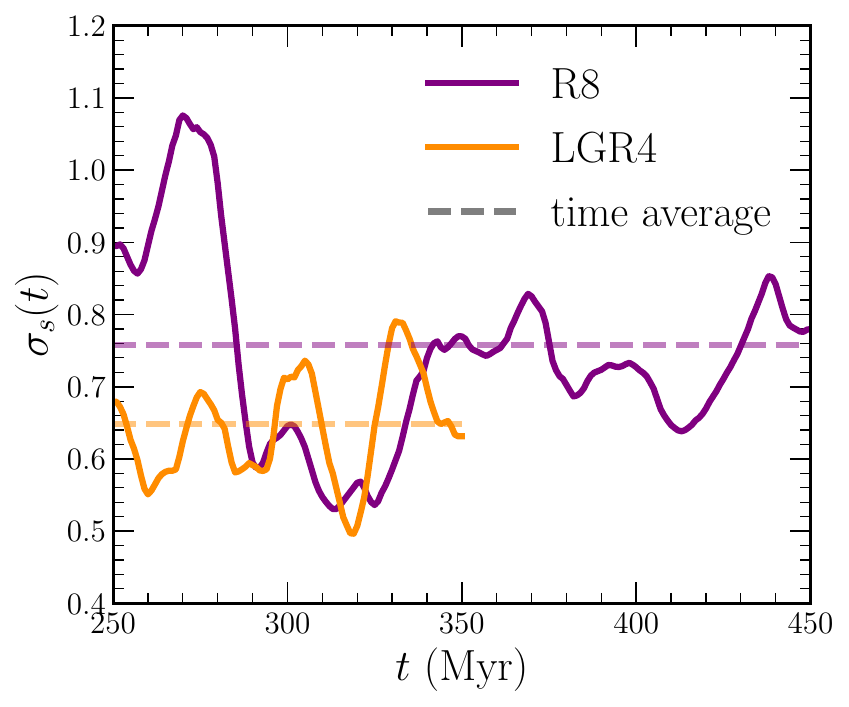}
    \caption{The standard deviation in $s=\ln(\Sigma(\mathbf{x},t)/\overline{\Sigma}(t))$ as a function of time for the R8 (purple) and LGR4 (orange) simulations. The light purple and orange dashed lines show the value calculated from the time-averaged distributions shown in \autoref{fig:s_pdfs} respectively.}
    \label{fig:sigma_s_t}
\end{figure}

In addition to studying the histogram of $s$ across the entire ensemble of snapshots, we may also measure the distribution of $s$ in any individual snapshot; we find that the distribution of $s$ is consistent with a Gaussian at each time, albeit with a varying value of $\sigma_s$ (and a correspondingly varying value of $\mu_s$). We attribute this time-dependence to the variations produced by feedback, shear, and turbulence in each simulation. \autoref{fig:sigma_s_t} illustrates the standard deviation $\sigma_s$ as a function of time: the value of $\sigma_s$ in each simulation (solid curves) oscillates about its time average (light dashed lines), with variations of up to $30-50\%$. The time variation appears quasi-periodic with a characteristic timescale comparable to the vertical fluid crossing time of $\sim 20\,$Myr, though variation on the vertical gravitational oscillation period\footnote{Note that fluid elements experience forces from thermal and magnetic as well as gravitational terms, so this timescale is distinct from the vertical fluid crossing time.} of $\sim 80\,$Myr is also evident in the R8 simulation, similar to the behavior observed in global properties such as the SFR, effective vertical velocity dispersion, and scale height, as evident in Figure 1 of \cite{Kim2023}.

Thus far, we have focused on the pdf of $s$ values measured at the grid scale $\Delta x$, i.e., at the highest resolution resolvable in each simulation. However, as is well-known in the cosmological literature (e.g., \citealt{PressSchechter1974, Peacock1999}) and has been applied more recently to studies of ISM structure in the context of star formation (e.g., \citealt{HennebelleChabrier2008, Hopkins2012a, Hopkins2012b}), important information regarding the structure of the field $s(\bx, t)$ can be gained by measuring its properties on larger scales $R > \Delta x$. We can investigate these scales by applying smoothing filters to the field as follows. We define the smoothed field as having Fourier components
\begin{equation}
    \label{eq:sRk_def}
    s^{(R)}_\bk(t) \equiv W^{(R)}(\bk) s_{\bk}(t),
\end{equation}
i.e., we weight each component $s_\bk$ of the unsmoothed field with a window function $W^{(R)}$, and produce the smoothed field by summing up these components as in \autoref{eq:fourier_details}. The window functions are chosen to be real functions of $\bk$ normalized so that $\int \md^2\bx \, W^{(R)}(\bx) = 1$ and $W^{(R)}(\bx = 0) = 1/(\pi R^2)$ for all $R$. This smoothing corresponds to the typical window convolution operation for a continuous and periodic field $Q$, 
\begin{equation}
    Q^{(R)}(\bx) = \int \md^2\bx' \, Q(\bx')W^{(R)}(\bx-\bx'),
\end{equation}
except in our case the field is discrete, and only satisfies periodic boundary conditions in the sheared coordinates.

The standard deviation of the smoothed field $s^{(R)}$ is given by 
\begin{align}
    \label{eq:sigma_s_fromfield}
    \sigma_s(R) & \equiv \left[\langle(s^{(R)})^2\rangle - \langle s^{(R)}\rangle^2\right]^{1/2} \\
    & = \left[\sum_{\bk} P_s(\bk) W^{(R)}(\bk)^2 - \frac{1}{4}\sigma_s^4\right]^{1/2}, 
\end{align}
where the angle brackets indicate an ensemble average (not necessarily a time-average in this context), and $P_s(\bk)$ is the power spectrum of fluctuations defined from the field $s(\bx)$ (see \autoref{eq:power_def}). In the second line, we have applied \autoref{eq:mean_s} to rewrite the second term---note that the mean of the field does not change under smoothing, $\langle s^{(R)}\rangle =\mu_s$. After choosing some reference scale $R_\mathrm{ref}$ (e.g., the grid scale of the simulations), we may rearrange this expression to write
\begin{align}
    \label{eq:sigma_s_R_power}
    \sigma_s(R) = & \bigg[\sigma_s(R_\mathrm{ref})^2 \nn \\
    & + \sum_{\bk} P_s(\bk)[W^{(R)}(\bk)^2 - W^{(R_\mathrm{ref})}(\bk)^2]\bigg]^{1/2}.
\end{align}
We will show in \autoref{sec:surface_density_spectra} that the power spectrum $P_s$ can be approximately modeled as an isotropic power law in wavevectors $\bk$, and so it is appropriate to use window functions that are isotropic in $\bk$ as well. Since $\Delta x/L \ll 1$, we can approximate the sum over wavevectors as an integral; under the assumption of isotropy, we find
\begin{equation}
    \label{eq:power_window_sum_approx}
    \sum_{\bk}P_s(\bk)W^{(R)}(\bk)^2 \approx \frac{L^2}{2\pi} \int_{2\pi/L}^{\infty} \md k \, kP_s(k) W^{(R)}(k)^2.
\end{equation}

Here, we will focus on two commonly used window functions: one that is sharp in $k$-space, given by
\begin{equation}
    \label{eq:window_sharpk}
    W^{(R)}_{k}(\bk) = \Theta\left(1 - \frac{kR}{2}\right) \leftrightarrow W^{(R)}_k(\bx) = \frac{1}{\pi R x}J_1\left(\frac{2x}{R}\right),
\end{equation}
and another that is sharp in real space, given by
\begin{equation}
    \label{eq:window_sharpreal}
    W^{(R)}_x(\bx) = \frac{1}{\pi R^2}\Theta\left(1 - \frac{x}{R}\right) \leftrightarrow W^{(R)}_x(\bk) = \frac{2}{kR}J_1(kR).
\end{equation}
In these expressions, $\Theta$ is the Heaviside step function, $J_n$ is a Bessel function of the first kind, and we have also provided the Fourier transforms of each window function for reference. The sharp $k$-space filter $W^{(R)}_k$ has properties amenable for excursion-set theory studies of the density field (see, e.g., \citealt{Bond1991}), while the sharp real-space filter is more easily applied to observations.

\begin{figure*}
    \centering
    \includegraphics[width=0.95\textwidth]{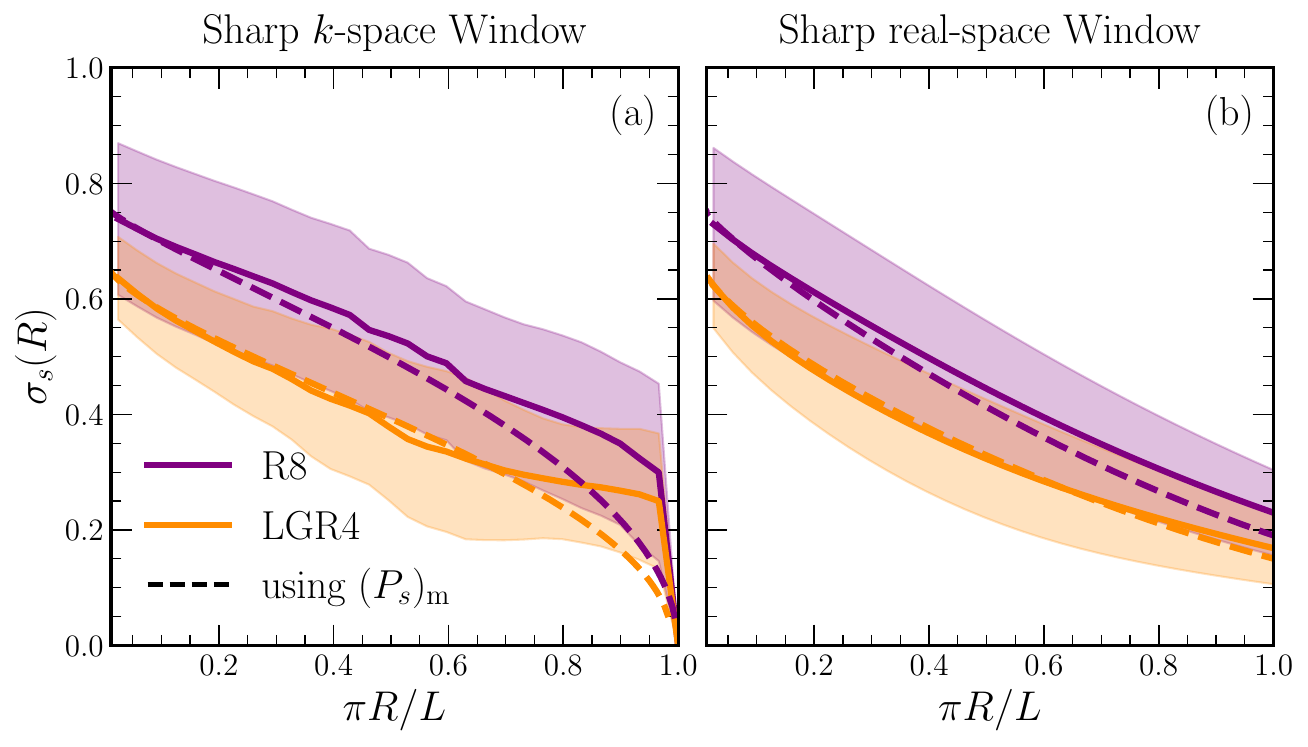}
    \caption{The standard deviation in $s \equiv \ln(\Sigma/\overline{\Sigma})$ as a function of scale for the R8 (purple) and LGR4 (orange) simulations. The left panel shows the standard deviation calculated after smoothing with a window function that is sharp in $k$-space (\autoref{eq:window_sharpk}), while the right panel uses a window function that is sharp in real space (\autoref{eq:window_sharpreal}). The dashed curves in each panel are calculated assuming the model isotropic power-law spectrum measured from fits to each simulation (see \autoref{eq:power_Sigma_spatial} and \autoref{tab:power_k_measurements} below).}
    \label{fig:sigma_s_R}
\end{figure*}

\autoref{fig:sigma_s_R} shows the standard deviation measured directly from the smoothed field snapshots $s^{(R)}$ for each simulation (again with R8 in purple and LGR4 in orange) as a function of the smoothing scale $R$, with the time-average shown as a solid curve and the $\pm 1\sigma$ variation around the time-average shown as a shaded region. In the left panel we smooth the field using $W^{(R)}_k$, while in the right panel we use $W^{(R)}_x$. Both panels show similar amplitudes of $\sigma_s$ at each scale $R$, although the differences in the functional form of the window lead to slight deviations between them\footnote{For instance, the measurement for $\sigma_s(R)$ derived from the sharp $k$-space window is less smooth and goes to zero at $R = L/\pi$ simply because it is more sensitive to the finite $k$-space resolution than the sharp real-space window, for which the weights assigned to each cell in $k$-space can take on values besides $0$ or $1$.}. For both simulations, we see that the widths of the shaded regions in \autoref{fig:sigma_s_R} are only weakly dependent on $R$, suggesting that feedback processes, shear, and turbulence contribute to time-variation across all scales similarly.

For comparison, in \autoref{fig:sigma_s_R} we also overplot dashed curves calculated using \autoref{eq:sigma_s_R_power} (taking the reference scale $R_\mathrm{ref}$ to be the grid scale $\Delta x$ of each simulation) and the approximation \autoref{eq:power_window_sum_approx} with the simplified isotropic power-law model for $P_s$ described in \autoref{sec:surface_density_spectra} below. For the sharp $k$-space window $W^{(R)}_k$, this can be done analytically: substituting, we find
\begin{align}
    \sigma_s(R) = \bigg[ & \sigma_s(\Delta x)^2 - \frac{P_{s, 0}}{(2\pi)^{n_s-1}(n_s-2)} \nn \\
    & \times \bigg(\bigg(\frac{\pi R}{L}\bigg)^{n_s-2} - \bigg(\frac{\pi\Delta x}{L}\bigg)^{n_s-2}\bigg)\bigg]^{1/2},
\end{align}
where $\sigma_s(\Delta x)$ is the standard deviation of the grid-scale pdf shown in \autoref{fig:s_pdfs}, and $P_{s, 0}$ and $n_s$ parametrize the model spectrum (see \autoref{eq:power_Sigma_spatial} and \autoref{tab:power_k_measurements}). Unfortunately, no simple closed-form expression results when applying the sharp real-space window $W^{(R)}_x$. In any case, we see from \autoref{fig:sigma_s_R} that the measured variance is consistent with the model across all except perhaps the largest smoothing scales. However, as we will describe below, at these scales the power spectrum in the simulation begins to slightly deviate from our fitted model, so such a discrepancy is to be expected.

%%%%%%%%%%%%%%%%%%%%%%%%%%%%%%%%%%%%%%%%%%%%%%%%%%%%%%%%%%%%
\subsection{Two-Point Spatial Surface Density Statistics}
\label{sec:surface_density_spectra}
%%%%%%%%%%%%%%%%%%%%%%%%%%%%%%%%%%%%%%%%%%%%%%%%%%%%%%%%%%%%

\begin{figure*}
    \centering
    \includegraphics[width=0.95\textwidth]{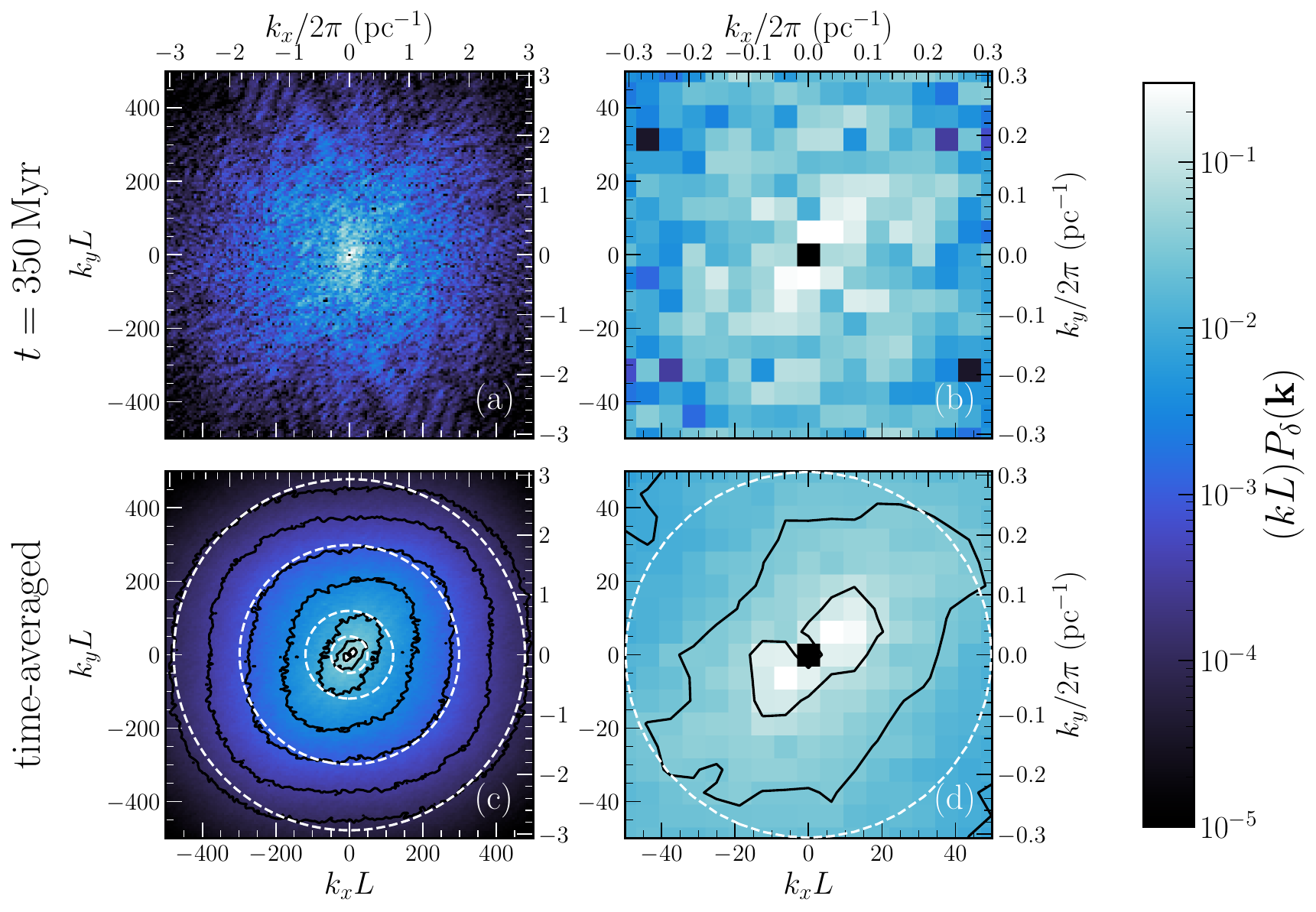}
    \caption{Contour plots of $(kL)|\delta_{\bk}|^2$ measured for a single snapshot (top row), and the corresponding ensemble average over all snapshots $(kL)P_\delta(\bk)$ (bottom row), for the R8 simulation. In the bottom row we also show level sets of $(kL)P_\delta(\bk)$ spaced by 0.5\,dex with black curves, as well as several white dashed circles to guide the eye; a perfectly isotropic spectrum would have black level sets coinciding with this white circles. Panels (b) and (d) show the same data as panels (a) and (c), but zoomed in to the innermost region; the single white dashed circle in panel (d) is the same as the innermost circle in panel (c).}
    \label{fig:P_delta_2D}
\end{figure*}

In this section, we focus on spatial power spectra, defined as in \autoref{eq:power_def}, for the $\delta$ and $s$ fields. In \autoref{fig:P_delta_2D}, we present the spectra of the linear fluctuation $\delta$ in the R8 simulations. We show the quantity $(kL)|\delta_\bk|^2$ for a single snapshot in the top row and its ensemble-average (here assumed to be the same as a time-average) $(kL)P_\delta(\bk) \equiv (kL)\langle|\delta_\bk|^2\rangle$ in the bottom row; the total power is proportional to the integral of these quantities over $k = |\bk|$. Although $(kL)|\delta_\bk|^2$ is very noisy for a single snapshot, its time-average has fairly smooth level sets (indicated by the black contours in panels (c) and (d), which are spaced by 0.5\,dex). Comparing these contours to circles of constant $k$ (overplotted as white dashed curves), we see that the power may be reasonably well-approximated as isotropic at small scales (large $k$). At the largest scales, as evidenced by the tilt in the contours in panel (d), the fluctuation power somewhat prefers trailing structures (those with the same sign of $k_x$ and $k_y$), as expected for a sheared flow.

Although we have presented \autoref{fig:P_delta_2D} for just the $\delta$ field in the R8 simulation, we note that analogous plots of $P_s(\bk)$ for the R8 simulation, and for both $P_\delta(\bk)$ and $P_s(\bk)$ for the LGR4 simulations, look qualitatively the same. In particular, they all show a near-isotropic field with a slight preference for trailing waves. Even for the scales that deviate most strongly from isotropy, the variation in the power at a fixed $k$ but varied $(k_x, k_y)$ is well within $\sim 0.25\,$dex, and so for the remainder of this paper, for simplicity, we will approximate all fluctuation power spectra to be a function of the magnitude of the wavevector, $k$, alone (but see \autoref{sec:nongaussianity_anisotropy} for further discussion).

\begin{figure*}
    \centering
    \includegraphics[width=0.95\textwidth]{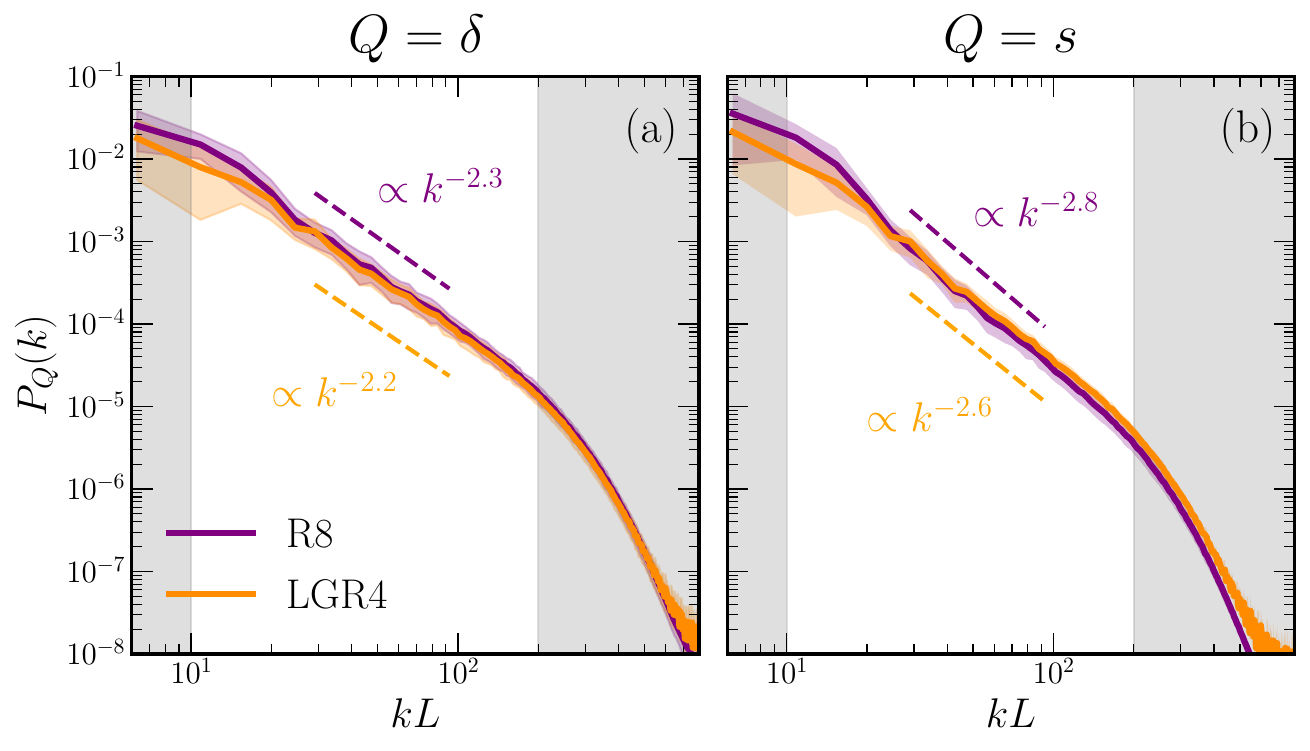}
    \caption{Spatial power spectra of the linear fluctuations (left) and logarithmic fluctuations (right) for the R8 (purple) and LGR4 (orange) simulations. The solid curves of each color are the time-averaged spectra averaged over the azimuthal angle in $k$-space, while shaded regions show $\pm 1\sigma$ variations around each curve due to anisotropy (see, e.g., \autoref{fig:P_delta_2D}) and time-dependence. The left and right shaded grey regions in each panel are excluded from fitting the power-law slope (indicated by the colored dashed curves) due to box-size and resolution effects respectively.}
    \label{fig:power_k}
\end{figure*}

In \autoref{fig:power_k} we plot the power spectra in $\delta$ (left) and $s$ (right) for the R8 (purple) and LGR4 (orange) simulations as a function of $k$, calculated by averaging the spectra $P_\delta(\bk)$ and $P_s(\bk)$ over contours of constant $k$ (e.g., the white dashed circles in \autoref{fig:P_delta_2D}). The solid curves all indicate the measured power spectra, while the shaded regions around them represent $\pm 1\sigma$ deviations in the values due to time-dependence in the fluctuation Fourier components and anisotropy. We see that the shaded regions are wider at smaller $k$, demonstrating the increased anisotropy present there, but the fractional variation remains small, illustrating that the approximation of isotropy is reasonable.

For both $P_\delta$ and $P_s$, the power spectra in both simulations are well-fit by power laws across a wide range of scales. We measure power law indices by fitting the slope of the curves over the range $kL \in [10, 200]$; for smaller $kL$ box-size effects and anisotropy may become important, while for larger $kL$ there is contamination due to the proximity to the grid scale $\Delta x$\footnote{Incorporating smaller values of $kL$ in the fit does not substantially alter the results---the error bars on the slope are simply increased due to the larger fractional error in power measured for the largest scales, as indicated in the shaded region of \autoref{fig:power_k} and described above.}. We demonstrate in \autoref{sec:convergence} that our results for the largest scales, which carry the most power, are not affected by the simulation spatial resolution. Thus, our model for the power spectra takes the form
\begin{equation}
    \label{eq:power_Sigma_spatial}
    (P_Q)_\mathrm{m}(k) = P_{Q,0}(kL)^{-n_Q}
\end{equation}
for both $Q = \delta$ and $Q = s$. For reference, the two-point correlation function $\xi_Q(\mathbf{r}) \equiv \langle Q(\bx)Q(\bx+\mathbf{r})\rangle$ corresponding to this spectrum is
\begin{align}
    \label{eq:correlation_function}
    (\xi_Q)_\mathrm{m}(r) & = \sum_{\bk} (P_Q)_\mathrm{m}(k) e^{i\bk\cdot \mathbf{r}} \nn \\
    & \approx \frac{P_{Q, 0}}{2\pi}\left(\frac{r}{L}\right)^{n_Q - 2}\int_{2\pi r/L}^{\infty} \md u \, u^{1-n_Q} J_0(u),
\end{align}
where in the second line we have approximated the sum with an integral as in \autoref{eq:power_window_sum_approx}. We determine the normalization coefficients $P_{Q,0}$ for each fluctuation and each simulation by ensuring that the model's total fluctuation power matches the simulation's total fluctuation power. Quantitatively, this means we set $P_{Q, 0}$ such that  
\begin{equation}
    \label{eq:P_normalization}
    \langle Q^2(t)\rangle = \frac{L^2}{2\pi}\int_{2\pi/L}^{\infty}\md k \, k P_{Q,0} (k L)^{-n_Q}
\end{equation}
for $Q=\delta$ and $Q=s$. \autoref{tab:power_k_measurements} shows the results of this fitting process. 

\begin{table}
    \centering
    \begin{tabular}{ccccc}
        \hline
        Sim & $P_{\delta,0}$ & $n_\delta$ & $P_{s,0}$ & $n_s$ \\
        (1) & (2) & (3) & (4) & (5) \\
        \hline
        \hline
        R8 & 2.4$\pm$1.4 & 2.3$\pm$0.1 & 12.7$\pm$4.6 & 2.8$\pm$0.1 \\
        LGR4 & 1.7$\pm$0.6 & 2.2$\pm$0.1 & 5.0$\pm$1.7 & 2.6$\pm$0.1 \\
        \hline
    \end{tabular}
    \caption{Measurements for power-law fits to the power spectra shown in \autoref{fig:power_k}, with columns as follows: (1) the simulation name, (2) the normalization for the linear fluctuation spectrum, (3) the power law index for the linear fluctuation spectrum, (4) the normalization for the logarithmic fluctuation spectrum, and (5) the power law index for the logarithmic fluctuation spectrum. For definitions of the parameters, see \autoref{eq:power_Sigma_spatial}. Note that the errors reported on $P_{Q,0}$ correspond to the errors on $n_Q$ propagated through \autoref{eq:P_normalization} for both $Q = \delta$ and $Q = s$.}
    \label{tab:power_k_measurements}
\end{table}

We see from \autoref{fig:power_k} and \autoref{tab:power_k_measurements} that despite the stark differences in the galactic environments and feedback strengths between the R8 and LGR4 simulations, their dimensionless fluctuation power spectra as functions of $kL$ have power law indices consistent with each other and even share similar normalizations. Because the power law spectral indices are sufficiently steep ($n_\delta, n_s > 2$), the power in these fluctuations is dominated by the largest scales. We speculate on the origin of the similarity across environments, as well as the broader implications of the steep power-law structure of the field in \autoref{sec:origins_implications}.

%%%%%%%%%%%%%%%%%%%%%%%%%%%%%%%%%%%%%%%%%%%%%%%%%%%%%%%%%%%%
\subsection{Two-Point Spatio-temporal Surface Density Statistics}
\label{sec:surface_density_spatiotemporal}
%%%%%%%%%%%%%%%%%%%%%%%%%%%%%%%%%%%%%%%%%%%%%%%%%%%%%%%%%%%%

So far, we have focused on the spatial features of the fluctuations, and averaged the relevant statistics over time as a proxy for an ensemble average. However, we are also able to probe the spatio-temporal structure of the fluctuations by studying the time-dependence of each $\bk$-mode (although by doing so, we no longer have an ensemble of snapshots over which to average). In particular, we can expand the spatial Fourier components of any quantity $Q$ (see \autoref{eq:fourier_details}) as a Fourier series in time\footnote{Because we are analyzing outputs over a finite temporal baseline $t_\mathrm{sim}$ (see \autoref{tab:sim_parameters}), we have $\omega \equiv 2\pi n / t_\mathrm{sim}$ for $n \in \mathbb{Z}$.} using
\begin{equation}
    \label{eq:fourier_time}
    Q_\bk(t) = \sum_\omega Q_\bk(\omega) e^{i\omega t}.
\end{equation}
The relevant two-point statistic in this context is then $|Q_\bk(\omega)|^2$, which we will refer to as the spatio-temporal ``power'' (despite the lack of ensemble averaging) throughout this section. To isolate the variation in this quantity due to the time-dependence, we can divide by the spatial power spectrum $P_Q(k)$ measured in \autoref{fig:power_k}. With this normalization, if the fluctuation field was nearly time-independent, then all the power would be clustered around $\omega = 0$ and independent of $k$, while if the fluctuations were completely uncorrelated in time, then the power would be uniform across both $k$ and $\omega$.

\begin{figure}
    \centering
    \includegraphics[width=0.48\textwidth]{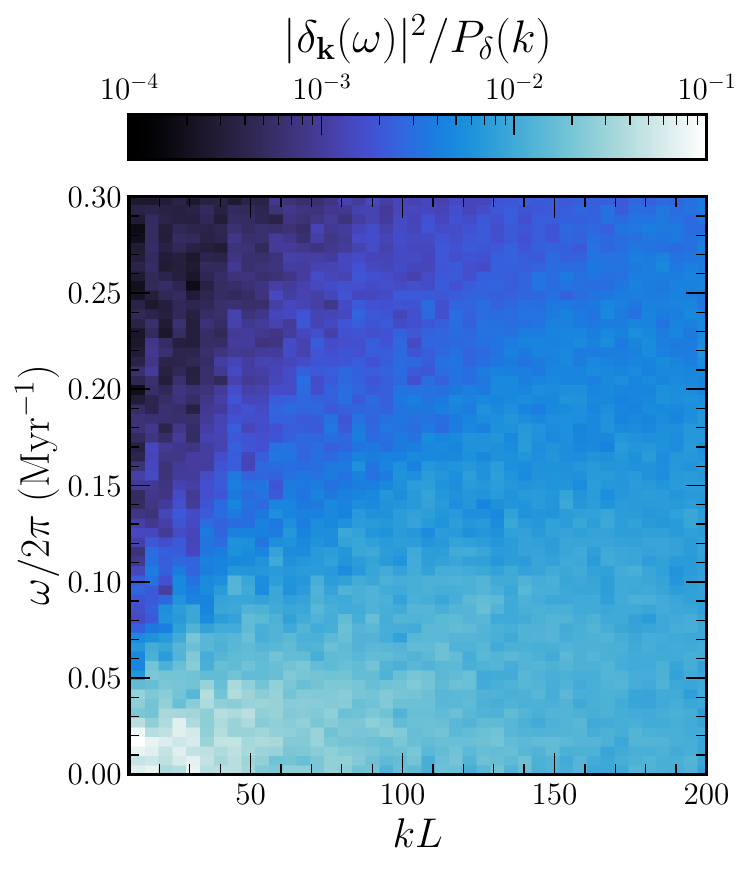}
    \caption{The spatio-temporal power spectrum of linear fluctuations, $|\delta_\bk(\omega)|^2$, normalized by the measured spatial power spectrum $P_\delta(k)$ shown in \autoref{fig:power_k} for the R8 simulation.}
    \label{fig:delta_omega_k_R8}
\end{figure}

In \autoref{fig:delta_omega_k_R8} we illustrate the structure of this normalized spatio-temporal power as a function of both $k$ and $\omega$ for the linear fluctuations ($Q = \delta$) in the R8 simulation; analogous plots for the logarithmic fluctuations ($Q = s$) or the LGR4 simulations show similar features. Qualitatively, it is clear that modes with different $k$ have different temporal structure. However, across all $\bk$-modes, the power peaks at $\omega = 0$ and decreases monotonically with $\omega$ thereafter. This indicates that the time-dependence of an individual $\bk$-mode's amplitude is decorrelated from other modes with distinct $k$---such correlation would lead to a peak at a nonzero characteristic frequency. Nonetheless, there are characteristic timescales in the problem: as we will illustrate below, these can be extracted from the width of the temporal power spectra.

To build a quantitative model for this spatio-temporal structure, we make the ansatz that the differing frequency-dependence of each $\bk$-mode is due to an effective ``dispersion relation'' for the underlying dynamics, i.e., that the functional behavior will be more uniform across $k$ when regarded as a function of
\begin{equation}
    \label{eq:omegatilde_def}
    \omegatilde \equiv \frac{\omega}{\omega_0(k)},
\end{equation}
where $\omega_0(k)$ is a scale-dependent characteristic frequency. We find that a good ansatz is
\begin{equation}
    \label{eq:omega_k}
    \omega_0(k)^2 = \frac{1}{\tau_0^2} + \veff^2 k^2;
\end{equation}
physically, $\tau_0$ represents a structural lifetime for the largest scales, while $\veff$ is an effective sound speed. Next, we decompose the spatio-temporal power as
\begin{equation}
    \label{eq:spatiotemporal_decomposition}
    |Q_\bk(\omega)|^2 = P_Q(k)\times \frac{\psi_Q(\omegatilde)}{\omega_0(k)t_\mathrm{sim}}
\end{equation}
for a function $\psi_Q(\omegatilde)$ that encodes the temporal structure of each $\bk$-mode (while $P_Q$ encodes the amplitude of the mode). Note that we have included an explicit extra factor of $\omega_0(k)t_\mathrm{sim}$ in the definition of $\psi_Q$ in comparison to the quantity plotted in \autoref{fig:delta_omega_k_R8} so that the normalization of $\psi_Q$ is set by
\begin{equation}
    \label{eq:psiQ_normalization}
    \int\frac{\md\omegatilde}{2\pi}\psi_Q(\omegatilde) = 1.
\end{equation}
This guarantees that the total \textit{spatio-temporal} power when integrated over both $\omega$ and $k$-space matches the total \textit{spatial} power integrated over $k$-space. In the model specified by \autoref{eq:omega_k} and \autoref{eq:spatiotemporal_decomposition}, the only free parameters associated with the temporal structure are therefore $\tau_0$ and $\veff$. We fit the values of these parameters from the simulation data by minimizing the variance between the values of the quantity $\psi_Q(\omegatilde)$ for different $k$ at each fixed $\omegatilde$ value.

\begin{table}
    \centering
    \begin{tabular}{ccc}
    \hline
    Sim & $\veff$ (km/s) & $\tau_0$ (Myr) \\
    (1) & (2) & (3) \\
    \hline
    \hline
    R8 & 12$\pm$2 & 5$\pm$2 \\
    LGR4 & 10$\pm$2 & 5$\pm$2 \\
    \hline
    \end{tabular}
    \caption{Measurements of parameters in the model for the temporal power spectra (see definitions in \autoref{eq:omega_k}) as fit to the simulation data shown in \autoref{fig:power_omegatilde}; the results are the same for both $Q = \delta$ and $Q = s$. Columns are as follows: (1) the simulation name, (2) the effective sound speed for small structures, and (3) the effective lifetime of large structures.}
    \label{tab:power_omega_measurements}
\end{table}

The results of this fitting procedure are summarized in \autoref{tab:power_omega_measurements}. We find that the best-fit values of $\tau_0$ and $\veff$ do not depend on whether the fluctuation in consideration is $Q = \delta$ or $Q = s$, and so we report one set of parameters for each simulation. In both simulations we measure $\tau_0\approx 5\,$Myr, which is similar to previous measures of the lifetime of dense structures in TIGRESS simulations \citep{Mao2020}, and is also similar to dispersal timescales under the action of feedback as measured in simulations focused on individual clouds \citep[e.g.][]{KimKimOstriker2018,KimOstrikerFilippova2021}. We also measure $\veff\approx 10-12\,\mathrm{km/s}$, which corresponds well with the effective velocity dispersion (including magnetic, thermal, and turbulent contributions) for warm and cold gas in both simulations as reported by \cite{Kim2023} (see their Table 5). Both $\tau_0$ and $\veff$ are somewhat weakly constrained by our analysis, but our constraints could certainly be strengthened with a proper ensemble of simulations over which to average the quantity $|Q_\bk(\omega)|^2$, finer time resolution in simulation outputs, and/or a longer temporal baseline of outputs.

\begin{figure*}
    \centering
    \includegraphics[width=0.95\textwidth]{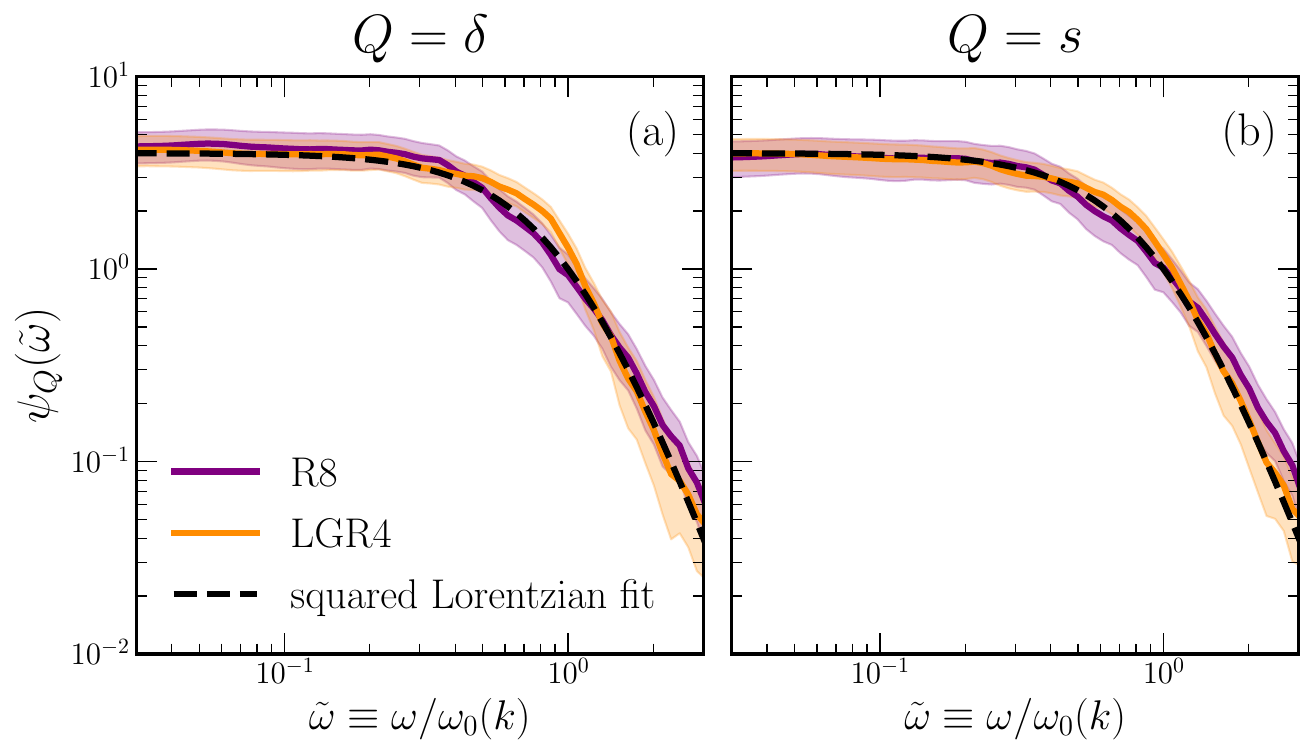}
    \caption{The spatio-temporal spectra for the linear (left, panel (a)) and logarithmic (right, panel (b)) surface density fluctuations for the R8 (purple) and LGR4 (orange) simulations with the spatial spectrum measured in \autoref{fig:power_k} factored out to produce the quantity $\psi_Q$ (see \autoref{eq:spatiotemporal_decomposition}), which encapsulates the frequency structure. We plot $\psi_Q$ as a function of the normalized frequency $\omegatilde \equiv \omega/\omega_0(k)$ (see \autoref{eq:omega_k}). The solid curves show an average of $\psi_Q(\omegatilde)$ over $k$ weighted by each $k$-value's contribution to the power $P_Q(k)$, and the shaded region shows a $\pm 1\sigma$ range around that average. The dashed curve is a fit using the ``squared-Lorentzian model'' (see \autoref{eq:Lorentzian_model}).}
    \label{fig:power_omegatilde}
\end{figure*}

In \autoref{fig:power_omegatilde}, we show the resulting curves of $\psi_Q(\omegatilde)$ for the linear (panel (a)) and logarithmic (panel (b)) fluctuations in the R8 (purple) and LGR4 (orange) simulations: the solid curves are spatial-power-weighted averages over $k$, while the shaded region indicates a $\pm 1\sigma$ range around this weighted average. The relatively small extent of the $\pm 1\sigma$ region indicates that the differences in temporal structure among distinct $\bk$-modes are well-captured by a ``dispersion relation'' of the form of \autoref{eq:omega_k}. Once again, despite the significant difference in galactic environment modeled by the R8 and LGR4 simulations, the two-point statistics presented in the Figure are remarkably similar. This suggests that the ISM's response to feedback processes that shape the spatio-temporal structure are statistically consistent across different simulations, once normalized by the appropriate ``dispersion relation'' for the environment.

We find that the measured curves $\psi_Q(\omegatilde)$ presented in \autoref{fig:power_omegatilde} are well-fit by a ``squared Lorentzian'' function of the form
\begin{equation}
    \label{eq:Lorentzian_model}
    (\psi_Q)_\mathrm{m}(\omegatilde) = \frac{4}{(1 + \omegatilde^2)^2};
\end{equation}
we overplot this fit as a black dashed curve in the Figure. The model \autoref{eq:Lorentzian_model} captures the plateau in power at the lowest $\omegatilde$ values accurately, and falls within the $\pm 1 \sigma$ interval around the $k$-averaged curve throughout the relevant $\omegatilde$ range, although it slightly underpredicts the power at the highest $\omegatilde$ in the R8 simulation and around $\omegatilde\approx 1$ in the LGR4 simulation.

The squared Lorentzian model is especially convenient because it corresponds to a two-sided exponential function for the fluctuation amplitudes in the time domain (in the limit where $\omega_0(k)t_\mathrm{sim}$ is large, which applies here). One interpretation\footnote{Of course, many other amplitude models $|Q_\bk(t)|_\mathrm{m}$ can yield the same spatio-temporal power spectrum---we just present this one heuristic example in order to build intuition.} of the temporal structure given by the model is therefore that the amplitudes of each $\bk$-mode are of the form
\begin{equation}
    \label{eq:Q_k_t_model}
    |Q_\bk(t)|_\mathrm{m} \propto \sum_{j} e^{-\omega_0(k)|t - t_j|},
\end{equation}
where each $t_j$ is a ``peak time'' for the $j$-th structure at scale $k$ drawn uniformly from $[0, t_\mathrm{sim}]$. The Fourier transforms of individual terms in \autoref{eq:Q_k_t_model} are proportional to $(\omega_0(k)^2 + \omega^2)^{-1}$, so that the total power from  a set of such fluctuations would yield a squared Lorentzian for $\psi_Q$, as in the model of \autoref{eq:Lorentzian_model}. Thus, we may interpret the quantity $2\omega_0(k)^{-1}$ as a characteristic lifetime for a structure at scale $k$, which ranges from $2\tau_0 \sim 10\,$Myr at the largest scales to $2/(k\veff) \sim 0.3\,\mathrm{Myr}\times (k/(2\pi/10\,\mathrm{pc}))^{-1}$ at the smallest scales. Physically, these limits are simply the statement that at scales $\lambda\lesssim 2\pi \veff \tau_0 \sim 300\,\mathrm{pc}$, the lifetime of a structure is of order the dynamical timescale associated with MHD waves and trans-sonic turbulence, while for larger scales, structure lifetimes are determined by feedback processes. We also note that these structure lifetimes are substantially shorter than the shear timescale $\sim (q\Omega)^{-1}\sim 30\,$Myr, so turbulence and feedback play a more significant role in dispersing structures than shear.

Overall, compiling the results presented throughout \autoref{sec:surface_density}, our model for the spatio-temporal power spectrum of the fluctuations takes the form 
\begin{align}
    \label{eq:spatiotemporal_model_combined}
    |Q_\bk(\omega)|_\mathrm{m}^2 = & P_{Q,0}(kL)^{-n_Q} \nn \\
    & \times \frac{4}{\omega_0(k) t_\mathrm{sim}}\left(1 + \left(\frac{\omega}{\omega_0(k)}\right)^2\right)^{-2},
\end{align}
with parameters as listed in \autoref{tab:power_k_measurements} (for both $Q=\delta\equiv \Sigma/\overline{\Sigma} - 1$ and $Q=s=\ln(1+\delta)$) and \autoref{tab:power_omega_measurements}. 

We emphasize that \autoref{eq:spatiotemporal_model_combined} provides a model for the \textit{amplitudes} of individual fluctuation components. In order to realize a single snapshot (or a temporal sequence of snapshots), the relative \textit{phases} of individual components must also be assigned. If the true phases were random, then since $s$ is well-described by a Gaussian pdf (see \autoref{fig:s_pdfs}), realizations of the surface density field could be generated e.g., by using standard algorithms to produce Gaussian random fields (GRFs) by setting $Q = s$ in \autoref{eq:spatiotemporal_model_combined}. We outline two methods for generating surface density fields in this way in \autoref{sec:generating_fluctuations}. However, as we shall show in \autoref{sec:nongaussianity_anisotropy}, this procedure does \textit{not} in fact succeed in recovering the true surface density, because the assumption of random phases is invalid for the nonlinear structure in the ISM.

For application to the influence of ISM perturbations on secular evolution of the stellar distribution function, however, it may be the case that only the relative power of different $(\omega,k)$ components (and not their relative phases) is relevant \citep{BinneyLacey1988}. As we will describe in the following section, setting $Q = \delta$ in the expression above provides a direct input for the volume density and gravitational potential fluctuation power spectra (see, e.g., \autoref{eq:power_phi_Sigma} below), which may be used in analytical studies of the perturbing influence of the ISM.

%%%%%%%%%%%%%%%%%%%%%%%%%%%%%%%%%%%%%%%%%%%%%%%%%%%%%%%%%%%%
%%%%%%%%%%%%%%%%%%%%%%%%%%%%%%%%%%%%%%%%%%%%%%%%%%%%%%%%%%%%
\section{Volume Density and Gravitational Potential}
\label{sec:rho_phi}
%%%%%%%%%%%%%%%%%%%%%%%%%%%%%%%%%%%%%%%%%%%%%%%%%%%%%%%%%%%%
%%%%%%%%%%%%%%%%%%%%%%%%%%%%%%%%%%%%%%%%%%%%%%%%%%%%%%%%%%%%

Having analyzed the (2D) surface density in each simulation, we now turn to characterizing the (3D) volume density and gravitational potential. Without loss of generality, we can decompose the volume density $\rho$ into its average at each $z, t$ and a fluctuation $\delta\rho$, i.e.
\begin{equation}
    \rho(\bx, z, t) \equiv \overline{\rho}(z, t) + \delta\rho(\bx, z, t),
\end{equation}
where as before the overline indicates averaging over the planar variables $\bx$. Similarly decomposing the potential,
\begin{equation}
    \phi(\bx, z, t) \equiv \overline{\phi}(z, t) + \delta\phi(\bx, z, t),
\end{equation}
the Poisson equation splits into
\begin{equation}
    \label{eq:poisson}
    \frac{\partial^2\overline{\phi}}{\partial z^2} = 4\pi G \overline{\rho} \,\,\,\,\,\,\,\,\,\mathrm{and}\,\,\,\,\,\,\,\,\, \nabla^2\delta\phi = 4\pi G \delta\rho,
\end{equation}
for the mean and and fluctuating quantities respectively. Note that the mean density and potential are exactly the $|\mathbf{k}|=k = 0$ components of planar Fourier expansions of each quantity, while the $k>0$ contributions represent fluctuations. Characterizing either the mean or fluctuating component of $\rho$ or $\phi$ yields a characterization of the corresponding component of the other. We will find it most convenient to characterize the mean density $\overline{\rho}$ directly; in contrast, we will focus on analyzing the fluctuations in the potential $\delta\phi$ to serve as a direct input for stellar dynamical studies.

\begin{figure}
    \centering
    \includegraphics[width=0.48\textwidth]{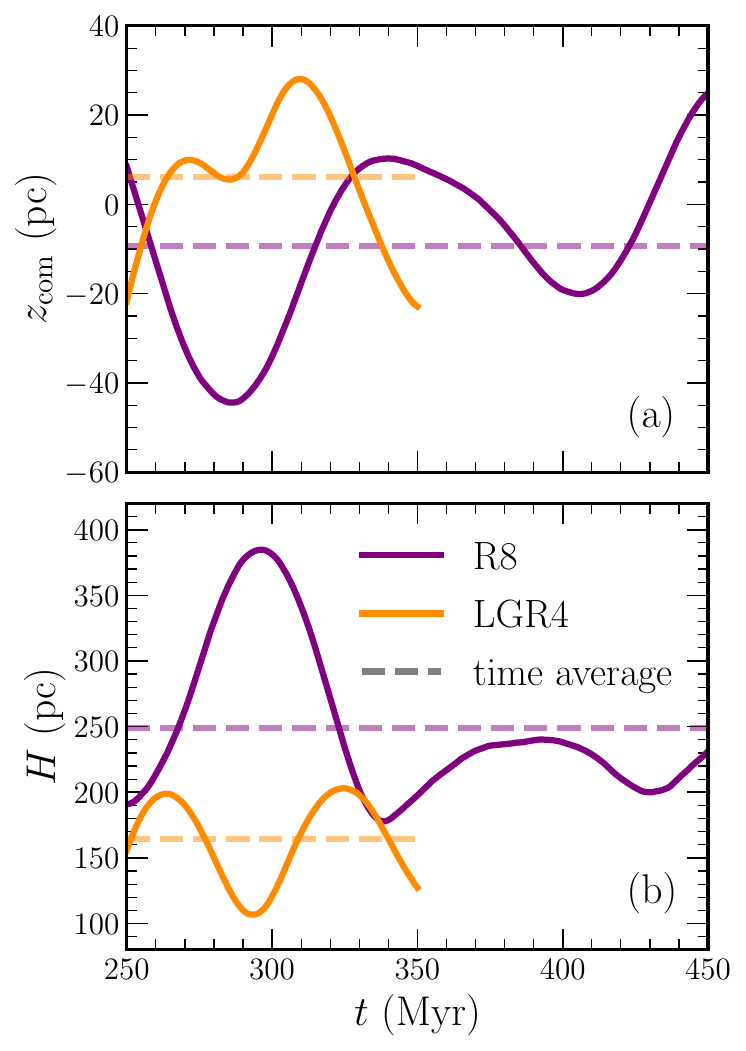}
    \caption{The gas center of mass $\zcom$ (panel a, \autoref{eq:zcom}) and root-mean-square thickness $H$ (panel b, \autoref{eq:H}) as a function of time for the R8 (purple) and LGR4 (orange) simulations. The light purple and orange dashed lines show the time-averaged values for reference.}
    \label{fig:zcom_H}
\end{figure}

\begin{table}
    \centering
    \begin{tabular}{cccc}
    \hline
     & $\langle H \rangle$ & $(\langle \zcom^2 \rangle - \langle\zcom\rangle^2)^{1/2}$ & $(\langle H^2 \rangle - \langle H \rangle^2)^{1/2}$ \\
    Sim & (pc) & (pc) & (pc) \\
    (1) & (2) & (3) & (4) \\
    \hline
    \hline 
    R8 & 249 & 18 & 59\\
    LGR4 & 165 & 14 & 31 \\
    \hline
    \end{tabular}
    \caption{A summary of the mean and amplitude of oscillations in $\zcom$ and $H$ shown in \autoref{fig:zcom_H}, with columns as follows: (1) the simulation name, (2) the average value of $H$, (3) the standard deviation of $\zcom$ over time, and (4) the standard deviation of $H$ over time.}
    \label{tab:zcom_H_values}
\end{table}

To extend our analysis of the 2D fluctuations from the previous section to the 3D quantities $\rho$ and $\phi$, we must develop a model for the vertical structure of gas. Such models are conveniently parametrized by their first two central moments: their vertical center of mass position,
\begin{equation}
    \label{eq:zcom}
    \zcom(t) \equiv \frac{\int \md z \, \overline{\rho}(z, t)z}{\int \md z \, \overline{\rho}(z, t)},
\end{equation}
and their root-mean-square thickness, 
\begin{align}
    \label{eq:H}
    H(t) & \equiv \left(\frac{\int \md z \, \overline{\rho}(z, t)[z-\zcom(t)]^2}{\int \md z \, \overline{\rho}(z, t)}\right)^{1/2}.
\end{align}
To illustrate the behavior of these quantities we present \autoref{fig:zcom_H}, which shows $\zcom$ (panel (a)) and $H$ (panel (b)) as functions of time for both the R8 (purple) and LGR4 (orange) simulations, with the time-averaged values of each quantity shown in a lighter dashed line.

Although the midplane of the gravitational potential set by the dark matter halo and the stellar disk (\autoref{eq:external_potential}) is fixed to $z = 0$, panel (a) shows that the center of mass $\zcom(t)$ of the gas oscillates about this plane by several tens of pc in both simulations, as a result of asymmetry in feedback processes. Similarly, if there were only microphysical pressure supporting the disk against vertical gravity, in a steady-state the thickness would settle to a constant equilibrium value, but panel (b) demonstrates that $H$ varies by almost a factor of two in both simulations, due to the important role of large-scale, time-varying turbulence in supporting the disk \citep[see][and references therein]{ostriker2022}\footnote{Although both $\zcom(t)$ and $H$ have significant variations, simulations (at lower resolution than those shown here) show that a well-defined quasi-steady-state is maintained over nearly 1\,Gyr timescales -- see, e.g., Figure 1 of \citealt{Kim2023}.}. In both simulations, oscillations of $z_\mathrm{com}$ and $H$ occur on vertical gravitational oscillation periods of $\sim 80\,$Myr. For reference, we record the time-averaged value of the thickness and typical amplitude of the oscillations of both $\zcom$ and $H$ (in the form of their standard deviations as a function of time) in \autoref{tab:zcom_H_values}. In the R8 simulation, a sharp increase in $H$ and a decrease in $\zcom$ is apparent around $t\approx 300\,$Myr; this is due to a significant burst in feedback that occurs at that time.

We will use these properties of the vertical structure to develop models for the density and potential; as before, we denote model quantities with a subscript $\mathrm{m}$ to distinguish them from those measured in the simulations. We make two empirically motivated ans\"atze to build the models:
\begin{enumerate}[label=(\roman*)]
    \item We model the density as separable in space, i.e., we assume that we can factor it as 
        \begin{equation}
            \label{eq:separable_density}
            \rho_\mathrm{m}(\bx, z, t) \equiv \overline{\rho}_\mathrm{m}(z, t)(1 + \delta(\bx, t)),
        \end{equation}
    where $\delta$ is the linear surface density fluctuation (\autoref{eq:delta_definition}). Here, $\overline{\rho}_\mathrm{m}$ is a model for the vertical density profile which we will fit to the measured planar-averaged density $\overline{\rho}$, and we are assuming that this model for the vertical structure extends also to the fluctuations, so that
        \begin{equation}
            \label{eq:deltarho_model}
            \delta\rho_\mathrm{m}(\bx, z, t) = \overline{\rho}_\mathrm{m}(z, t) \delta(\bx, t).
        \end{equation}
    That is, the model simply ``inflates'' all surface density structures to have the same vertical profile. Note that because the purpose of our models is to connect the fluctuations in the surface density field to those in the volume density and potential, we will restrict our approximations to the vertical quantities, and use the real, measured planar quantities (e.g., $\delta(\bx, t)$ and $\overline{\Sigma}(t)$) throughout.
    \item We assume that the time-variability in the vertical structure of the gas disk is entirely dominated by the variation in $\zcom$ and $H$, in the combination
        \begin{equation}
            \label{eq:ztilde}
            \ztilde \equiv \frac{z - \zcom(t)}{H(t)},
        \end{equation}
    so that when fitting our model we will assume e.g., $\overline{\rho}_\mathrm{m} = \overline{\rho}_\mathrm{m}(\ztilde(t))$, with no additional explicit time-dependence.
\end{enumerate}
We find that these simplifying approximations give a reasonable description of the vertical structure, but they are certainly not comprehensive; we address some of their shortcomings in the following subsections when relevant.

In \autoref{sec:vertical_k0} we begin by developing models for the vertical density and potential averaged over the in-plane variables ($k = 0$), and in \autoref{sec:vertical_kgtr} we extend these models to characterize the potential fluctuations as a function of horizontal position $\mathbf{x}$ ($k>0$).

%%%%%%%%%%%%%%%%%%%%%%%%%%%%%%%%%%%%%%%%%%%%%%%%%%%%%%%%%%%%
\subsection{Mean Vertical Structure}
\label{sec:vertical_k0}
%%%%%%%%%%%%%%%%%%%%%%%%%%%%%%%%%%%%%%%%%%%%%%%%%%%%%%%%%%%%

Without loss of generality, we can always write the mean volume density as
\begin{equation}
    \label{eq:zeta_def}
    \overline{\rho}(z, t) = \frac{\overline{\Sigma}(t)}{2h(t)}\zeta(z, t),
\end{equation}
where $\zeta$ is a dimensionless function describing the vertical profile with $\zeta(z = \zcom(t), t) = 1$, and for normalization we must have
\begin{equation}
    \label{eq:h_def}
    h(t) \equiv \frac{1}{2}\int \md z \, \zeta(z, t),
\end{equation}
since $\overline{\rho}$ must integrate to the mean surface density $\overline{\Sigma}$. Our model for the mean vertical density will take the form
\begin{equation}
    \label{eq:rhobar_model}
    \overline{\rho}_\mathrm{m}(z, t) = \frac{\overline{\Sigma}(t)}{2h_\mathrm{m}(t)}\zeta_\mathrm{m}(\ztilde),
\end{equation}
where we will fit a function $\zeta_\mathrm{m}$ to the curve $\zeta$ measured in the simulations (and $h_\mathrm{m}$ follows from that fit as in \autoref{eq:h_def}).

\begin{figure*}
    \centering
    \includegraphics[width=0.95\textwidth]{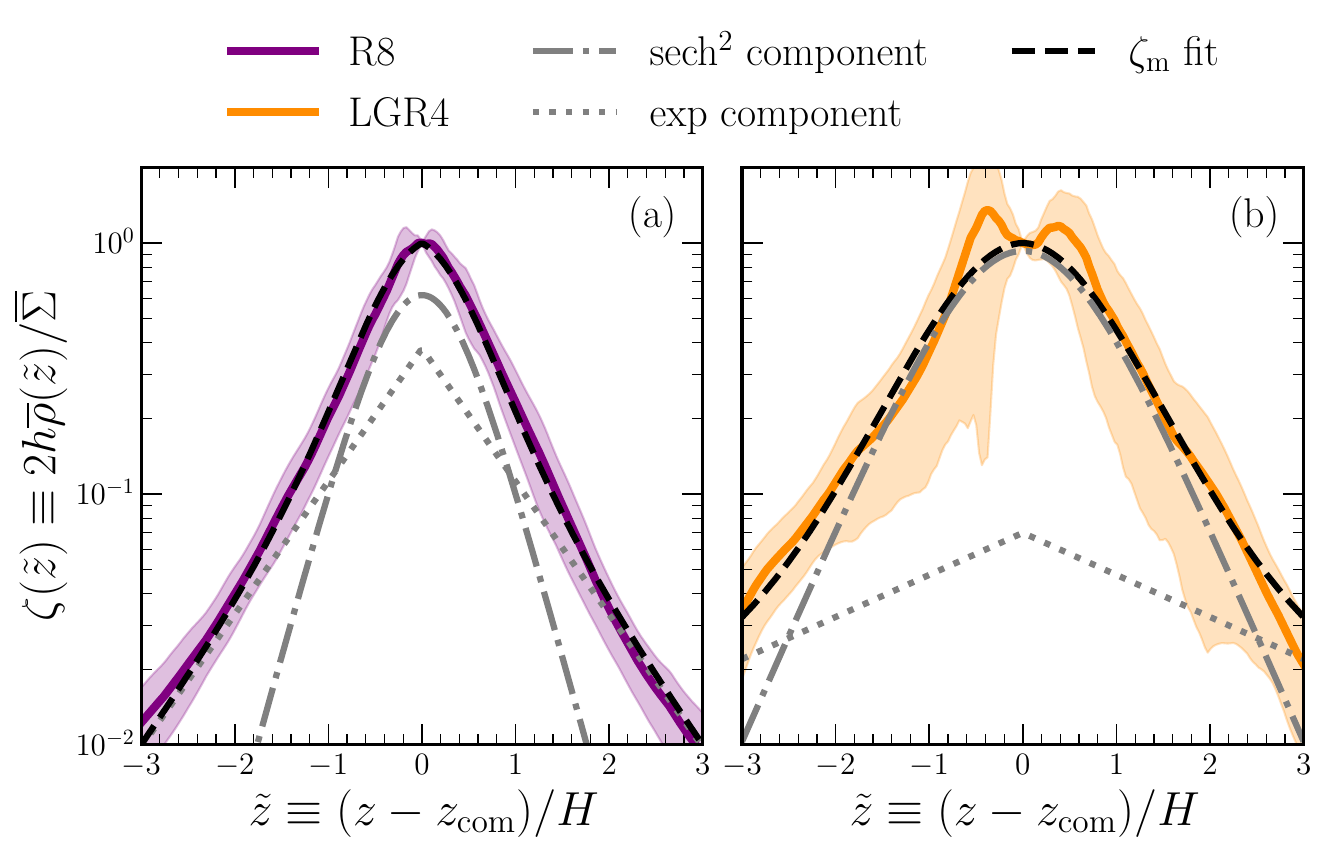}
    \caption{The time average of the vertical profile $\zeta$ (defined in \autoref{eq:zeta_def}) for the horizontally averaged ($k = 0$) density in the R8 (purple, panel (a)) and LGR4 (orange, panel (b)) simulations. The shaded regions indicate $\pm 1\sigma$ error bars around the mean, which is indicated by the solid curve. In each panel, the fitted model $\zeta_\mathrm{m}$ constructed from a weighted combination of sech$^2$ and exponential components is shown by a black dashed curve, with the individual sech$^2$ and exponential components shown in grey dot-dashed and dotted curves respectively (see \autoref{eq:zeta_model}).}
    \label{fig:zeta}
\end{figure*}

In \autoref{fig:zeta} we plot the vertical profiles $\zeta \equiv 2h\overline{\rho}/\overline{\Sigma}$ as a function of $\ztilde$ measured in the R8 (purple) and LGR4 (orange) simulations, with the time averages shown as solid curves and the $\pm 1\sigma$ range about each average shown as a shaded region. For the R8 simulation, the relatively small deviations from the time-averaged profile indicates that there is indeed little vertical time-dependence outside of that encapsulated in $\ztilde$. In the LGR4 simulation, the stronger feedback produces more significant variability in moments of the vertical profile beyond $\zcom(t)$ and $H(t)$, resulting in greater snapshot-to-snapshot variation from the mean profile. The stronger feedback is also evident in the local maxima in the time-averaged curve at $|\ztilde|\approx 0.5$, which are produced when gas is consistently blown away from the midplane.

To model these distributions, we fit functions of the form
\begin{equation}
    \label{eq:zeta_model}
    \zeta_\mathrm{m}(\ztilde) = (1-w)\sech^2\left(\frac{\ztilde}{\alpha_\mathrm{s}}\right) + we^{-|\ztilde|/\alpha_\mathrm{e}},
\end{equation}
i.e., a weighted mixture of $\sech^2$ and exponential profiles, where the parameter $w$ sets the relative weight of each component, $\alpha_\mathrm{s}$ sets the relative thickness of the $\sech^2$, and $\alpha_\mathrm{e}$ sets the relative thickness of the exponential. For reference, for this form of $\zeta_\mathrm{m}$, the thickness of the gas layer is characterized by (see \autoref{eq:h_def})
\begin{equation} 
    \label{eq:h_sech2exp}
    h_\mathrm{m}(t) = \big[(1-w)\alpha_\mathrm{s} + w\alpha_\mathrm{e}\big]H(t) \equiv \alpha_w H(t).
\end{equation}
The functional form of \autoref{eq:zeta_model} is physically well motivated. First, exponential density profiles are produced assuming hydrostatic equilibrium for an isothermal gas in a constant gravitational field, as we expect to be true at large $|\ztilde|$ (since only the hottest gas can extend far beyond the midplane, and the gravitational field of an infinite plane is independent of height). Second, $\sech^2$ density profiles are produced assuming hydrostatic equilibrium for an isothermal gas when the gas produces the confining gravitational field, as we expect to be true at small $|\ztilde|$ \citep{Spitzer1942}\footnote{The gas disk's vertical structure also responds to the stellar disk's vertical gravity, which can lead to a more Gaussian-like gas profile if the stellar disk is much thicker than that of the gas, but that is not the case for the current models.}. Both the exponential and $\sech^2$ profiles are commonly applied in fitting observations (see, e.g., \citealt{JohannessonPorterMoskalenko2018}), and are especially convenient because they allow for analytic solutions to the Poisson equation. In particular, the mean potential corresponding to the density profile specified by \autoref{eq:rhobar_model} and \autoref{eq:zeta_model} is
\begin{align}
    \label{eq:phi_k0}
    \overline{\phi}_\mathrm{m}(z, t) = \phi_0(t) & \bigg[(1-w)\alpha_\mathrm{s}^2\ln\left(\cosh\left(\frac{\ztilde}{\alpha_\mathrm{s}}\right)\right) \nn \\
    & + w\alpha_\mathrm{e}^2(e^{-|\ztilde|/\alpha_\mathrm{e}} - 1) + w\alpha_\mathrm{e}|\ztilde|\bigg],
\end{align}
where
\begin{align}
    \label{eq:phi_normalization}
    \phi_0(t) & \equiv \frac{2\pi}{\alpha_w}G\overline{\Sigma}(t)H(t) \nn \\
    & =\frac{54 \,(\mathrm{km/s})^2}{\alpha_w} \,\left(\frac{\overline{\Sigma}(t)}{10\,\Msun\mathrm{\, pc}^{-2}}\right)\left(\frac{H(t)}{200\,\mathrm{pc}}\right).
\end{align}
The characteristic velocities associated with the mean potential of the gas disk are $\sqrt{\phi_0} \sim 10-15\,\mathrm{km/s}$ for the present models.

\autoref{tab:zeta_measurements} summarizes the best-fit parameters $(w, \alpha_\mathrm{s}, \alpha_\mathrm{e})$ for each simulation, and also provides the ratio of the effective thicknesses to the scale heights $h_\mathrm{m}/H \equiv \alpha_w$ for each model. The fits are overplotted as black dashed curves in \autoref{fig:zeta}, with the contributions of the individual $\sech^2$ and exponential components in grey dot-dashed and dotted curves respectively. The top row of \autoref{fig:phi0_phik_z} below compares the measured mean vertical potential $\overline{\phi}$ to the model $\overline{\phi}_\mathrm{m}$ produced using \autoref{eq:phi_k0}.

\begin{table}
    \centering
    \begin{tabular}{ccccc}
    \hline
    Sim & $w$ & $\alpha_\mathrm{s}$ & $\alpha_\mathrm{e}$ & $\alpha_w$ \\
    (1) & (2) & (3) & (4) & (5) \\
    \hline
    \hline 
    R8 & 0.38$\pm$0.02 & 0.64$\pm$0.01 & 0.82$\pm$0.02 & 0.71$\pm$0.01 \\
    LGR4 & 0.07$\pm$0.06 & 1.02$\pm$0.02 & 2.6$\pm$1.8 & 1.1$\pm$0.16 \\
    \hline
    \end{tabular}
    \caption{The model parameters for the fits to the $\zeta(\ztilde)$ profiles shown in \autoref{fig:zeta}, with columns as follows: (1) the simulation name, (2) the weight of the exponential term $w$, (3) the scale parameter of the $\sech^2$ term $\alpha_\mathrm{s}$, (4) the scale parameter of the exponential term $\alpha_\mathrm{e}$, and (5) the ratio of the model's effective thickness to the mass-weighted RMS disk thickness, $\alpha_w = h/H$.}
    \label{tab:zeta_measurements}
\end{table}

We find that the fit is excellent across all $\ztilde$ for the R8 simulation. For LGR4, while the peaks at $|\ztilde| \approx 0.5$ are not captured, the model is entirely consistent with the time-averaged density and potential profiles across most other $\ztilde$ values. In both simulations, the profile is weighted more heavily toward the $\sech^2$ component (the fitted $w < 1/2$), and as anticipated the exponential component only begins to dominate several scale-heights above the midplane.

%%%%%%%%%%%%%%%%%%%%%%%%%%%%%%%%%%%%%%%%%%%%%%%%%%%%%%%%%%%%
\subsection{Potential Fluctuation Vertical Structure}
\label{sec:vertical_kgtr}
%%%%%%%%%%%%%%%%%%%%%%%%%%%%%%%%%%%%%%%%%%%%%%%%%%%%%%%%%%%%

Expanding the gravitational potential and surface density fluctuations in $\bk$ as in \autoref{eq:fourier_details}, from the Poisson equation (\autoref{eq:poisson}), we find that the potential fluctuations for our model separable density given in \autoref{eq:separable_density} have Fourier components
\begin{equation}
    \label{eq:phi_Sigma_k_def}
    (\delta\phi_\bk)_\mathrm{m}(z, t) = -\frac{\pi G \overline{\Sigma}(t) \delta_\bk(t)}{kh_\mathrm{m}(t)} \int \md z' e^{-k|z-z'|}\zeta_\mathrm{m}(z', t),
\end{equation}
where $\delta_\bk$ are the Fourier components of the linear surface density fluctuation (\autoref{eq:delta_definition}). Substituting the model vertical profile $\zeta_\mathrm{m}$ given in \autoref{eq:zeta_model} and applying our ansatz that the time-variability is dominated by changes in $\zcom(t)$ and $H(t)$, we can recast this expression as
\begin{equation}
    \label{eq:phi_Sigma_k}
    (\delta\phi_\bk)_\mathrm{m}(z, t) = -\phi_0(t) \frac{\delta_\bk(t)}{\ktilde} \chi_\mathrm{m}(\ktilde, \ztilde),
\end{equation}
where we have defined a dimensionless wavevector
\begin{equation}
    \label{eq:ktilde}
    \ktilde \equiv kH(t),
\end{equation}
and absorbed the model vertical profile into a new function
\begin{align}
    \label{eq:chi_def}
    \chi_\mathrm{m}(\ktilde, \ztilde) \equiv \frac{1}{2} & (1-w)\alpha_\mathrm{s}\int\md u \, e^{-\ktilde|\ztilde-\alpha_\mathrm{s} u|} \sech^2(u) \nn \\
    & + w\alpha_\mathrm{e} \left(\frac{(\alpha_\mathrm{e}\ktilde) e^{-|\ztilde|/\alpha_\mathrm{e}} - e^{-\ktilde|\ztilde|}  }{(\alpha_\mathrm{e}\ktilde)^2 - 1}\right).
\end{align}
We will use the exact expression in \autoref{eq:chi_def} when evaluating our models throughout the remainder of this paper, but for convenient reference, an approximate analytic expression with the correct limiting behavior is
\begin{align}
    \label{eq:chi_approx}
    \chi_\mathrm{m}(\ktilde, \ztilde) \approx & (1-w) \alpha_\mathrm{s}\frac{\sech^2(\ztilde/\alpha_\mathrm{s})}{\left(\sech^{2p}(\ztilde/\alpha_\mathrm{s}) + (\alpha_\mathrm{s}\ktilde)^{p}\right)^{1/p}} \nn \\
    & + w\alpha_\mathrm{e} \left(\frac{(\alpha_\mathrm{e}\ktilde) e^{-|\ztilde|/\alpha_\mathrm{e}} - e^{-\ktilde|\ztilde|}  }{(\alpha_\mathrm{e}\ktilde)^2 - 1}\right),
\end{align}
where the parameter $p > 0$ controls the transition between the limits. Setting $p = 3/2$ produces an approximation that is accurate to within $1\%$ when $\ztilde = 0$ and $20\%$ across all relevant model parameters and $(\ktilde, \ztilde)$ values; setting $p = 1$ instead yields errors of at most $12\%$ when $\ztilde = 0$ and at most $25\%$ over the relevant parameters. Choosing $p = 1$ for simplicity, the potential fluctuations within the disk ($|\ztilde| \ll 1$) are then given approximately by 
\begin{equation}
    \label{eq:midplane_phik_approx}
    (\delta\phi_\mathbf{k})_\mathrm{m} \approx - \phi_0(t)\frac{\delta_\mathbf{k}(t)}{\ktilde}\left(\frac{(1-w)\alpha_\mathrm{s}}{1 +\alpha_\mathrm{s}\ktilde} +\frac{w\alpha_\mathrm{e}}{1+\alpha_\mathrm{e}\ktilde}\right).
\end{equation}

Overall, our model allows us to relate the power spectrum of potential fluctuations to the power spectrum of linear surface density fluctuations we studied in \autoref{sec:surface_density_spectra}, 
\begin{equation}
    \label{eq:power_phi_Sigma}
    (P_{\delta\phi})_\mathrm{m}(k, z, t) = \phi_0(t)^2 \frac{P_\delta(k, t)\chi_\mathrm{m}^2(\ktilde, \ztilde)}{\ktilde^{2}},
\end{equation}
and we anticipate that the quantity $(P_{\delta\phi})_\mathrm{m}/\phi_0^2$ will have the same temporal dependence as the surface density fluctuations, as discussed in \autoref{sec:surface_density_spatiotemporal}.

\begin{figure*}
    \centering
    \includegraphics[width=0.95\textwidth]{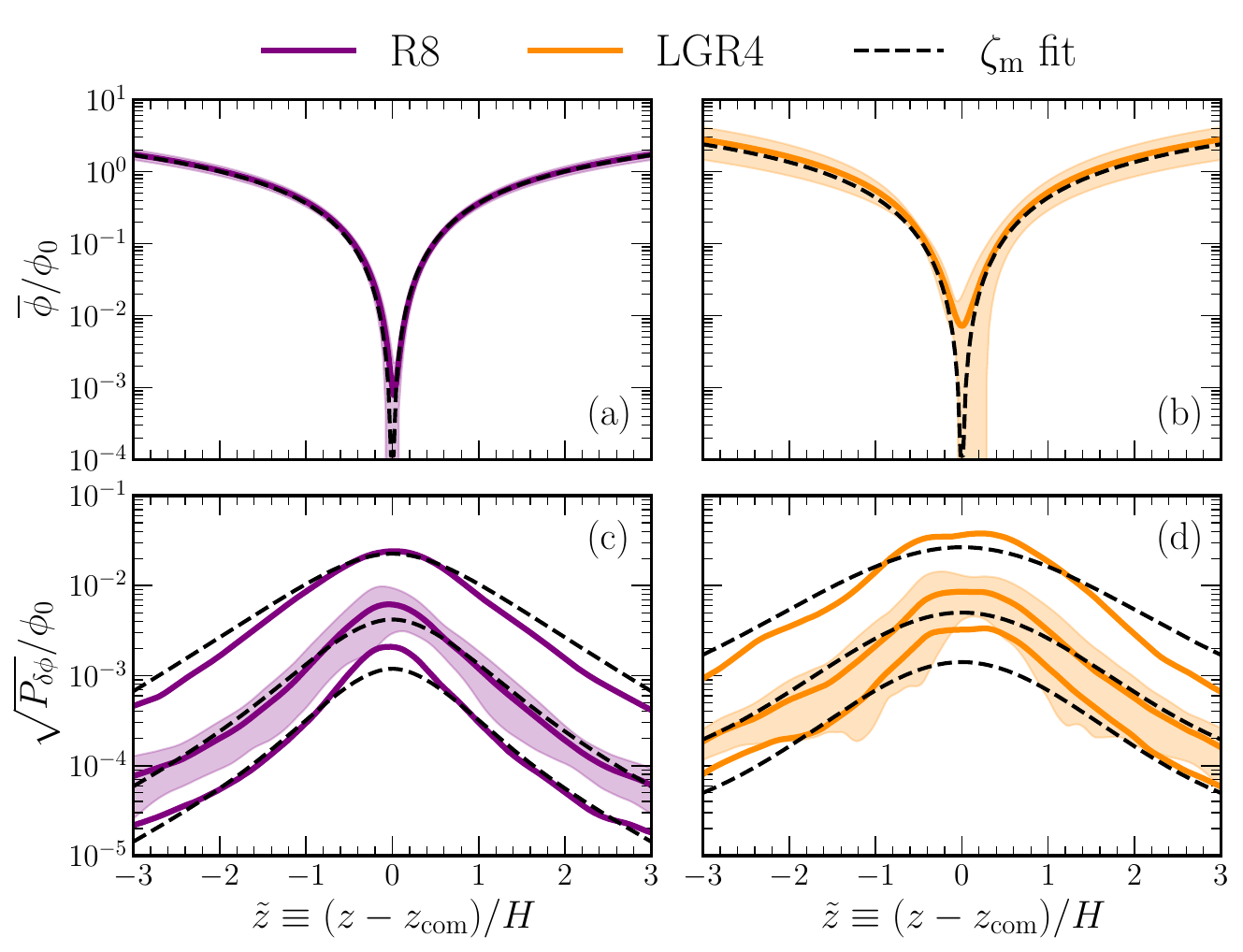}
    \caption{The $k = 0$ component of the potential fluctuations as a function of $\ztilde$ (top row) as well as selected $k > 0$ components as a function of $\ztilde$ at fixed $\ktilde = \{2, 4, 6\}$ (bottom row, with $\ktilde$ increasing from top to bottom in each panel) for the R8 (left, purple) and LGR4 (right, orange) simulations. For the $k > 0$ components, we show the square root of the power spectrum of potential fluctuations, which is the root-mean-square of each $\delta\phi_\bk$. We normalize both rows by $\phi_0$; see \autoref{eq:phi_normalization} (and note the different vertical axis limits). For clarity, we only show the standard deviation (shaded region) around the time-averaged quantity (solid curve) for one value of $\ktilde$, but the fractional variations are comparable across all the curves. In each panel, the black dashed curve shows the result assuming a separable density with the fitted vertical profile shown in \autoref{fig:zeta}, which leads to the vertical models given in \autoref{eq:phi_k0} and \autoref{eq:power_phi_Sigma}.}
    \label{fig:phi0_phik_z}
\end{figure*}

The bottom row of \autoref{fig:phi0_phik_z} shows the measured root-mean-square of the potential fluctuation Fourier amplitudes, $\sqrt{P_{\delta\phi}}$ (bottom row), normalized by $\phi_0$, as functions of $\ztilde$ for the R8 (purple, left column) and LGR4 (orange, right column) simulations. For the potential fluctuations, we show slices at $\ktilde = 2$, $\ktilde = 4$, and $\ktilde = 6$ (top, middle, and bottom curves respectively); these correspond approximately to $kL \sim$ 10, 20, and 30 and were chosen as representative values capturing the first $\sim 1$\,dex of surface density fluctuation power (see \autoref{fig:power_k}). We show the time-averaged values as solid curves, and for clarity we only indicate the $\pm 1\sigma$ deviations around the average as shaded regions around one $\ktilde$ value; the fractional variations are similar for each $\ktilde$. The relatively small fractional variations about the mean for the R8 simulation indicate that much of the time-dependence is contained in the evolution of the gas disk's center of mass and thickness; as in \autoref{fig:zeta}, the LGR4 simulation exhibits larger deviations from the mean. In both simulations, comparing the mean to the fluctuations, we note that the typical amplitude of each fluctuation component is significantly less than the amplitude of the mean for all $|\ztilde| \gtrsim 1$. This indicates that while the potential fluctuations are extremely dynamically important near the midplane (where they can be key drivers of the evolution of vertical structure in the stellar disk), they may be negligible beyond one scale height away from it.

Overplotted as black dashed curves in all panels are the predictions of the model, given by \autoref{eq:phi_k0} for the mean and \autoref{eq:power_phi_Sigma} for the fluctuations. As expected given the excellent fit to the mean vertical profile shown in \autoref{fig:zeta}, the model matches the measured mean vertical potential quite accurately. For each $\ktilde$ value of the fluctuations, the model curve typically falls within the $\pm1\sigma$ range around the measured time-average curve, although for the smallest $\ktilde$ values, the power is somewhat overpredicted at $|\ztilde| \gtrsim 1$, and for the larger $\ktilde$ values, the power is somewhat underpredicted for $|\ztilde| \lesssim 1$. These discrepancies can be attributed to the fact that the underlying gas volume density fluctuations are actually more concentrated near the midplane than the mean vertical density profile. In the stellar-dynamical context, this difference may be consequential for vertical scattering processes experienced by stars in the thin disk, with small vertical excursions from the midplane compared to the gas scale height. Therefore, we describe an extension to our model allowing for the fluctuation vertical profile to differ from the mean in \autoref{sec:zetagtr}. Nevertheless, we find that even our simple model with a single function $\zeta_\mathrm{m}$ for both the $k = 0$ and $k > 0$ components captures many of the essential features of the spectrum of potential fluctuations.

\begin{figure*}
    \centering
    \includegraphics[width=0.95\textwidth]{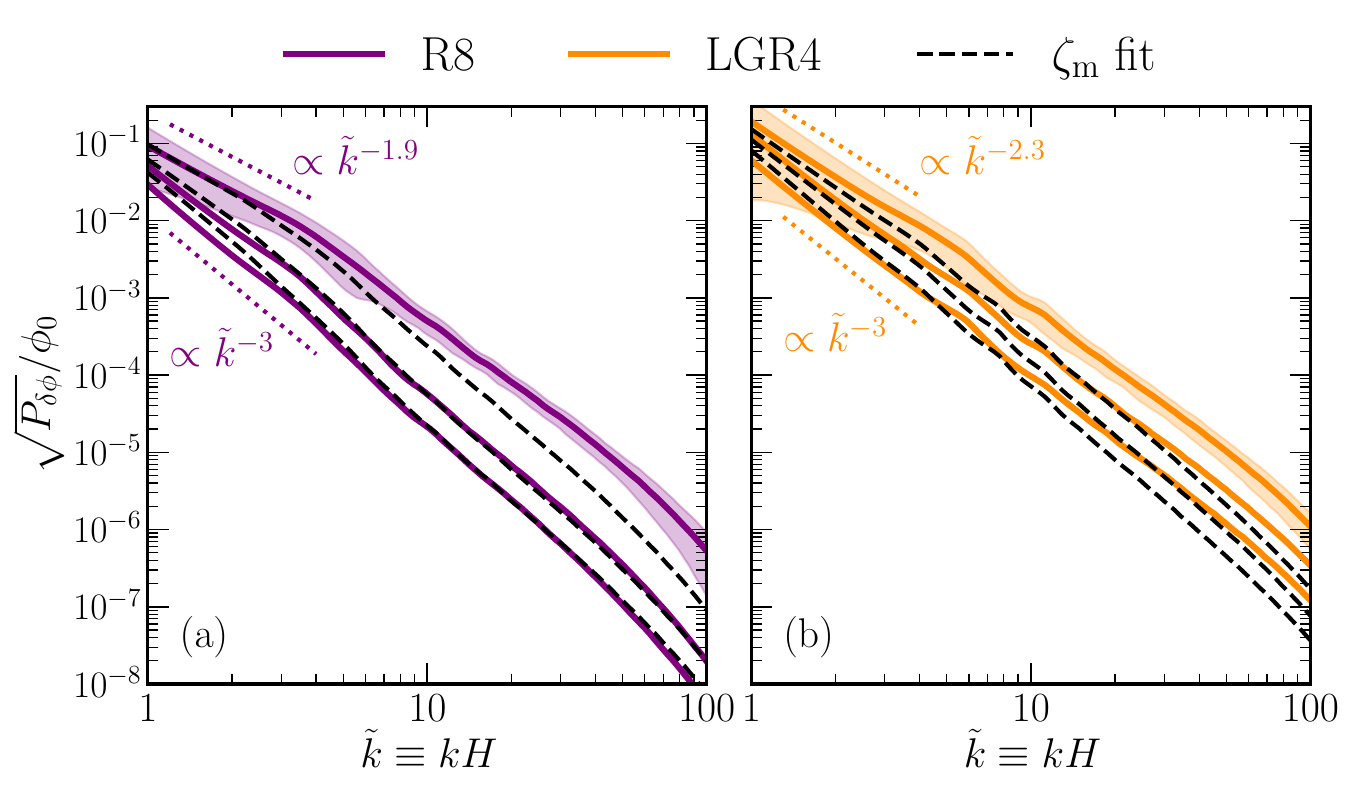}
    \caption{The square root of the power spectrum of potential fluctuations as a function of $\ktilde$ at fixed $\ztilde = \{0, 1, 1.5\}$ (increasing from top to bottom in each panel) for the R8 (left, purple) and LGR4 (right, orange) simulations. We normalize by $\phi_0$; see \autoref{eq:phi_normalization}. This is the same quantity shown in panels (c) and (d) of \autoref{fig:phi0_phik_z}, except now as a function of $\ktilde$ instead of $\ztilde$. For clarity, we only show the standard deviation (shaded region) around the time-averaged quantity (solid curve) for one value of $\ktilde$, but the fractional variations are comparable across all the curves. In each panel, the black dashed curve shows the result assuming a separable density profile with the fitted vertical profile shown in \autoref{fig:zeta}, according to \autoref{eq:power_phi_Sigma}. The dotted curves indicate power laws of slope indicated by the text in each panel for reference.}
    \label{fig:phik_k}
\end{figure*}

\autoref{fig:phik_k} shows the same root-mean-square potential fluctuation amplitudes as the bottom row of \autoref{fig:phi0_phik_z}, but this time as a function of $\ktilde$ for the midplane $\ztilde = 0$ (top curve in each panel), as well as $\ztilde = 1$ and $\ztilde = 1.5$ (middle and bottom curves respectively). Once again, we show the $\pm 1\sigma$ range about the time-average quantity for just one curve for clarity, but the fractional variations about the mean are comparable for all $\ztilde$ values. We again overplot the predictions from the model given by \autoref{eq:power_phi_Sigma} as black dashed curves. We indicate the approximate power law scalings measured from the simulation at low $\ktilde$ on each panel and plot dotted lines of the corresponding slope for comparison.

For the R8 simulation, the model is accurate across a wide range of $\ktilde$ for $|\ztilde| \geq 1$, and is reasonably accurate at the midplane for small values of $\ktilde$ as well, though the model amplitude drops off much more steeply with $\ktilde$ than the simulation. For the LGR4 simulation, the models are reasonably accurate at smaller $\ktilde$ values, but once again decay much more rapidly with $\ktilde$ than the simulation curve. From \autoref{eq:power_phi_Sigma} and the approximation presented in \autoref{eq:chi_approx}, we anticipate that at low $|\ztilde|$, the amplitude steepens from $\propto k^{-(n_\delta/2 + 1)} \approx k^{-2.1}$ to $\propto k^{-(n_\delta/2 + 2)} \approx k^{-3.1}$ as $\ktilde$ increases, while at larger $|\ztilde|$, the amplitude is $\propto k^{-(n_\delta/2 + 2)} \approx k^{-3.1}$ across all $\ktilde$ values. This steepening at low $|\ztilde|$ and the lack thereof at larger $|\ztilde|$ is apparent in both the R8 and LGR4 simulations, and in each case the model slopes are approximately consistent with (though somewhat steeper than) the simulations. The origin of the difference in slope between the model and the simulation at the midplane is the approximation that the surface density power is representative of the midplane density: in reality, a slice of the midplane density has a slightly shallower power spectrum than $P_\delta$, because small-scale features generated by turbulence and feedback at the midplane are smoothed over when integrated vertically to produce a surface density map. Note that the extension allowing the fluctuations to be more concentrated around the midplane described in \autoref{sec:zetagtr} does improve the match to the curves over a somewhat wider range of $\ktilde$ values, though the difference at large $\ktilde$ persists. However, because the vast majority of the power is contained in the fluctuations at small $\ktilde$ (note the larger dynamic range shown on the vertical axis compared to \autoref{fig:phi0_phik_z}) where the absolute difference in fluctuation amplitude is not yet very large, our model still yields accurate results.

\begin{figure*}
    \centering
    \includegraphics[width=0.95\textwidth]{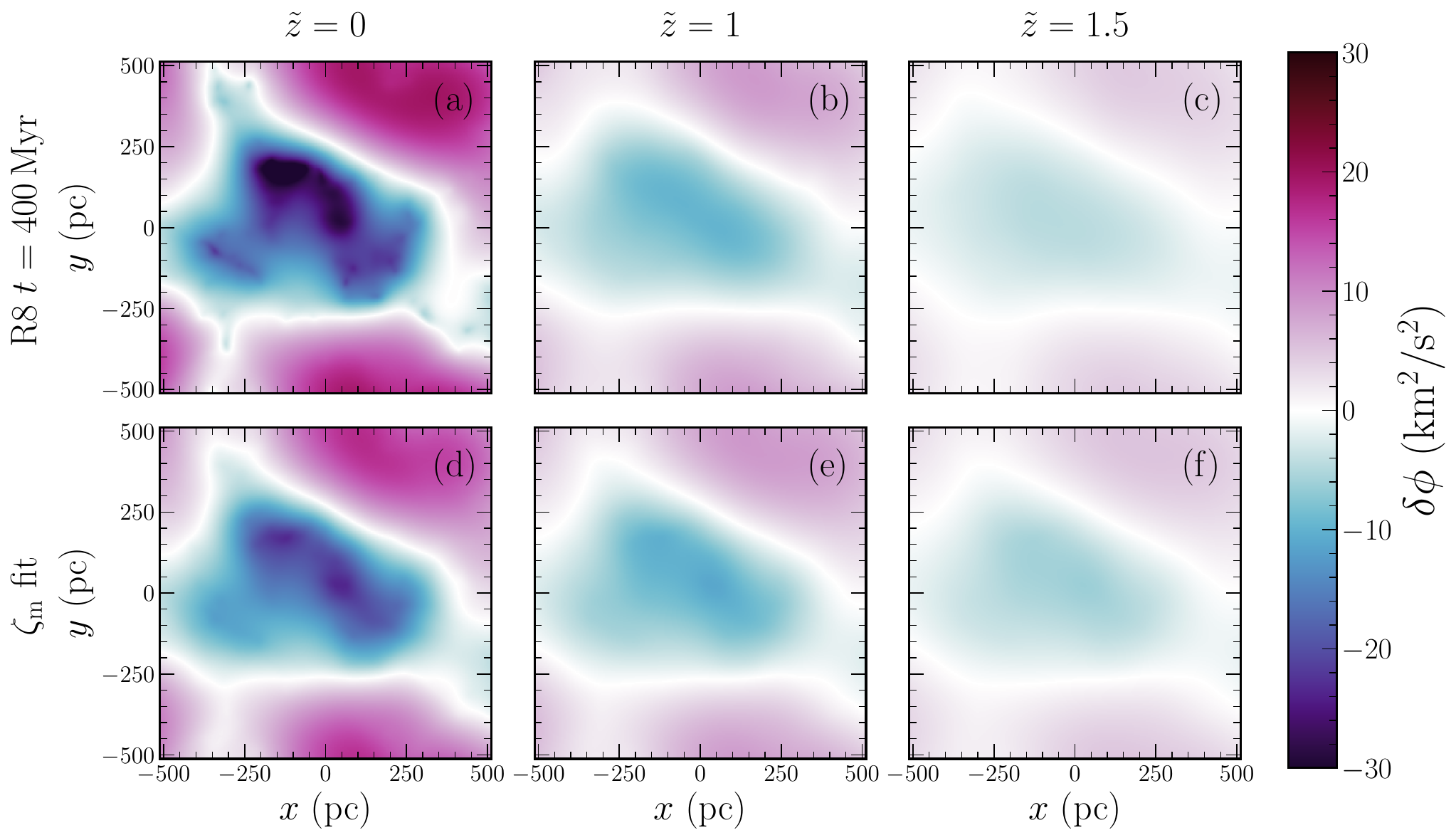}
    \caption{The potential fluctuation $\delta\phi$ as a function of $\bx$ at several different $\ztilde$ values for a snapshot of the R8 simulation (top row) and as reconstructed using \autoref{eq:phi_Sigma_k} from the surface density fluctuations from the same snapshot time, assuming a separable density profile with the fitted vertical profile shown in \autoref{fig:zeta}.}
    \label{fig:phi_reconstruction_comparison_xy}
\end{figure*}

\begin{figure*}
    \centering
    \includegraphics[width=0.95\textwidth]{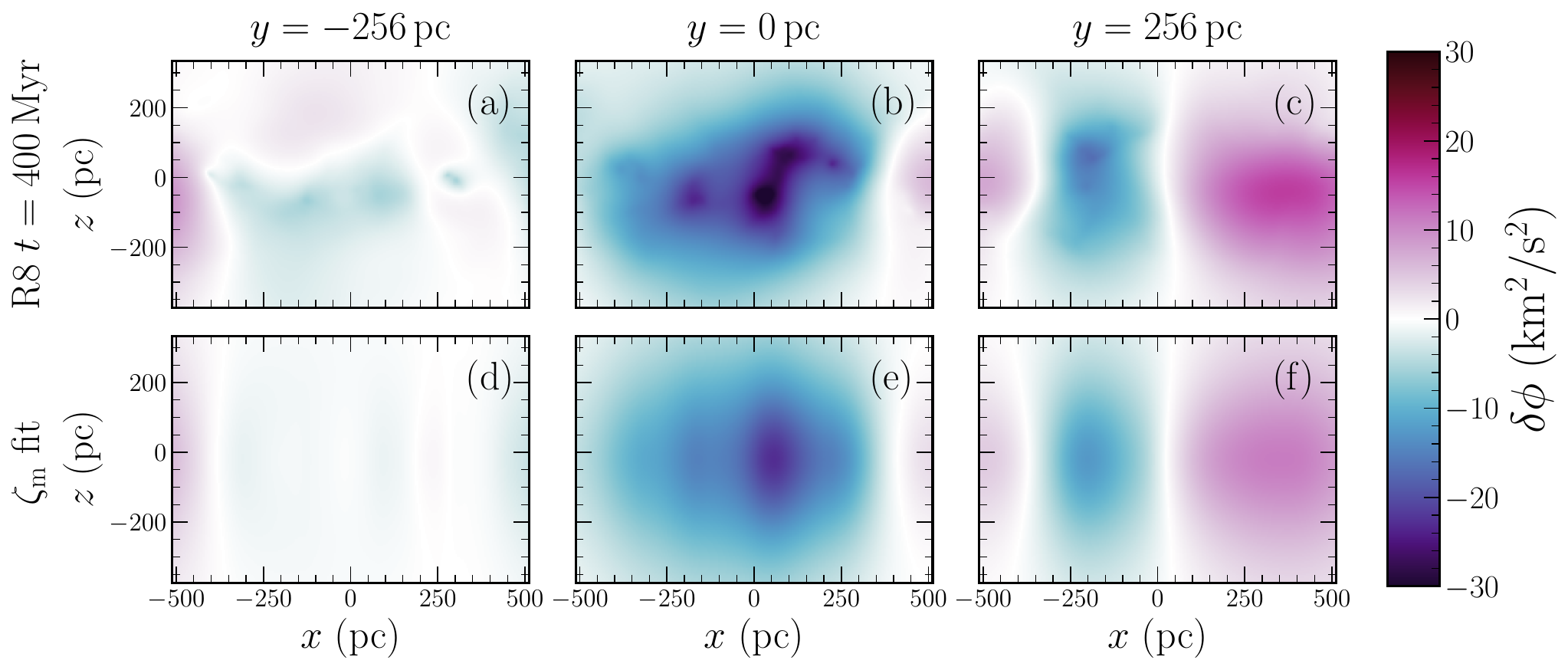}
    \caption{Analogous to \autoref{fig:phi_reconstruction_comparison_xy}, but now comparing the potential fluctuations $\delta\phi$ from a snapshot of the R8 simulation (top row) with the reconstruction \autoref{eq:phi_Sigma_k} (bottom row) as a function of $(x, z)$ for several different $y$ values to illustrate the vertical structure.}
    \label{fig:phi_reconstruction_comparison_xz}
\end{figure*}

Thus far, we have been comparing our model to the simulation outputs by studying only the amplitudes of each Fourier component of the fluctuations. However, another consequence of \autoref{eq:phi_Sigma_k} is that the phase of $(\delta\phi_\bk)_\mathrm{m}$ is entirely set by the phase of $\delta_\bk$, since all the other factors are real. One simple way to assess whether the separable model is adequate in this regard is to reconstruct a potential fluctuation field from the measured $\delta_\bk$ at a given snapshot (by summing up the components given by \autoref{eq:phi_Sigma_k} using \autoref{eq:fourier_details}) and comparing to the actual field at that time. If the phases of $\delta_\bk$ and $\delta\phi_\bk$ are well-correlated, then the reconstruction at each value of $z$ should have a similar morphology to the real field, but if the phases are not well-correlated, it may look entirely different.

\autoref{fig:phi_reconstruction_comparison_xy} compares the measured $\delta\phi$ field for a snapshot of the R8 simulation (top row) with the reconstruction based on \autoref{eq:phi_Sigma_k} (bottom row) for vertical slices at $\ztilde = 0$ (the midplane, left column) and $\ztilde = 1$ and 1.5 (middle and right columns respectively). Note that these are the same $\ztilde$ values shown in \autoref{fig:phik_k}. We see that the reconstruction is very accurate for the $\ztilde \geq 1$ panels, and captures the essential large-scale features in the midplane, although as anticipated from \autoref{fig:phik_k}, the real fluctuation field shows stronger small-scale (high $\ktilde$) features in the midplane. \autoref{fig:phi_reconstruction_comparison_xz} shows slices of the same snapshot and reconstruction, but now as a function of $(x, z)$ at fixed (arbitrary) $y$ values. The range of $z$ values shown corresponds to $|\ztilde| \leq 1.5$. Although there are slight asymmetries in the real field that cannot be produced in a separable model, the reconstruction is still reasonably similar, and the vertical extent of the fluctuations is comparable to the real snapshot.

Ultimately, we have demonstrated that the simple separable model described above gives a good description of the potential fluctations as a function of both space and time. Together with the results regarding the linear surface density fluctuations presented in \autoref{sec:surface_density}, \autoref{eq:power_phi_Sigma} provides a direct input for stellar dynamical studies of e.g., diffusion processes driven by external perturbations \citep{BinneyLacey1988}, which we will explore in a future work (Modak et al. in prep).

%%%%%%%%%%%%%%%%%%%%%%%%%%%%%%%%%%%%%%%%%%%%%%%%%%%%%%%%%%%%
%%%%%%%%%%%%%%%%%%%%%%%%%%%%%%%%%%%%%%%%%%%%%%%%%%%%%%%%%%%%
\section{Discussion}
\label{sec:discussion}
%%%%%%%%%%%%%%%%%%%%%%%%%%%%%%%%%%%%%%%%%%%%%%%%%%%%%%%%%%%%
%%%%%%%%%%%%%%%%%%%%%%%%%%%%%%%%%%%%%%%%%%%%%%%%%%%%%%%%%%%%

In this work, we have characterized the statistics of the (2D) surface density fluctuation fields of the ISM, and subsequently demonstrated that the structures of the (3D) volume density and gravitational potential fields are well-approximated by a separable model in which the 3D fluctuation fields essentially inherit the statistical structure of the 2D field (see \autoref{eq:separable_density} and \autoref{eq:deltarho_model}), with a specified vertical profile (see \autoref{eq:zeta_model}). Thus, our results can be summarized using our characterization of the surface density fluctuation spatio-temporal spectra in \autoref{eq:spatiotemporal_model_combined}. With a realization of the surface density fluctuation $\delta(\mathbf{x},t) \equiv \Sigma(\mathbf{x},t)/\overline{\Sigma}(t)  -1$ in hand, realizations of the corresponding volume density, vertical gravitational potential, and gravitational potential fluctuations can then be produced using \autoref{eq:separable_density}, \autoref{eq:phi_k0}, and \autoref{eq:phi_Sigma_k} respectively, and center-of-mass motion (\autoref{eq:zcom}) or thickness variation (\autoref{eq:H}) can be specified as desired, calibrated to the values reported in \autoref{tab:zcom_H_values}.

Physically, the content of \autoref{eq:spatiotemporal_model_combined} is that
\begin{enumerate}[label=(\roman*)]
    \item modes at a scale $k$ have a characteristic lifetime $\sim 2\omega_0(k)^{-1}$ (see \autoref{eq:omega_k}), which is longer (set by feedback) for large-scale modes and shorter (set by the dynamical time using the effective sound speed) for small-scale modes, and
    \item fluctuation power is distributed across these modes as a steep power law with index $n_\delta \approx -2.2$ for linear surface density fluctuations $\delta$ (or $n_s\approx -2.8$ for logarithmic surface density fluctuations $s\equiv \ln(1+\delta)$), so that a majority of the power is concentrated at the largest spatial scales.
\end{enumerate}
Below, we speculate on the implications and possible origins of the power-law spatial structure (\autoref{sec:origins_implications}), discuss how our results may be compared with observations (\autoref{sec:observations}), and finally comment on caveats and possible generalizations of our characterization (\autoref{sec:nongaussianity_anisotropy}).

%%%%%%%%%%%%%%%%%%%%%%%%%%%%%%%%%%%%%%%%%%%%%%%%%%%%%%%%%%%%
\subsection{Implications and Origins of Power-Law Spectra}
\label{sec:origins_implications}
%%%%%%%%%%%%%%%%%%%%%%%%%%%%%%%%%%%%%%%%%%%%%%%%%%%%%%%%%%%%

An immediate consequence of the power-law spatial spectra of the surface density field is that ISM substructures are not well-described as compact ``clouds'' with well-defined sizes and masses. In contrast to these classical models, the spectra we measure are relatively featureless, suggestive of a fractal, hierarchical structure. As an example of an application where this difference leads to qualitatively dissimilar results, we note that in the stellar-dynamical context, where it is natural to classify fluctuations by their scale (e.g., \citealt{galactokinetics}), the large-scale ISM fluctuations that carry the most of the power drive scattering processes distinct from those of small-scale clouds. We will describe the implications of this fluctuation structure for stellar orbital heating and migration processes in Galactic disks in a future work (Modak et al. in prep).

%One possible method of producing such a featureless spectrum is via a scale-free substructure density profile, which need not have finite mass or a characteristic size: if the volume density of such a structure scales as $\rho(r) \propto r^{-(3-n_\delta/2)}$ where $r$ is the distance from its center, then we could reproduce $P_\delta\propto k^{-n_\delta}$ with a homogeneous and isotropic distribution of many such fluctuations. Using $n_\delta\approx 2.3$, these density profiles would have $\rho\propto r^{-1.85}$, somewhat shallower than the classical Larson-Penston \citep{Larson1969, Penston1969} solution. The deviations may be attributed to the fact that the gas in the simulation exists in several distinct phases with both magnetic and turbulent pressure playing an important role in the dynamics (see e.g., Figures 4 and 9 in \citealt{Kim2023}) while the Larson-Penston solution assumes isothermality and primarily thermal pressure support.

Previous analysis of TIGRESS simulations demonstrated that only a small fraction of the gas is strongly self-gravitating \citep{Mao2020}. The structure in the density and surface density, therefore, largely reflects the complex, multi-scale interactions that develop in a multiphase, magnetized medium in which kinetic energy injection is correlated on large scales (due to the clustering of supernovae), large-scale background shear is imposed by the balance of Coriolis forces and tidal gravity, and angular momentum and vertical gravity tend to limit significant radial and vertical excursions.

Most analyses of turbulence in simulations, whether for highly idealized setups (e.g., isothermal, with driving in $\bk$-space at large scales) or for setups that include more of the physics of the multiphase ISM, with injection of energy by supernovae (e.g., \citealt{Padoan2016, Beattie2025b}), focus on velocity power spectra. Solenoidal components of the velocity have power spectra similar to the theoretical predictions of \cite{Iroshnikov1964, Kraichnan1965} or \cite{Kolmogorov1941}, while compressive components follow the steeper \cite{Burgers1948} spectrum. While the perturbations in density are driven by the compressive components of the velocity field, the density and velocity spectra are expected to match only for pure acoustic waves. For very subsonic turbulent flows, simulations have recovered density spectra with $4\pi k^2 P_{\mathrm{3D},\delta}(k) \propto k^{-2.2}$ \citep[e.g.,][]{Gotoh2001,Pan2016}, close to the expected index $-7/3$ for the pressure spectrum in incompressible turbulence \citep{Batchelor1951}. At higher Mach number $\cal M$,  the density spectrum in simulations becomes flatter, approaching  $4\pi k^2P_{\mathrm{3D},\delta} \propto k^{-1}$ at ${\cal M} \sim 10$ \citep[e.g.][]{Ryu2005, Kritsuk2007, Konstandin2016, Pan2016}. Thus, for isotropic turbulence in a single-phase medium, the index of the 3D power spectrum is in the range $\sim -4$ to $-3$, taking into account the additional factors of $k$ from the Jacobian.

In the discussion above, we have explicitly written $P_{\mathrm{3D}}$ to denote power spectra measured in 3D (and averaged over angles for $k=(k_x^2 + k_y^2 + k_z^2)^{1/2}$). For an isotropic spectrum, integrating vertically to obtain the 2D projection simply selects $k_z=0$, with $P(k_x,k_y) \propto P_{\mathrm{3D}}(k=(k_x^2+k_y^2)^{1/2})$ up to a normalization factor. The spectra we find, $P_\delta \propto k^{-2.3}$ to $k^{-2.2}$, are considerably flatter than would be expected if we were simply projecting from the isotropic, single-phase models described above. The difference likely has more to do with the multiphase character of the gas than the fact that we are considering turbulence in a disk, since the ratio between the vertical dimension of the disk and the horizontal dimension in the simulation is modest, $2H/L \sim 0.6$. However, the cold gas is collected in small clouds with densities two orders of magnitude larger than those of the warm diffuse gas, while the density in superbubbles that fill a large fraction of the domain is two or more orders of magnitude below that of the warm gas.

\subsection{Comparison with Observations}
\label{sec:observations}
%%%%%%%%%%%%%%%%%%%%%%%%%%%%%%%%%%%%%%%%%%%%%%%%%%%%%%%%%%%%

Our surface density fluctuation pdfs and power spectra offer a natural avenue for comparison with observations. This is most straightforward in external galaxies that are nearly face on, and in principle can make use of both gas and dust tracers for the column density. However, a key distinction between analyzing the observations and our simulations is that we characterize the density field directly, while observations are of an emission field that is not necessarily a linear tracer of density.

Velocity-integrated 21 cm emission is the most direct tracer of gas surface density in regions that are dominated by atomic gas. For two-point statistics, power spectra of \ion{H}{1} column densities have long been used to characterize the ISM, covering scales $10 - 10^4\,$pc in a range of nearby systems. For the SMC, \cite{Stanimirovic1999} found an \ion{H}{1} spectrum close to $P_{\delta} \propto k^{-3}$, somewhat steeper than our results. \cite{Elmegreen2001} found a steeper spectrum at small scales than large scales in the power spectrum of the LMC, suggesting that the transition marks scales comparable to the disk thickness. Based on analyses of several dwarfs and spirals, \cite{Dutta2009, Dutta2013} also found a transition from a steeper spectrum at small scales to a shallower one at large scales. Making use of a large survey of dwarfs for the LITTLE THINGS project, measured (velocity-integrated) power spectra reported by \cite{Zhang2012} were in the range $P_{\delta} \propto k^{-1.4}$ to $k^{-4.4}$ (without breaks), centered near $k^{-3}$,  which is similar to (though a bit steeper than) our results. \cite{Grisdale2017} analyzed power spectra from \ion{H}{1} in several massive disk galaxies, and found a range $P_{\delta} \propto k^{-1.6}$ to $k^{-2.8}$, with half having indices of $-2.1$ or $-2.2$ with no break, quite similar to our results. For M31 and M33, \cite{Koch2020} found indices $-2.7$ and $-2.1$, respectively, when fitting power spectra of \ion{H}{1} maps at scales exceeding $\sim 200$\,pc. 

While there have been some suggestions in the literature of breaks in the power spectrum appearing at scales corresponding to the disk thickness (i.e. $kH\sim \pi$), \cite{Grisdale2017} and \cite{Koch2020} both conclude from detailed modeling that previous inferences of breaks in power spectra can be accounted for by the beam scale (or PSF) in observations. We do not observe a clear break in the power law index for our measured spectra, although there may be a hint of a slightly shallower slope at the largest scales we measure (approaching the box size of $\sim 1$\,kpc). In order to test definitively whether or not there is a break in the power spectrum (representing a transition in effective dimension for structures smaller than or larger than the gas scale height), TIGRESS simulations in boxes larger than those we analyze here would be required.

Dust emission also offers a promising means to analyze the detailed structure of the ISM, both from earlier multi-wavelength IR maps in individual nearby galaxies, and very recently with the advent of JWST surveying a large number of targets uniformly. \cite{Koch2020}, analyzing power spectra of dust surface density maps in four Local Group galaxies (created by \citealt{Utomo2019} from multi-wavelength \textit{Herschel} data), concluded that a single power law with index between $-2$ and $-2.5$ describes the observations for the SMC, the LMC, and M31 well, while M33 has a shallower index. \cite{Koch2020} also measured power spectra in individual mid-to-far IR \textit{Herschel} and \textit{Spitzer} bands,  finding indices similar to those for dust maps at long wavelength ($100-500\,\mu$m) and significantly shallower spectra at 24\,$\mu$m, which they concluded was due to local heating in star-forming regions.

Mid-IR emission maps from JWST for large spirals as well as dwarfs include both thermal dust-dominated and PAH-dominated bands, and go to very high spatial resolution. The analysis of 2D power spectra from 13 galaxies by \cite{LindThomsenSneppenSteinhardt2025} produced a mean index $\sim -2$ in PAH-dominated bands, similar to the spectrum we find. Slopes measured in bands dominated by thermal dust were shallower ($\sim -1 $ to $-1.5$), however.  \cite{Elmegreen2025} found shallow slopes (index $\sim -1$) in both thermally dominated and PAH-dominated bands for a sample that included both spirals and dwarfs.  While the latter work found no evidence for power spectrum breaks, the former did; since the sample overlaps, it is not clear what the origin of the discrepancy is.

It is important to note that one should not expect single-band dust emission maps, whether for thermal-dominated or PAH-dominated bands in the mid-IR, to be directly comparable to surface density maps, since the emission represents a convolution of the dust heating rate (proportional to the local radiation field) and dust mass, with additional dependence on dust temperature for thermal bands. Given the nonuniform heating of the ISM, mid-IR emission is not a linear tracer of dust surface density, and this nonlinearity could significantly alter the measured statistics. Using numerical simulations with ray-tracing radiative transfer from star-forming regions, \cite{Linzer2024} demonstrated that the mid-IR dust emission is proportional to a dust heating parameter $\mathcal{N}_d(\bx) \equiv \int\md z\, \mathcal{J}(\bx, z)\,n_H(\bx, z)$ where the normalized FUV intensity $\mathcal{J}$ has large spatial variations. In order to convert mid-IR maps to true maps of surface density, it is necessary to effectively divide out by the heating rate; fortunately, simulations show that $\mathcal{J}$ can be reasonably well estimated based on the measured locations and bolometric luminosities of star forming regions, combined with simple attenuation approximations \citep{Linzer2024}.

We leave a direct comparison between observed and simulated mid-IR maps to the future. We note, however, some additional hints that the observed fluctuation fields are consistent with the measurements in our simulations. In particular, the mid-IR emission pdfs measured by \cite{Pathak2024} have the same approximately log-normal form with a power law tail as the FUV fields measured in \cite{Linzer2024}, with log-normal standard deviations comparable to those measured in \autoref{fig:sigma_s_R} at their resolution of $\sim 100$\,pc. With the surface density taking a log-normal form, as we find, the dust heating parameter and resulting mid-IR emission would also be approximately log-normal (away from strong radiation sources) since a product of log-normal variables is log-normally distributed. This suggests that the one-point statistics of the fluctuations in the TIGRESS-NCR simulations (see \autoref{fig:s_pdfs}) may be similar to those in PHANGS, although a quantitative test of this will require significant additional modeling.

%%%%%%%%%%%%%%%%%%%%%%%%%%%%%%%%%%%%%%%%%%%%%%%%%%%%%%%%%%%%
\subsection{Characterizing Anisotropy and Non-Gaussianity}
\label{sec:nongaussianity_anisotropy}
%%%%%%%%%%%%%%%%%%%%%%%%%%%%%%%%%%%%%%%%%%%%%%%%%%%%%%%%%%%%

\begin{figure*}
    \centering
    \includegraphics[width=0.95\textwidth]{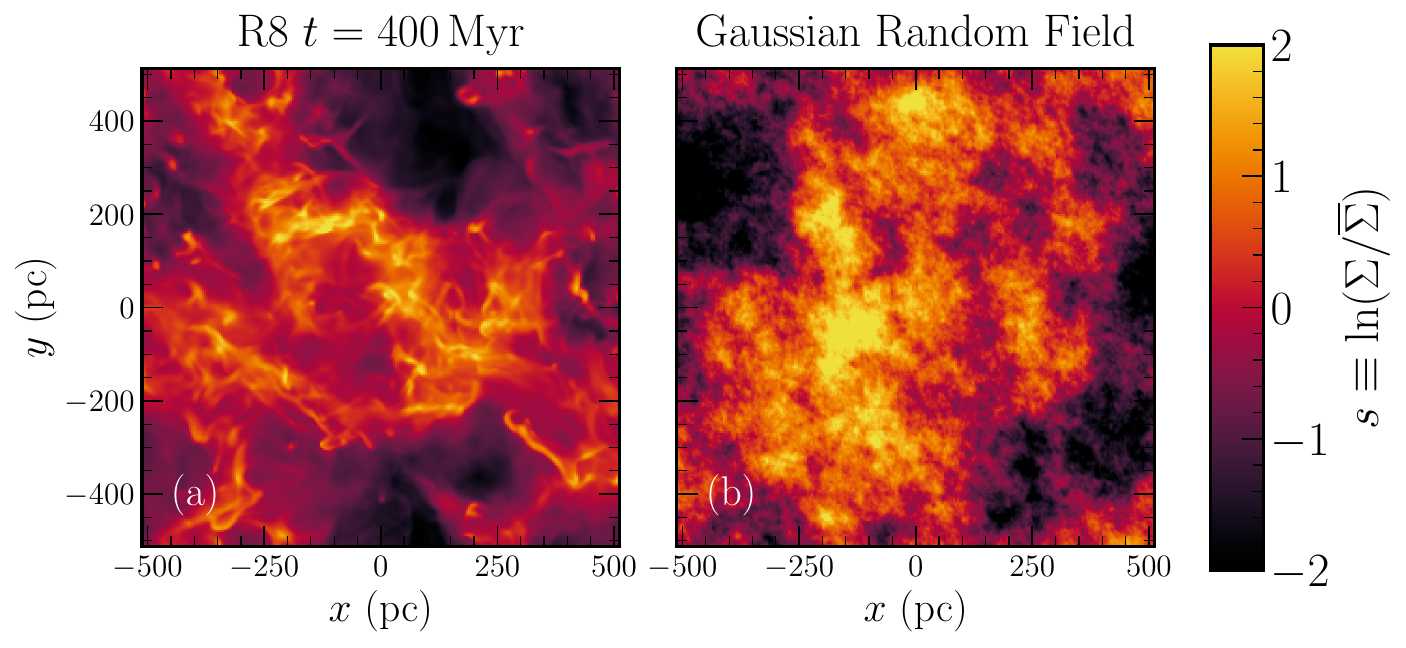}
    \caption{A comparison between a snapshot of the logarithmic surface density fluctuation $s$ for the TIGRESS R8 simulation (panel (a)) and a snapshot of a GRF generated using the model power spectrum following the method described in \autoref{sec:generating_GRFs_direct} with parameters from \autoref{tab:power_k_measurements} and \autoref{tab:power_omega_measurements} (panel (b)).}
    \label{fig:GRF_comparison}
\end{figure*}

Given our model as specified in \autoref{eq:spatiotemporal_model_combined}, a natural question is whether it fully encapsulates the structure of the fluctuation field. In particular, if a GRF with the specified power spectrum and \textit{random} phases is generated, how do its properties differ from the simulated fluctuations? While the temporal phases of each $\bk$-mode are reasonably well-described as uncorrelated (based on e.g., the lack of a nonzero peak frequency in \autoref{fig:delta_omega_k_R8} and \autoref{fig:power_omegatilde}), the spatial structure at a fixed time is more complex to describe. \autoref{fig:GRF_comparison} illustrates the difference between a snapshot of the actual logarithmic fluctuation $s$ in the R8 simulation (panel (a)), versus a snapshot as produced in a realization of a GRF with our model power spectrum using random phases for all Fourier components as described in \autoref{sec:generating_GRFs_direct} (panel (b)). While we see that the characteristic magnitudes (e.g., maxima and minima) of $s$ are similar across both fields, with similar overarching sizes of voids and overdensities, qualitative differences in the field structure are also evident. We attribute these differences to two approximations made in our models, which we describe below.

Most significantly, in modeling the fluctuations as a GRF, it is assumed that the field is fully characterized by its two-point statistics, so that, e.g., the bispectrum is identically zero and the phases of individual Fourier components are uncorrelated. However, this is not the case in the real fluctuation field---the presence of filamentary features in the overdensities is a signature of phase correlation. More generally, any system with significant nonlinearity will have correlation in the phases of harmonics. While characterizing higher order spectra is outside the scope of this paper, we note that several methods of doing so have been developed, especially in the cosmological context. In particular, algorithms to detect and characterize filaments have been successfully applied to ISM structure as mapped in continuum and lines (e.g., \citealt[][and references therein]{KochRosolowsky2015, Hacar2023}). Additionally, the ``line correlation function'' (related to the three-point structure of the phases) introduced in \cite{Obreschkow2013} in the cosmological context has also been demonstrated to capture the filamentary structure, and the estimator developed in \cite{WolstenhulmeBonvinObreschkow2015} could be used to measure this statistic for ISM fluctuation fields as well. Measures of the three-point correlation function of ISM structures have also been developed in \cite{Portillo2018}, and such statistics could be studied equally effectively in the TIGRESS-NCR simulations. Finally, alternatives to higher order correlation functions and spectra such as wavelet scattering transforms have also been developed in the context of the ISM \citep{Allys2019, Saydjari2021} and have been demonstrated to capture non-Gaussian information.

A second reason for the difference of the two panels in \autoref{fig:GRF_comparison} is that the model in \autoref{eq:power_Sigma_spatial} treats the surface density fluctuation power as isotropic in $\bk$-space. While this is true for intermediate to large $k$, for the smallest $k$ values that carry the most power, in reality the structures are primarily trailing: ellipses fitted to the contours shown in \autoref{fig:P_delta_2D} monotonically decrease in eccentricity from $e \approx 0.9$ at the largest scales to  $e \approx 0.2$ at the smallest scales. We note that the pitch angle (measured as the orientation of the major axis of the fitted ellipses) also varies with scale. These deviations from isotropy ($e = 0$) contribute to the larger scatter around the mean power spectrum present at the smallest $k$ values (see \autoref{fig:power_k}). Nonetheless, isotropic models provide a reasonable description of the statistics, and are convenient in analytic applications due to their simplicity.

Potentially, it may be possible to develop a 
theoretical model for the spectrum that specifies its characteristic anisotropy and tilt as a function of scale.  Because the self-gravity of the gas is modest (the typical Toomre $Q$ parameter ranges from $\sim 1.2-3$ for the models studied here), mode amplitudes primarily reflect the magnetohydrodynamic response of the sheared system to turbulence excited by feedback, though self-gravitating swing amplification may still play a role (see, e.g., \citealt{HeinemannPapaloizou2009, HeinemannPapaloizou2012}). Given the strong nonlinearities and challenge of properly specifying excitation for a highly inhomogeneous system, however, development of rigorous theory is nontrivial.  We defer such an analysis to future work.

The importance of both the anisotropy and non-GRF-nature of ISM fluctuations is likely to vary depending on the application. For instance, current observational characterizations of power spectra in dust and gas emission, as discussed in \autoref{sec:observations}, are presented as a function of the magnitude $k$ alone. In the stellar-dynamical context, the power spectrum of the potential entirely characterizes the dynamical influence of the fluctuations in quasilinear diffusion theory \citep{BinneyLacey1988, chavanis2023secular}, and so the difference between the evolution of orbits under the real fluctuation field and a realization of a GRF with our model power spectrum may not be significant. We will assess the accuracy of this latter claim in a future work (Modak et al. in prep).

%%%%%%%%%%%%%%%%%%%%%%%%%%%%%%%%%%%%%%%%%%%%%%%%%%%%%%%%%%%%
%%%%%%%%%%%%%%%%%%%%%%%%%%%%%%%%%%%%%%%%%%%%%%%%%%%%%%%%%%%%
\section{Summary}
\label{sec:summary}
%%%%%%%%%%%%%%%%%%%%%%%%%%%%%%%%%%%%%%%%%%%%%%%%%%%%%%%%%%%%
%%%%%%%%%%%%%%%%%%%%%%%%%%%%%%%%%%%%%%%%%%%%%%%%%%%%%%%%%%%%

In this paper, we have characterized the structure of the ISM's surface density, volume density, and gravitational potential in the TIGRESS-NCR suite's R8 (solar-neighborhood-like) and LGR4 (higher surface density environment) simulations. Our main results can be summarized as follows:
\begin{enumerate}[label=(\roman*)]
    \item The pdf of the logarithmic surface density fluctuation $s \equiv \ln(\Sigma/\overline{\Sigma})$ is well approximated as a Gaussian (\autoref{eq:gaussian_s} and \autoref{fig:s_pdfs}). When smoothed over larger scales, the variance is reduced, as expected given the power spectrum of the fluctuations (\autoref{eq:sigma_s_R_power}, \autoref{fig:sigma_s_R}).
    \item The spatial power spectrum of both $s$ and the linear surface density fluctuation $\delta \equiv (\Sigma - \overline{\Sigma})/\overline{\Sigma}$ are steep power laws, with fluctuation power concentrated at the largest scales (\autoref{fig:power_k}, \autoref{eq:power_Sigma_spatial}, and \autoref{tab:power_k_measurements}). The power law indices we measure are similar to those measured in observed power spectra of gas and dust surface density fluctuations in external galaxies (\autoref{sec:observations}).
    \item The spatio-temporal power spectrum of both $s$ and $\delta$ indicate that structures of different spatial scales have different characteristic lifetimes (\autoref{eq:omega_k} and \autoref{tab:power_omega_measurements}), but across all scales, the power at a given frequency normalized by that scale's characteristic lifetime is well-described as a squared Lorentzian (\autoref{eq:Lorentzian_model} and \autoref{fig:power_omegatilde}).
    \item The volume density is well-described as separable into an instantaneous mid-plane density proportional to the surface density, and a vertical profile that depends on time only through variations in the gas layer's center of mass position $\zcom(t)$ and thickness $H(t)$ (\autoref{eq:separable_density}, \autoref{eq:ztilde}, and \autoref{eq:rhobar_model}). The vertical profile is well-described as a mixture of $\sech^2$ and exponential functions (\autoref{eq:zeta_model}, \autoref{tab:zeta_measurements}, and \autoref{fig:zeta}).
    \item The gravitational potential's structure is well-described by a mean (independent of the in-plane coordinates) component that follows directly from the model for the vertical profile (\autoref{eq:phi_k0}, \autoref{fig:phi0_phik_z}), and a gravitational potential fluctuation power spectrum proportional to the surface density fluctuation spectrum (\autoref{eq:power_phi_Sigma}, \autoref{fig:phi0_phik_z}, and \autoref{fig:phik_k}).
    \item To generate a GRF realization for surface density fluctuations, as detailed in \autoref{sec:generating_fluctuations}, \autoref{eq:spatiotemporal_model_combined} may be used (with $Q = s$ and parameters from \autoref{tab:power_k_measurements} and \autoref{tab:power_omega_measurements}). To generate a realization of the corresponding volume density or gravitational potential, after specifying the time-dependence of $\zcom$ and $H$ (\autoref{fig:zcom_H}, \autoref{tab:zcom_H_values}), the vertical profile model summarized in (iv) may be used in conjunction with \autoref{eq:phi_Sigma_k}. Because spatial phases are correlated, it cannot be expected that maps of surface density generated in this way would resemble real maps. However, gravitational potential fluctuations generated in this way are all that is needed for applications to quasilinear evolution of the stellar distribution function.
\end{enumerate}
Ultimately, these characterizations will allow for direct comparison with observations, while also providing a key input for a wide range of (semi-)analytic stellar dynamical studies that wish to incorporate the influence of a realistic ISM.

\begin{acknowledgments}
We thank Sanghyuk Moon, Chang-Goo Kim, Nickolas Kokron, Nora Linzer, Jacob Nibauer, David Thilker, Jenny Greene, and Jeremy Goodman for helpful conversations. We also thank the referee for a thorough and insightful report. S.M. acknowledges support from the National Science Foundation Graduate Research Fellowship under Grant No.\ DGE-2039656. C.H. is supported by the John N. Bahcall Fellowship Fund at the Institute for Advanced Study. Partial support for this work was provided by grant 510940 from the Simons Foundation to E.C.O. 
\end{acknowledgments}

%% to get the citations to show in the compiled file do the following:
%%
%% pdflatex sample631.tex
%% bibtext sample631
%% pdflatex sample631.tex
%% pdflatex sample631.tex

\appendix

%%%%%%%%%%%%%%%%%%%%%%%%%%%%%%%%%%%%%%%%%%%%%%%%%%%%%%%%%%%%
%%%%%%%%%%%%%%%%%%%%%%%%%%%%%%%%%%%%%%%%%%%%%%%%%%%%%%%%%%%%
\section{Demonstrating Convergence with Simulation Resolution}
\label{sec:convergence}
%%%%%%%%%%%%%%%%%%%%%%%%%%%%%%%%%%%%%%%%%%%%%%%%%%%%%%%%%%%%
%%%%%%%%%%%%%%%%%%%%%%%%%%%%%%%%%%%%%%%%%%%%%%%%%%%%%%%%%%%%

The convergence of the TIGRESS-NCR R8 and LGR4 simulations with respect to spatial resolution was demonstrated in \cite{Kim2023} for basic ISM global properties (mean surface density, SFR, etc.). So, here, we will simply demonstrate convergence of the specific spatial statistics we consider in this work, namely the logarithmic surface density fluctuation pdf $p_s$ (see \autoref{fig:s_pdfs}), the standard deviation of those pdfs as a function of smoothing scale $\sigma_s(R)$ (see \autoref{fig:sigma_s_R}), and the spatial power spectra $P_Q(k) \equiv \langle |Q_\bk(t)|^2 \rangle$ for both $Q = \delta$ and $Q = s$ (see \autoref{fig:power_k}). We will focus our comparison on the R8 simulations; the LGR4 simulations show similarly converged results. The fiducial R8 simulation studied in the main text is the highest resolution in the suite, with a spatial resolution of $\Delta x = 4\,$pc; we will compare it to a simulation with identical initial conditions and a resolution of $\Delta x = 8\,$pc. We use snapshots at the same time intervals (separated by $\Delta t = 1\,$Myr) for both simulations.

\begin{figure}
    \centering
    \includegraphics[width=0.48\textwidth]{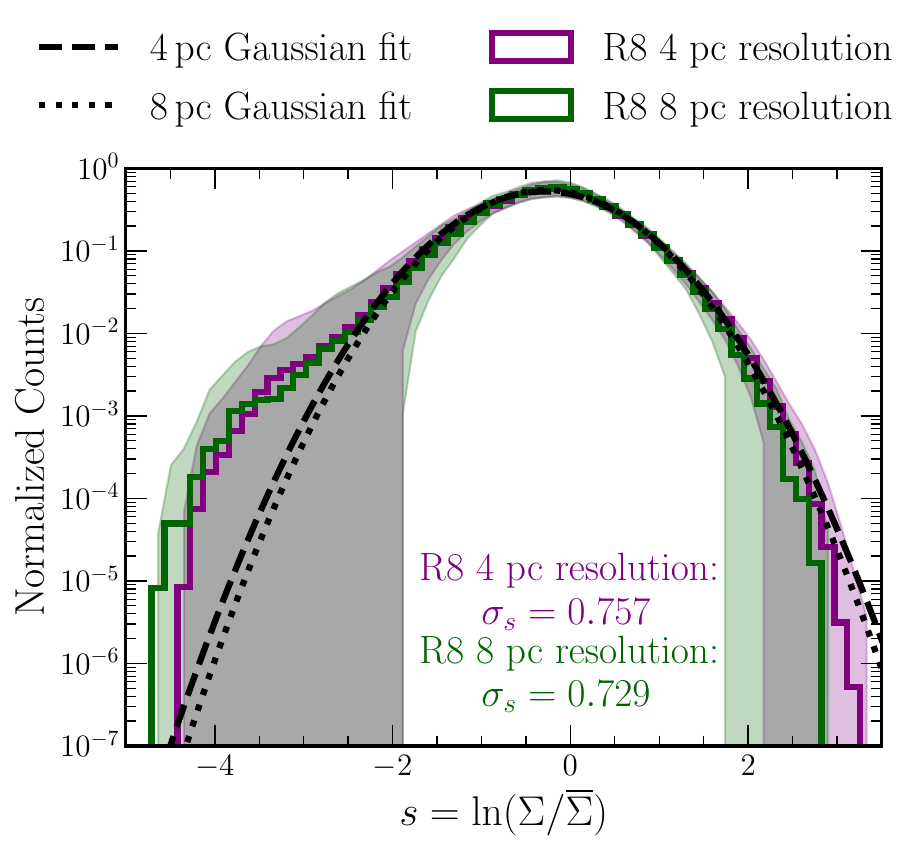}
    \caption{Analogous to \autoref{fig:s_pdfs}, but comparing the R8 4\,pc resolution (purple) and R8 8\,pc resolution (green) simulations.}
    \label{fig:s_pdfs_convergence}
\end{figure}

In \autoref{fig:s_pdfs_convergence} we plot the distribution of the logarithmic surface density fluctuation $p_s$ measured at each simulation's respective grid scale for the R8 8\,pc resolution (green) and 4\,pc resolution (purple, as in the main text) simulations. As in \autoref{fig:s_pdfs}, the solid curve indicates the time-averaged quantity while the shaded region around it is a $\pm 1\sigma$ range around the average. We see that the distributions are both consistent with Gaussian distributions described by \autoref{eq:gaussian_s} and \autoref{eq:mean_s}, with only small differences in the measured variances due to the different grid scales (which may be thought of as different smoothing scales for the same underlying field), as we discuss below.

\begin{figure*}
    \centering
    \includegraphics[width=0.95\textwidth]{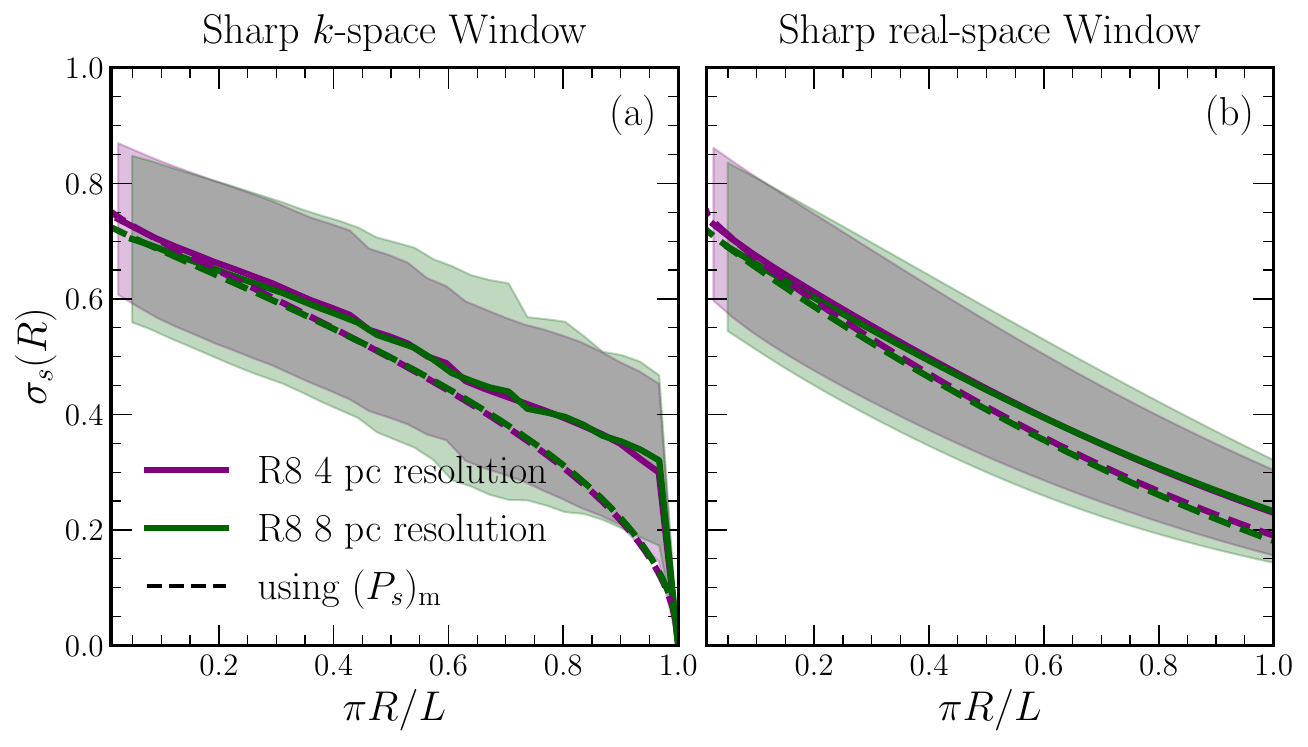}
    \caption{Analogous to \autoref{fig:sigma_s_R}, but comparing the R8 4\,pc resolution (purple) and R8 8\,pc resolution (green) simulations.}
    \label{fig:sigma_s_R_convergence}
\end{figure*}

In \autoref{fig:sigma_s_R_convergence}, analogous to  \autoref{fig:sigma_s_R}, we plot the standard deviation of the logarithmic surface density fluctuation field smoothed over a scale $R$, $\sigma_s(R)$, for a window function that is sharp in $k$-space (panel (a)) and sharp in real space (panel (b)), for the 4\,pc resolution (purple) and 8\,pc resolution (green) simulations. As expected for simulations converged in resolution, the standard deviation of the 4\,pc simulation smoothed over an 8\,pc scale is very close to that of the 8\,pc simulation's grid scale standard deviation (see the left-most portion of the $\sigma_s(R)$ curves in each panel). Indeed, across all smoothing scales, the curves lie nearly on top of each other, well within the $\pm 1\sigma$ shaded range. The variation around the time-averaged curve is larger for the lower resolution simulation despite using the same number of temporal snapshots, but this is to be expected since the total number of ``samples'' of the distribution is lower due to the smaller number of grid cells present.

\begin{figure*}
    \centering
    \includegraphics[width=0.95\textwidth]{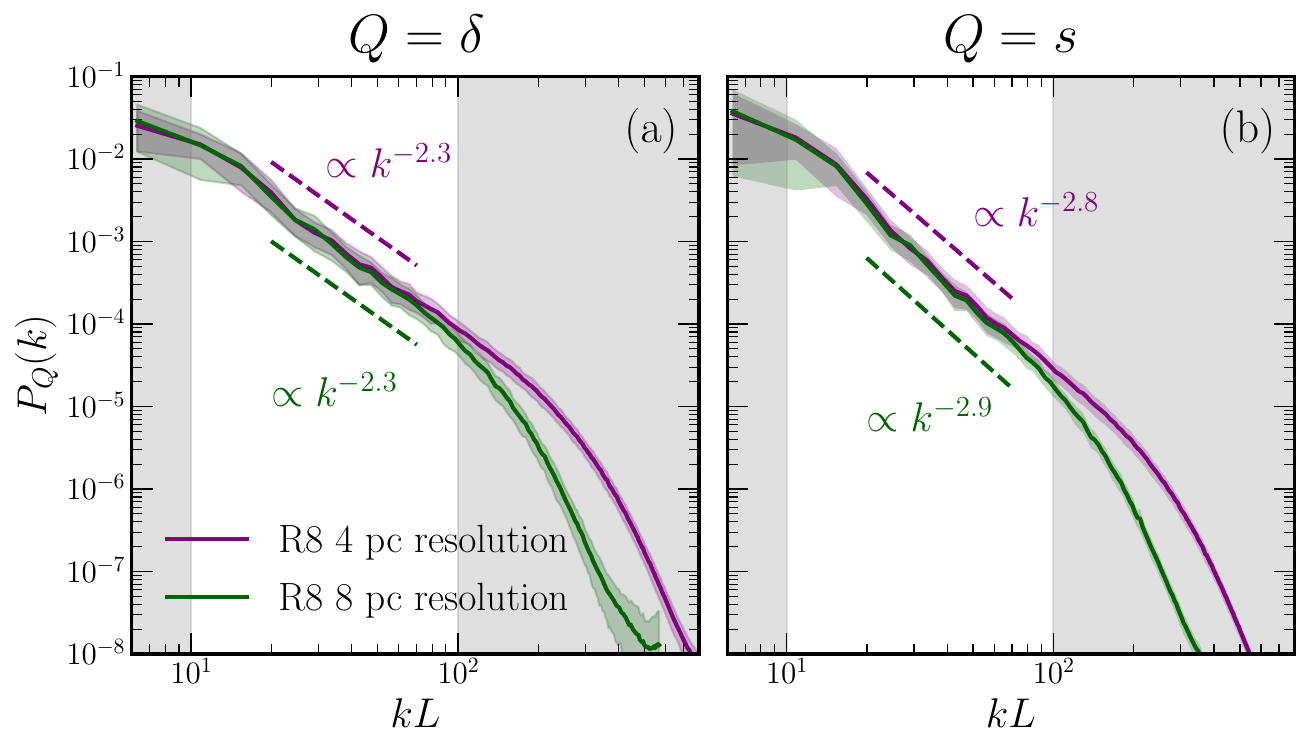}
    \caption{Analogous to \autoref{fig:power_k}, but comparing the R8 4\,pc resolution (purple) and R8 8\,pc resolution (green) simulations.}
    \label{fig:power_k_convergence}
\end{figure*}

Finally, in \autoref{fig:power_k_convergence}, analogous to \autoref{fig:power_k}, we plot the power spectra $P_Q(k)$ for $Q = \delta$ (panel (a)) and $Q = s$ (panel (b)) for the 4\,pc resolution (purple) and 8\,pc resolution (green) simulations. At the largest scales (smallest $k$), the spectra coincide very well---we measure consistent normalizations and power law indices. Note that the reduced spatial resolution in the 8\,pc simulation causes the spurious steepening of the spectrum to occur beginning at $kL \lesssim 100$ instead of $kL \approx 200$ for the 4\,pc simulation, so to measure power law indices for that simulation we restrict to a smaller range than in the main text, $10 \leq kL \leq 100$. To clearly identify the power law index over at least 1\,dex, therefore, requires a simulation resolution of at least $\sim 8\,$pc, and we might expect the power law to extend to even higher $k$ in simulations with resolution finer than $4\,$pc. Regardless of whether the power law extends to smaller scales, the key conclusion that the power spectrum is featureless and most of the fluctuation power is present at the largest spatial scales continues to hold.

%%%%%%%%%%%%%%%%%%%%%%%%%%%%%%%%%%%%%%%%%%%%%%%%%%%%%%%%%%%%
%%%%%%%%%%%%%%%%%%%%%%%%%%%%%%%%%%%%%%%%%%%%%%%%%%%%%%%%%%%%
\section{Generating Realizations of Surface Density Fluctuations}
\label{sec:generating_fluctuations}
%%%%%%%%%%%%%%%%%%%%%%%%%%%%%%%%%%%%%%%%%%%%%%%%%%%%%%%%%%%%
%%%%%%%%%%%%%%%%%%%%%%%%%%%%%%%%%%%%%%%%%%%%%%%%%%%%%%%%%%%%

In this appendix, we describe two methods for generating realizations of an ISM surface density fluctuation field $Q(\bx, t)$ (where $Q = s$, $\delta$) using the model given in \autoref{eq:spatiotemporal_model_combined} for the spatio-temporal power spectrum of the fluctuations. Both methods described here produce a fluctuation field with spatio-temporal Fourier \textit{amplitudes} that match those of the simulated fluctuations (to the extent that the model matches the measured spectrum), but do not attempt to match the Fourier \textit{phases} of the simulated fluctuation field. Since the fluctuations are highly nonlinear in amplitude, phases are correlated in a nontrivial way. As a result, model fluctuations generated by both methods lack many of the visual characteristics of the simulated ISM, like filamentary features (see also \autoref{sec:nongaussianity_anisotropy} and  \autoref{fig:GRF_comparison}). However, these model fluctuations still prove useful for capturing the ISM-driven orbital heating/migration, as we show in Modak et al. (in prep.). This is what one expects from quasilinear diffusion theory, where phase information can be neglected \citep{BinneyLacey1988, HamiltonFouvry2024}.

%%%%%%%%%%%%%%%%%%%%%%%%%%%%%%%%%%%%%%%%%%%%%%%%%%%%%%%%%%%%
\subsection{Producing a Spatio-temporal Gaussian Random Field}
\label{sec:generating_GRFs_direct}
%%%%%%%%%%%%%%%%%%%%%%%%%%%%%%%%%%%%%%%%%%%%%%%%%%%%%%%%%%%%

The most direct method for generating a time series of fluctuation snapshots given their spatio-temporal power spectrum is to approximate the fluctuations as a GRF. A field produced in this way will have a Gaussian pdf, so it is most physically appropriate to use this method for a field $Q = s$. The standard algorithm for producing a realization of a GRF with power spectrum $P_Q(k, \omega)$ on a finite grid of size $N_x$, $N_y$, and $N_t$ in the $x$, $y$, and $t$ directions respectively is as follows:
\begin{enumerate}[label=(\roman*)]
    \item Sample $N_x \cdot N_y \cdot N_t$ values from a standard normal distribution, and reshape the values to the desired shape of the full sequence of snapshots of the fluctuation field, i.e., assign spatial and temporal coordinates to each sample to produce a field $f(\bx, t)$ with a Gaussian pdf.
    \item Fourier transform the resulting field in both space and time to produce $f_\bk(\omega)$.
    \item Multiply the field values $f_\bk(\omega)$ by $\sqrt{P_Q(k, \omega)}$.
    \item Invert the Fourier transform to arrive at a field realization $Q(\bx, t)$ with the desired spectrum $P_Q$.
\end{enumerate}
Implicit in the use of the Fourier transform in steps (ii) and (iv) is the assumption that the field satisfies periodic boundary conditions in space. However, the ISM fluctuations studied in this work instead satisfy \textit{shearing}-periodic boundary conditions as given in \autoref{eq:shearing_periodic}. Thus, the standard GRF realization algorithm must be modified as follows: 
\begin{enumerate}[label=(\roman*)]
    \item Follow steps (i)-(iii) as described above.
    \item Instead of performing the inverse Fourier transform in both space and time as described in (iv) above, first, only the temporal dimension should be inverse transformed, to produce $Q_\bk(t)$.
    \item Convert from the unsheared wavevectors $\bk$ to the sheared wavevectors $\bk'$ using the time-dependent relations given in \autoref{eq:sheared_kx} and \autoref{eq:sheared_ky}.
    \item Inverse Fourier transform the resulting field to arrive at the field $Q(\bx', t)$ in sheared spatial coordinates.
    \item Convert from the sheared coordinates $\bx'$ to the unsheared coordinates $\bx$ using the time-dependent relations given in \autoref{eq:sheared_x} and \autoref{eq:sheared_y} to arrive at a realization $Q(\bx, t)$.
\end{enumerate}
By separately performing the temporal and spatial transforms in this way, the spatio-temporal power spectrum remains unchanged, but the field now satisfies spatial shearing-periodic boundary conditions as desired. Note that the use of the time-dependent transformation from sheared to unsheared coordinates requires the choice of a temporal origin, and the snapshots will include a discontinuous jump in time-evolution at each ``nearest periodic time'' $t_n = nL_y/(q\Omega L_x)$, because the shearing transform depends on $(t-t_n)$ and $n = \lfloor q\Omega tL_x/L_y \rceil$ is a discontinuous function of time. However, in preliminary numerical tests of the dynamical influence of ISM fluctuations produced using this method, we find this point discontinuity to have no effect on the ensemble-averaged evolution of stellar orbits (Modak et al., in prep).

%%%%%%%%%%%%%%%%%%%%%%%%%%%%%%%%%%%%%%%%%%%%%%%%%%%%%%%%%%%%
\subsection{Spatial GRFs with Exponential Time-Dependence}
\label{sec:generating_GRFs_exp}
%%%%%%%%%%%%%%%%%%%%%%%%%%%%%%%%%%%%%%%%%%%%%%%%%%%%%%%%%%%%

The temporal discontinuity described above occurs because of the dependence on $t-t_n$, which is necessary to prevent fluctuations winding to very high azimuthal wavevectors $k_y$. However, this is simply a shortcoming of the GRF realization algorithm: in reality, fluctuation structures are disrupted by feedback processes on timescales much shorter than the winding time. An alternative method, which avoids these discontinuities, is to use the time-dependence specified directly in \autoref{eq:Q_k_t_model}. The procedure is as follows:
\begin{enumerate}[label=(\roman*)]
    \item Choose $N_p$ ``peak times'' $\{t_j\}$ uniformly at random from the interval $[-t_\mathrm{buffer}, t_\mathrm{sim}+t_\mathrm{buffer}]$ where $t_\mathrm{sim}$ is the desired temporal baseline for the fluctuations, and the extension of the interval by $t_\mathrm{buffer}$ on either side allows for modes to be rising or falling in amplitude even at the boundaries of the time interval of interest.
    \item For each peak time $t_j$, perform the following steps to generate a (purely spatial) GRF:
    \begin{enumerate}[label=(\alph*)]
        \item Sample $N_x \cdot N_y$ values from a standard normal distribution, and reshape the values to the desired shape of a single snapshot of the fluctuation field, i.e., assign spatial coordinates to each sample to produce an uncorrelated field $f^{(j)}(\bx)$.
        \item Fourier transform the resulting field in space to produce $f^{(j)}_\bk$.
        \item Multiply the field values $f_\bk$ by $\sqrt{P_Q(k)\omega_0(k)t_\mathrm{sim}/(4N_p)}$ to produce a realization of a single structure $Q^{(j)}_\bk$. In addition to the spatial power spectrum, the factor of $\omega_0(k)t_\mathrm{sim}/(4N_p)$ ensures the resulting superposition of structures is appropriately normalized.
    \end{enumerate}
    \item For each desired snapshot time $t$, combine the individual GRFs $Q^{(j)}_\bk$ as follows:
    \begin{enumerate}[label=(\alph*)]
        \item Multiply the values $Q^{(j)}_\bk$ by $e^{-\omega_0(k)|t-t_j|}$ to produce each structure's time-dependent contribution to the full fluctuation field, $Q^{(j)}_\bk(t)$.
        \item Convert from the unsheared wavevectors $\bk$ to the sheared wavevectors $\bk'$ using the time-dependent relations given in \autoref{eq:sheared_kx} and \autoref{eq:sheared_ky}, except using $t$ in place of $(t-t_n)$, e.g., $k_x' = k_x - q\Omega t k_y$, to produce each structure's contribution in sheared wavevectors, $Q^{(j)}_{\bk'}(t)$.
        \item Sum the contributions to arrive at $Q_{\bk'}(t) \equiv \sum_j Q^{(j)}_{\bk'}(t)$.
    \end{enumerate}
    \item Inverse Fourier transform the resulting field to arrive at the field $Q(\bx', t)$ in sheared spatial coordinates.
    \item Convert from the sheared coordinates $\bx'$ to the unsheared coordinates $\bx$ using the time-dependent relations given in \autoref{eq:sheared_x} and \autoref{eq:sheared_y}, except using $t$ in place of $(t-t_n)$ as in (iiib) above, to arrive at a realization $Q(\bx, t)$.
\end{enumerate}
This method correctly reproduces the model spatio-temporal power spectrum and does not include any temporal discontinuities. The price we pay is that we have introduced two new free parameters $N_p$ and $t_\mathrm{buffer}$. Empirically, we find that setting $N_p \sim t_\mathrm{sim}/\tau_0$ and $t_\mathrm{buffer} \sim 2\tau_0$ yields fluctuation fields with spatio-temporal power spectra that are well-converged to the values measured in the TIGRESS simulations; these choices ensure that on average, at least one large-scale structure is near its peak amplitude for each simulation snapshot (see \autoref{eq:omega_k}). We note that in preliminary numerical tests, we find no significant differences between the dynamical influences of ISM fluctuations produced using this method and the GRF method described in the previous subsection (Modak et al., in prep).

%%%%%%%%%%%%%%%%%%%%%%%%%%%%%%%%%%%%%%%%%%%%%%%%%%%%%%%%%%%%
%%%%%%%%%%%%%%%%%%%%%%%%%%%%%%%%%%%%%%%%%%%%%%%%%%%%%%%%%%%%
\section{Fitting a Fluctuation Vertical Profile}
\label{sec:zetagtr}
%%%%%%%%%%%%%%%%%%%%%%%%%%%%%%%%%%%%%%%%%%%%%%%%%%%%%%%%%%%%
%%%%%%%%%%%%%%%%%%%%%%%%%%%%%%%%%%%%%%%%%%%%%%%%%%%%%%%%%%%%

As detailed in \autoref{eq:deltarho_model}, the separable density model we apply throughout \autoref{sec:rho_phi} assumes that the vertical profile $\zeta$ of the $k = 0$ component of the density also applies to the fluctuations, while in reality, each Fourier component really possesses its own vertical profile $\zeta_\bk$ that need not be equal to the mean $\zeta_{k = 0}$. While the simple ``single-$\zeta$'' model we present in the main text corresponds reasonably well with what is measured in the simulations, below we describe a simple extension of the model allowing for the $k = 0$ and $k > 0$ components of the density to have distinct vertical profiles, which we will denote $\zeta$ and $\zeta^>$ respectively. Of course, further subdividing $\zeta^>$ into several individual $\zeta_\bk$ would improve the fit, but at the cost of significantly complicating the model, so we refrain from doing so here.

To be precise, our model for the fluctuations is
\begin{equation}
    \label{eq:deltarho_zetagtr}
    \delta\rho_\mathrm{m}(\bx, z, t) = \frac{\overline{\Sigma}(t)}{2h_\mathrm{m}^>(t)} \zeta_\mathrm{m}^>(\ztilde) \delta(\bx, t),
\end{equation}
where again we denote our model quantities with a subscript m and quantities measured in the simulation without a subscript; $\zeta_\mathrm{m}^>$ is our model for the $k > 0$ vertical structure, and the thickness $h_\mathrm{m}^>$ is calculated from $\zeta_\mathrm{m}^>$ using \autoref{eq:h_def} as usual. We fit the same mixture of a $\sech^2$ and exponential profile,
\begin{equation}
    \zeta_\mathrm{m}^>(\ztilde) = (1-w^>)\sech^2\left(\frac{\ztilde}{\alpha_\mathrm{s}^>}\right) + w^>e^{-|\ztilde|/\alpha_\mathrm{e}^>},
\end{equation}
but this time to the fluctuation power spectra, i.e., we fit a model of the form \autoref{eq:power_phi_Sigma} to the measured potential power spectrum $P_{\delta\phi}(\ktilde, \ztilde)$, except using $\zeta_\mathrm{m}^>$ in place of $\zeta_\mathrm{m}$ in e.g., \autoref{eq:phi_Sigma_k_def}.

\begin{figure*}
    \centering
    \includegraphics[width=0.95\textwidth]{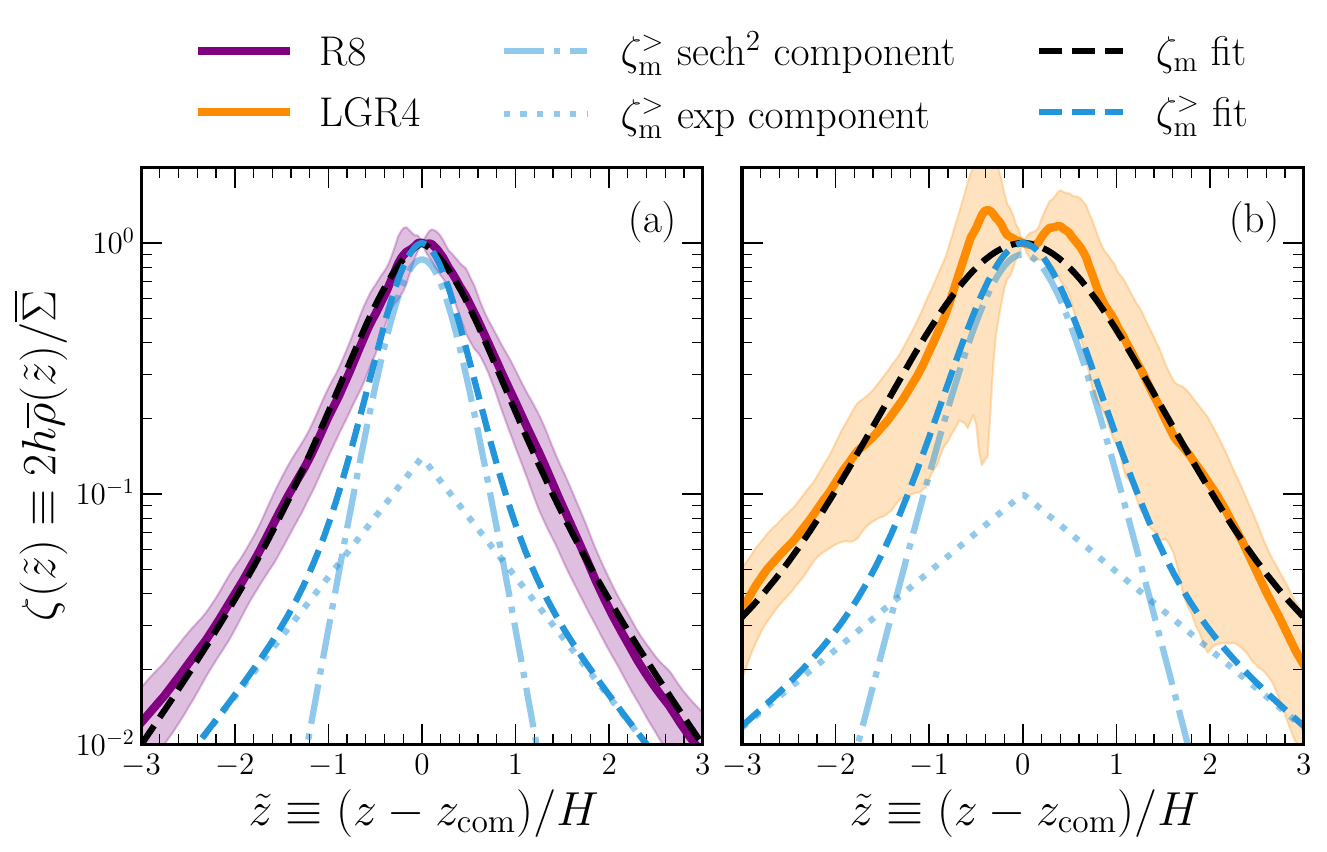}
    \caption{Analogous to \autoref{fig:zeta}, but showing the fits to the $k > 0$ potential fluctuations, $\zeta^>_\mathrm{m}$ (blue dashed curves in both panels), in addition to the original $\zeta_\mathrm{m}$ fit to only the $k = 0$ density profile, which is overplotted as a black dashed curve for comparison. The individual $\sech^2$ (blue dot-dashed curve) and exponential (blue dotted curve) components of the $\zeta_\mathrm{m}^>$ fit are also shown for reference.}
    \label{fig:zetagtr}
\end{figure*}

\begin{table}
    \centering
    \begin{tabular}{ccccc}
    \hline
    Sim & $w^>$ & $\alpha_\mathrm{s}^>$ & $\alpha^>_e$ & $\alpha_w^>$ \\
    (1) & (2) & (3) & (4) & (5) \\
    \hline
    \hline 
    R8 & 0.14$\pm$0.04 & 0.42$\pm$0.06 & 0.91$\pm$0.06 & 0.49$\pm0.06$ \\
    LGR4 & 0.11$\pm$0.09 & 0.76$\pm$0.06 & 1.2$\pm$0.5 & 0.81$\pm$0.09 \\
    \hline
    \end{tabular}
    \caption{Analogous to \autoref{tab:zeta_measurements}, but for the function $\zeta^>_\mathrm{m}$ derived from fits to the $k > 0$ potential fluctuations.}
    \label{tab:zetagtr_measurements}
\end{table}

We summarize the results of this fitting process in \autoref{tab:zetagtr_measurements}, and compare the fitted function $\zeta_\mathrm{m}^>$ to the model $\zeta_\mathrm{m}$ derived from the $k = 0$ density component, as well as the mean vertical profile $\zeta$ as measured from the simulation in \autoref{fig:zetagtr}. We see that the fluctuations are much more strongly concentrated around the midplane, as evident from the results of \autoref{sec:vertical_kgtr} and anticipated physically (since the feedback driving the fluctuations is strongest at the midplane).

\begin{figure*}
    \centering
    \includegraphics[width=0.95\textwidth]{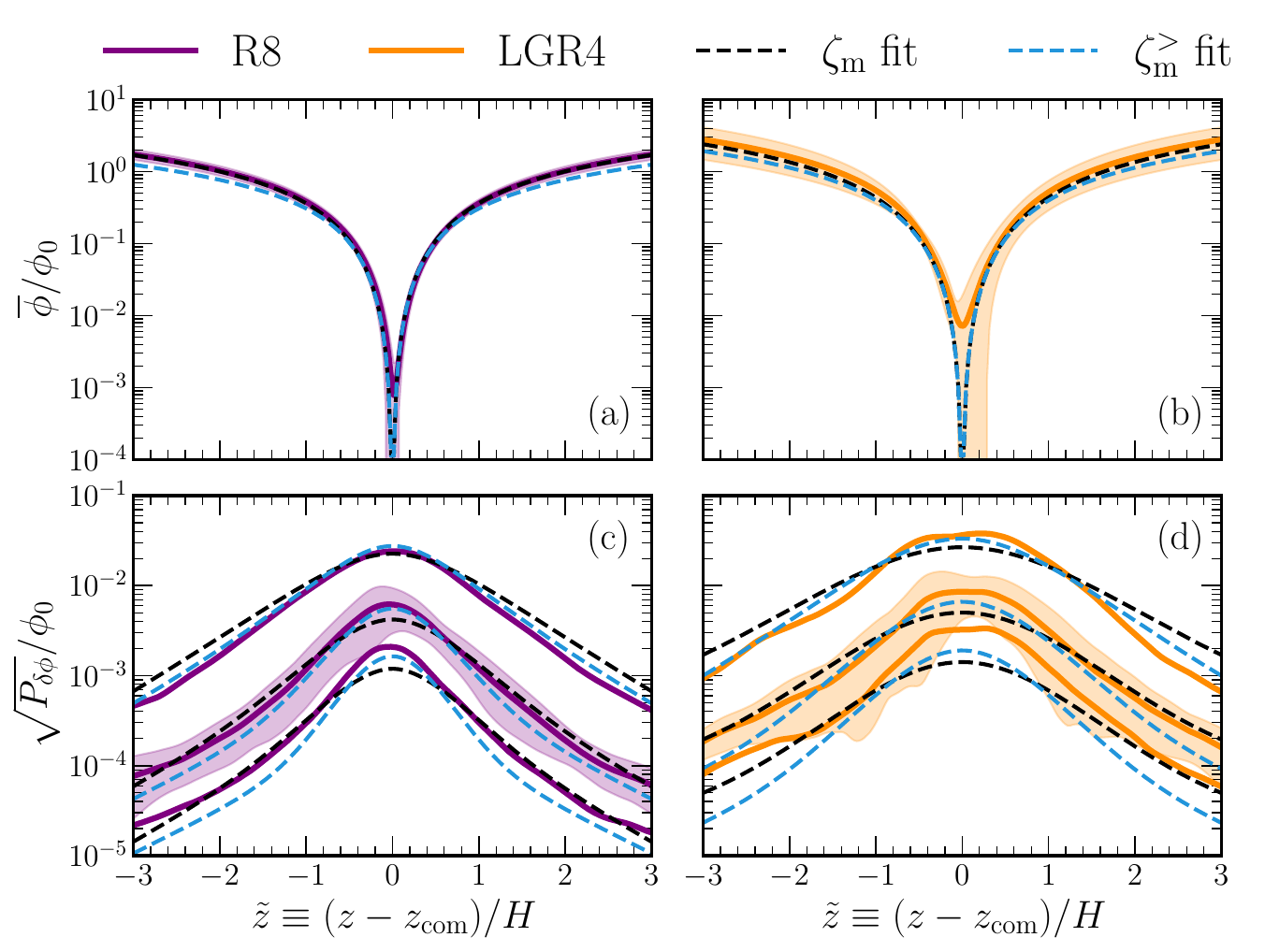}
    \caption{Analogous to \autoref{fig:phi0_phik_z}, but showing the fits to the $k > 0$ potential fluctuations, $\zeta^>_\mathrm{m}$ (blue dashed curves in both panels), in addition to the original $\zeta_\mathrm{m}$ fit to only the $k = 0$ density profile, which is overplotted as a black dashed curve for comparison.}
    \label{fig:phi0_phik_z_zetagtr}
\end{figure*}

\begin{figure*}
    \centering
    \includegraphics[width=0.95\textwidth]{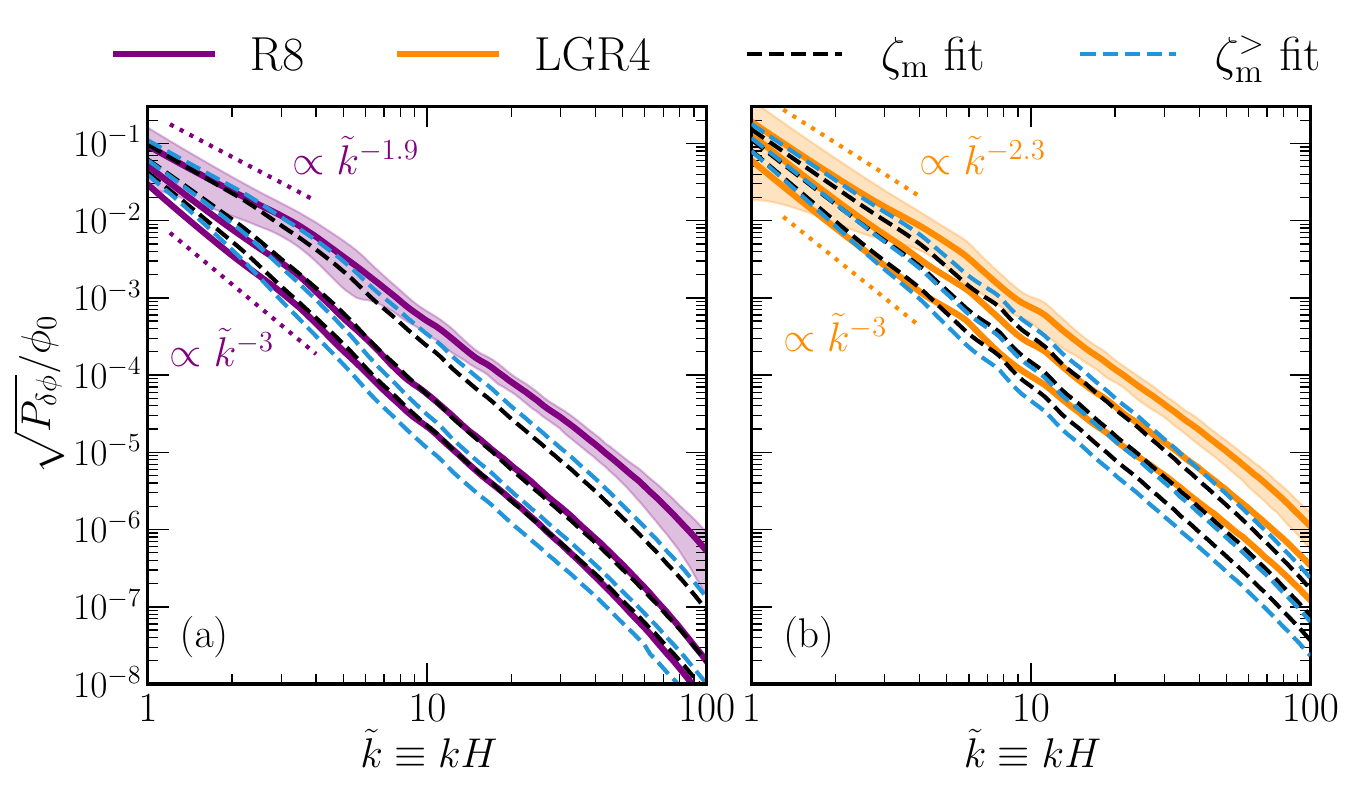}
    \caption{Analogous to \autoref{fig:phik_k}, but showing the fits to the $k > 0$ potential fluctuations, $\zeta^>_\mathrm{m}$ (shown in a light blue dashed curve in both panels), in addition to the original $\zeta_\mathrm{m}$ fit to only the $k = 0$ density profile, which is overplotted as a black dashed curve for comparison.}
    \label{fig:phik_k_zetagtr}
\end{figure*}

In \autoref{fig:phi0_phik_z_zetagtr} and \autoref{fig:phik_k_zetagtr} we compare the potential fluctuation power spectrum $P_{\delta\phi}$ measured in the simulation with both the $\zeta_\mathrm{m}$ and $\zeta_\mathrm{m}^>$ vertical profile models. In each figure, the mean vertical profile $\zeta_\mathrm{m}$ is shown as a black dashed curve as in the main text, and the fitted $\zeta_\mathrm{m}^>$ is shown as a light blue dashed curve. In \autoref{fig:phi0_phik_z_zetagtr}, we see that the fits are both consistent with the measurements to within $\pm 1 \sigma$, but highlight that fitting for the $k > 0$ vertical structure directly allows for a much closer match to the midplane amplitude of the fluctuations---for instance, the blue dashed curves are much closer to the simulation curves in panels (c) and (d). However, the $k > 0$ fit is somewhat less accurate than the $k = 0$ vertical structure fit at larger $|\ztilde|$. In \autoref{fig:phik_k_zetagtr}, we again see that at the midplane (top curve in each panel), the $\zeta_\mathrm{m}^>$ fit is closer to the simulation results, particularly at the smallest $k$ values that contain most of the power, although once again the fits suffer at larger $|\ztilde|$ (the lower two curves in each panel).

Ultimately, we have shown that the $\zeta_\mathrm{m}^>$ model is more well-suited as a characterization of the density and potential fluctuations near the midplane, and recommend its use for applications restricted to within approximately $|\ztilde| \lesssim 1$. However, because it requires several additional parameters and is somewhat less accurate away from the midplane than the simple ``single-$\zeta$'' model presented in the main text, it is not essential to use in general, and for applications spanning a wider range of vertical scales, using the fit $\zeta_\mathrm{m}$ of the main text is recommended.

\bibliography{bibliography}{}

\begin{thebibliography}{}
\expandafter\ifx\csname natexlab\endcsname\relax\def\natexlab#1{#1}\fi
\providecommand{\url}[1]{\href{#1}{#1}}
\providecommand{\dodoi}[1]{doi:~\href{http://doi.org/#1}{\nolinkurl{#1}}}
\providecommand{\doeprint}[1]{\href{http://ascl.net/#1}{\nolinkurl{http://ascl.net/#1}}}
\providecommand{\doarXiv}[1]{\href{https://arxiv.org/abs/#1}{\nolinkurl{https://arxiv.org/abs/#1}}}

\bibitem[{E. {Allys} {et~al.}(2019){Allys}, {Levrier}, {Zhang}, {Colling}, {Regaldo-Saint Blancard}, {Boulanger}, {Hennebelle}, \& {Mallat}}]{Allys2019}
{Allys}, E., {Levrier}, F., {Zhang}, S., {et~al.} 2019, \bibinfo{title}{{The RWST, a comprehensive statistical description of the non-Gaussian structures in the ISM},} \aap, 629, A115, \dodoi{10.1051/0004-6361/201834975}

\bibitem[{M. {Aumer} {et~al.}(2016{\natexlab{a}}){Aumer}, {Binney}, \& {Sch{\"o}nrich}}]{AumerBinneySchonrich2016a}
{Aumer}, M., {Binney}, J., \& {Sch{\"o}nrich}, R. 2016{\natexlab{a}}, \bibinfo{title}{{The quiescent phase of galactic disc growth},} \mnras, 459, 3326, \dodoi{10.1093/mnras/stw777}

\bibitem[{M. {Aumer} {et~al.}(2016{\natexlab{b}}){Aumer}, {Binney}, \& {Sch{\"o}nrich}}]{AumerBinneySchonrich2016b}
{Aumer}, M., {Binney}, J., \& {Sch{\"o}nrich}, R. 2016{\natexlab{b}}, \bibinfo{title}{{Age-velocity dispersion relations and heating histories in disc galaxies},} \mnras, 462, 1697, \dodoi{10.1093/mnras/stw1639}

\bibitem[{J. {Ballesteros-Paredes} {et~al.}(2007){Ballesteros-Paredes}, {Klessen}, {Mac Low}, \& {Vazquez-Semadeni}}]{Ballesteros2007}
{Ballesteros-Paredes}, J., {Klessen}, R.~S., {Mac Low}, M.~M., \& {Vazquez-Semadeni}, E. 2007, in Protostars and Planets V, ed. B.~{Reipurth}, D.~{Jewitt}, \& K.~{Keil}, 63, \dodoi{10.48550/arXiv.astro-ph/0603357}

\bibitem[{J. {Bally} {et~al.}(1987){Bally}, {Langer}, {Stark}, \& {Wilson}}]{Bally1987}
{Bally}, J., {Langer}, W.~D., {Stark}, A.~A., \& {Wilson}, R.~W. 1987, \bibinfo{title}{{Filamentary Structure in the Orion Molecular Cloud},} \apjl, 312, L45, \dodoi{10.1086/184817}

\bibitem[{J. {Bally} {et~al.}(1988){Bally}, {Stark}, {Wilson}, \& {Henkel}}]{Bally1988}
{Bally}, J., {Stark}, A.~A., {Wilson}, R.~W., \& {Henkel}, C. 1988, \bibinfo{title}{{Galactic Center Molecular Clouds. II. Distribution and Kinematics},} \apj, 324, 223, \dodoi{10.1086/165891}

\bibitem[{G.~K. {Batchelor}(1951){Batchelor}}]{Batchelor1951}
{Batchelor}, G.~K. 1951, \bibinfo{title}{{Pressure fluctuations in isotropic turbulence},} Proceedings of the Cambridge Philosophical Society, 47, 359, \dodoi{10.1017/S0305004100026712}

\bibitem[{J.~R. {Beattie} {et~al.}(2025){Beattie}, {Kolborg}, {Ramirez-Ruiz}, \& {Federrath}}]{Beattie2025b}
{Beattie}, J.~R., {Kolborg}, A.~N., {Ramirez-Ruiz}, E., \& {Federrath}, C. 2025, \bibinfo{title}{{So Long Kolmogorov: The Forward and Backward Turbulence Cascades in a Supernovae-driven, Multiphase Interstellar Medium},} \apj, 994, 193, \dodoi{10.3847/1538-4357/ae07cd}

\bibitem[{J. {Binney} \& C. {Lacey}(1988){Binney} \& {Lacey}}]{BinneyLacey1988}
{Binney}, J., \& {Lacey}, C. 1988, \bibinfo{title}{{The diffusion of stars through phase space},} \mnras, 230, 597, \dodoi{10.1093/mnras/230.4.597}

\bibitem[{L. {Blitz}(1993){Blitz}}]{Blitz1993}
{Blitz}, L. 1993, in Protostars and Planets III, ed. E.~H. {Levy} \& J.~I. {Lunine}, 125

\bibitem[{L. {Blitz} \& P. {Thaddeus}(1980){Blitz} \& {Thaddeus}}]{Blitz1980}
{Blitz}, L., \& {Thaddeus}, P. 1980, \bibinfo{title}{{Giant molecular complexes and OB associations. I. The Rosette molecular complex.},} \apj, 241, 676, \dodoi{10.1086/158379}

\bibitem[{A.~D. {Bolatto} {et~al.}(2008){Bolatto}, {Leroy}, {Rosolowsky}, {Walter}, \& {Blitz}}]{Bolatto2008}
{Bolatto}, A.~D., {Leroy}, A.~K., {Rosolowsky}, E., {Walter}, F., \& {Blitz}, L. 2008, \bibinfo{title}{{The Resolved Properties of Extragalactic Giant Molecular Clouds},} \apj, 686, 948, \dodoi{10.1086/591513}

\bibitem[{J.~R. {Bond} {et~al.}(1991){Bond}, {Cole}, {Efstathiou}, \& {Kaiser}}]{Bond1991}
{Bond}, J.~R., {Cole}, S., {Efstathiou}, G., \& {Kaiser}, N. 1991, \bibinfo{title}{{Excursion Set Mass Functions for Hierarchical Gaussian Fluctuations},} \apj, 379, 440, \dodoi{10.1086/170520}

\bibitem[{N. {Brucy} {et~al.}(2020){Brucy}, {Hennebelle}, {Bournaud}, \& {Colling}}]{Brucy2020}
{Brucy}, N., {Hennebelle}, P., {Bournaud}, F., \& {Colling}, C. 2020, \bibinfo{title}{{Large-scale Turbulent Driving Regulates Star Formation in High-redshift Gas-rich Galaxies},} \apjl, 896, L34, \dodoi{10.3847/2041-8213/ab9830}

\bibitem[{J. Burgers(1948)Burgers}]{Burgers1948}
Burgers, J. 1948, in  (Elsevier), 171--199, \dodoi{https://doi.org/10.1016/S0065-2156(08)70100-5}

\bibitem[{P.-H. Chavanis(2023)Chavanis}]{chavanis2023secular}
Chavanis, P.-H. 2023, \bibinfo{title}{The Secular Dressed Diffusion Equation,} Universe, 9, 68

\bibitem[{M. {Chevance} {et~al.}(2023){Chevance}, {Krumholz}, {McLeod}, {Ostriker}, {Rosolowsky}, \& {Sternberg}}]{Chevance2023}
{Chevance}, M., {Krumholz}, M.~R., {McLeod}, A.~F., {et~al.} 2023, in Astronomical Society of the Pacific Conference Series, Vol. 534, Protostars and Planets VII, ed. S.~{Inutsuka}, Y.~{Aikawa}, T.~{Muto}, K.~{Tomida}, \& M.~{Tamura}, 1, \dodoi{10.48550/arXiv.2203.09570}

\bibitem[{T.~M. {Dame} {et~al.}(1986){Dame}, {Elmegreen}, {Cohen}, \& {Thaddeus}}]{Dame1986}
{Dame}, T.~M., {Elmegreen}, B.~G., {Cohen}, R.~S., \& {Thaddeus}, P. 1986, \bibinfo{title}{{The Largest Molecular Cloud Complexes in the First Galactic Quadrant},} \apj, 305, 892, \dodoi{10.1086/164304}

\bibitem[{C.~L. {Dobbs} {et~al.}(2014){Dobbs}, {Krumholz}, {Ballesteros-Paredes}, {Bolatto}, {Fukui}, {Heyer}, {Low}, {Ostriker}, \& {V{\'a}zquez-Semadeni}}]{Dobbs2014}
{Dobbs}, C.~L., {Krumholz}, M.~R., {Ballesteros-Paredes}, J., {et~al.} 2014, in Protostars and Planets VI, ed. H.~{Beuther}, R.~S. {Klessen}, C.~P. {Dullemond}, \& T.~{Henning}, 3--26, \dodoi{10.2458/azu_uapress_9780816531240-ch001}

\bibitem[{B.~T. {Draine}(2011){Draine}}]{Draine2011}
{Draine}, B.~T. 2011, {Physics of the Interstellar and Intergalactic Medium}

\bibitem[{P. {Dutta} {et~al.}(2009){Dutta}, {Begum}, {Bharadwaj}, \& {Chengalur}}]{Dutta2009}
{Dutta}, P., {Begum}, A., {Bharadwaj}, S., \& {Chengalur}, J.~N. 2009, \bibinfo{title}{{The scaleheight of NGC 1058 measured from its HI power spectrum},} \mnras, 397, L60, \dodoi{10.1111/j.1745-3933.2009.00684.x}

\bibitem[{P. {Dutta} {et~al.}(2013){Dutta}, {Begum}, {Bharadwaj}, \& {Chengalur}}]{Dutta2013}
{Dutta}, P., {Begum}, A., {Bharadwaj}, S., \& {Chengalur}, J.~N. 2013, \bibinfo{title}{{Probing interstellar turbulence in spiral galaxies using H I power spectrum analysis},} \na, 19, 89, \dodoi{10.1016/j.newast.2012.08.008}

\bibitem[{B. {Elmegreen} {et~al.}(2025){Elmegreen}, {Adamo}, {Bajaj}, {Duarte-Cabral}, {Calzetti}, {Cignoni}, {Correnti}, {Gallagher}, {Grasha}, {Gregg}, {Johnson}, {Linden}, {Messa}, {Ostlin}, {Pedrini}, \& {Ryon}}]{Elmegreen2025}
{Elmegreen}, B., {Adamo}, A., {Bajaj}, V., {et~al.} 2025, \bibinfo{title}{{Power Spectra of JWST images of Local Galaxies: Searching for Disk Thickness},} The Open Journal of Astrophysics, 8, 21, \dodoi{10.33232/001c.130810}

\bibitem[{B.~G. {Elmegreen} {et~al.}(2001){Elmegreen}, {Kim}, \& {Staveley-Smith}}]{Elmegreen2001}
{Elmegreen}, B.~G., {Kim}, S., \& {Staveley-Smith}, L. 2001, \bibinfo{title}{{A Fractal Analysis of the H I Emission from the Large Magellanic Cloud},} \apj, 548, 749, \dodoi{10.1086/319021}

\bibitem[{B.~G. {Elmegreen} \& J. {Scalo}(2004){Elmegreen} \& {Scalo}}]{elmegreen2004}
{Elmegreen}, B.~G., \& {Scalo}, J. 2004, \bibinfo{title}{{Interstellar Turbulence I: Observations and Processes},} \araa, 42, 211, \dodoi{10.1146/annurev.astro.41.011802.094859}

\bibitem[{C.~F. {Gammie}(2001){Gammie}}]{Gammie2001}
{Gammie}, C.~F. 2001, \bibinfo{title}{{Nonlinear Outcome of Gravitational Instability in Cooling, Gaseous Disks},} \apj, 553, 174, \dodoi{10.1086/320631}

\bibitem[{P. {Goldreich} \& D. {Lynden-Bell}(1965){Goldreich} \& {Lynden-Bell}}]{GoldreichLyndenBell1965}
{Goldreich}, P., \& {Lynden-Bell}, D. 1965, \bibinfo{title}{{II. Spiral arms as sheared gravitational instabilities},} \mnras, 130, 125, \dodoi{10.1093/mnras/130.2.125}

\bibitem[{H. {Gong} \& E.~C. {Ostriker}(2013){Gong} \& {Ostriker}}]{GongOstriker2013}
{Gong}, H., \& {Ostriker}, E.~C. 2013, \bibinfo{title}{{Implementation of Sink Particles in the Athena Code},} \apjs, 204, 8, \dodoi{10.1088/0067-0049/204/1/8}

\bibitem[{M. {Gong} {et~al.}(2017){Gong}, {Ostriker}, \& {Wolfire}}]{Gong2017}
{Gong}, M., {Ostriker}, E.~C., \& {Wolfire}, M.~G. 2017, \bibinfo{title}{{A Simple and Accurate Network for Hydrogen and Carbon Chemistry in the Interstellar Medium},} \apj, 843, 38, \dodoi{10.3847/1538-4357/aa7561}

\bibitem[{T. {Gotoh} \& D. {Fukayama}(2001){Gotoh} \& {Fukayama}}]{Gotoh2001}
{Gotoh}, T., \& {Fukayama}, D. 2001, \bibinfo{title}{{Pressure Spectrum in Homogeneous Turbulence},} \prl, 86, 3775, \dodoi{10.1103/PhysRevLett.86.3775}

\bibitem[{K. {Grisdale} {et~al.}(2017){Grisdale}, {Agertz}, {Romeo}, {Renaud}, \& {Read}}]{Grisdale2017}
{Grisdale}, K., {Agertz}, O., {Romeo}, A.~B., {Renaud}, F., \& {Read}, J.~I. 2017, \bibinfo{title}{{The impact of stellar feedback on the density and velocity structure of the interstellar medium},} \mnras, 466, 1093, \dodoi{10.1093/mnras/stw3133}

\bibitem[{A. {Gurman} {et~al.}(2025){Gurman}, {Steinwandel}, {Hu}, \& {Sternberg}}]{Gurman2025}
{Gurman}, A., {Steinwandel}, U.~P., {Hu}, C.-Y., \& {Sternberg}, A. 2025, \bibinfo{title}{{The GHOSDT Simulations: I. Magnetic Support in Gas-rich Disks},} \apj, 984, 142, \dodoi{10.3847/1538-4357/adc814}

\bibitem[{A. {Hacar} {et~al.}(2023){Hacar}, {Clark}, {Heitsch}, {Kainulainen}, {Panopoulou}, {Seifried}, \& {Smith}}]{Hacar2023}
{Hacar}, A., {Clark}, S.~E., {Heitsch}, F., {et~al.} 2023, in Astronomical Society of the Pacific Conference Series, Vol. 534, Protostars and Planets VII, ed. S.~{Inutsuka}, Y.~{Aikawa}, T.~{Muto}, K.~{Tomida}, \& M.~{Tamura}, 153, \dodoi{10.48550/arXiv.2203.09562}

\bibitem[{C. {Hamilton} \& J.-B. {Fouvry}(2024){Hamilton} \& {Fouvry}}]{HamiltonFouvry2024}
{Hamilton}, C., \& {Fouvry}, J.-B. 2024, \bibinfo{title}{{Kinetic theory of stellar systems: A tutorial},} Physics of Plasmas, 31, 120901, \dodoi{10.1063/5.0204214}

\bibitem[{C. {Hamilton} {et~al.}(2026){Hamilton}, {Modak}, \& {Tremaine}}]{galactokinetics}
{Hamilton}, C., {Modak}, S., \& {Tremaine}, S. 2026, \bibinfo{title}{{Galactokinetics},} \apj, 997, 28, \dodoi{10.3847/1538-4357/ae17bb}

\bibitem[{J.~F. {Hawley} {et~al.}(1995){Hawley}, {Gammie}, \& {Balbus}}]{HawleyGammieBalbus1995}
{Hawley}, J.~F., {Gammie}, C.~F., \& {Balbus}, S.~A. 1995, \bibinfo{title}{{Local Three-dimensional Magnetohydrodynamic Simulations of Accretion Disks},} \apj, 440, 742, \dodoi{10.1086/175311}

\bibitem[{T. {Heinemann} \& J.~C.~B. {Papaloizou}(2009){Heinemann} \& {Papaloizou}}]{HeinemannPapaloizou2009}
{Heinemann}, T., \& {Papaloizou}, J.~C.~B. 2009, \bibinfo{title}{{The excitation of spiral density waves through turbulent fluctuations in accretion discs - I. WKBJ theory},} \mnras, 397, 52, \dodoi{10.1111/j.1365-2966.2009.14799.x}

\bibitem[{T. {Heinemann} \& J.~C.~B. {Papaloizou}(2012){Heinemann} \& {Papaloizou}}]{HeinemannPapaloizou2012}
{Heinemann}, T., \& {Papaloizou}, J.~C.~B. 2012, \bibinfo{title}{{A weakly non-linear theory for spiral density waves excited by accretion disc turbulence},} \mnras, 419, 1085, \dodoi{10.1111/j.1365-2966.2011.19763.x}

\bibitem[{P. {Hennebelle} \& G. {Chabrier}(2008){Hennebelle} \& {Chabrier}}]{HennebelleChabrier2008}
{Hennebelle}, P., \& {Chabrier}, G. 2008, \bibinfo{title}{{Analytical Theory for the Initial Mass Function: CO Clumps and Prestellar Cores},} \apj, 684, 395, \dodoi{10.1086/589916}

\bibitem[{M. {Heyer} \& T.~M. {Dame}(2015){Heyer} \& {Dame}}]{Heyer2015}
{Heyer}, M., \& {Dame}, T.~M. 2015, \bibinfo{title}{{Molecular Clouds in the Milky Way},} \araa, 53, 583, \dodoi{10.1146/annurev-astro-082214-122324}

\bibitem[{P.~F. {Hopkins}(2012{\natexlab{a}}){Hopkins}}]{Hopkins2012a}
{Hopkins}, P.~F. 2012{\natexlab{a}}, \bibinfo{title}{{An excursion-set model for the structure of giant molecular clouds and the interstellar medium},} \mnras, 423, 2016, \dodoi{10.1111/j.1365-2966.2012.20730.x}

\bibitem[{P.~F. {Hopkins}(2012{\natexlab{b}}){Hopkins}}]{Hopkins2012b}
{Hopkins}, P.~F. 2012{\natexlab{b}}, \bibinfo{title}{{The stellar initial mass function, core mass function and the last-crossing distribution},} \mnras, 423, 2037, \dodoi{10.1111/j.1365-2966.2012.20731.x}

\bibitem[{C.-Y. {Hu} {et~al.}(2021){Hu}, {Sternberg}, \& {van Dishoeck}}]{Hu2021}
{Hu}, C.-Y., {Sternberg}, A., \& {van Dishoeck}, E.~F. 2021, \bibinfo{title}{{Metallicity Dependence of the H/H$_{2}$ and C$^{+}$/C/CO Distributions in a Resolved Self-regulating Interstellar Medium},} \apj, 920, 44, \dodoi{10.3847/1538-4357/ac0dbd}

\bibitem[{P. {Hut} \& S. {Tremaine}(1985){Hut} \& {Tremaine}}]{HutTremaine1985}
{Hut}, P., \& {Tremaine}, S. 1985, \bibinfo{title}{{Have interstellar clouds disrupted the Oort comet cloud?},} \aj, 90, 1548, \dodoi{10.1086/113868}

\bibitem[{S. {Ida} {et~al.}(1993){Ida}, {Kokubo}, \& {Makino}}]{Ida1993}
{Ida}, S., {Kokubo}, E., \& {Makino}, J. 1993, \bibinfo{title}{{The Origin of Anisotropic Velocity Dispersion of Particles in a Disc Potential},} \mnras, 263, 875, \dodoi{10.1093/mnras/263.4.875}

\bibitem[{P.~S. {Iroshnikov}(1964){Iroshnikov}}]{Iroshnikov1964}
{Iroshnikov}, P.~S. 1964, \bibinfo{title}{{Turbulence of a Conducting Fluid in a Strong Magnetic Field},} \sovast, 7, 566

\bibitem[{G. {J{\'o}hannesson} {et~al.}(2018){J{\'o}hannesson}, {Porter}, \& {Moskalenko}}]{JohannessonPorterMoskalenko2018}
{J{\'o}hannesson}, G., {Porter}, T.~A., \& {Moskalenko}, I.~V. 2018, \bibinfo{title}{{The Three-dimensional Spatial Distribution of Interstellar Gas in the Milky Way: Implications for Cosmic Rays and High-energy Gamma-ray Emissions},} \apj, 856, 45, \dodoi{10.3847/1538-4357/aab26e}

\bibitem[{R. {Kannan} {et~al.}(2020){Kannan}, {Marinacci}, {Simpson}, {Glover}, \& {Hernquist}}]{Kannan2020}
{Kannan}, R., {Marinacci}, F., {Simpson}, C.~M., {Glover}, S. C.~O., \& {Hernquist}, L. 2020, \bibinfo{title}{{Efficacy of early stellar feedback in low gas surface density environments},} \mnras, 491, 2088, \dodoi{10.1093/mnras/stz3078}

\bibitem[{C.-G. {Kim} {et~al.}(2023){Kim}, {Kim}, {Gong}, \& {Ostriker}}]{Kim2023}
{Kim}, C.-G., {Kim}, J.-G., {Gong}, M., \& {Ostriker}, E.~C. 2023, \bibinfo{title}{{Introducing TIGRESS-NCR. I. Coregulation of the Multiphase Interstellar Medium and Star Formation Rates},} \apj, 946, 3, \dodoi{10.3847/1538-4357/acbd3a}

\bibitem[{C.-G. {Kim} \& E.~C. {Ostriker}(2017){Kim} \& {Ostriker}}]{KimOstriker2017}
{Kim}, C.-G., \& {Ostriker}, E.~C. 2017, \bibinfo{title}{{Three-phase Interstellar Medium in Galaxies Resolving Evolution with Star Formation and Supernova Feedback (TIGRESS): Algorithms, Fiducial Model, and Convergence},} \apj, 846, 133, \dodoi{10.3847/1538-4357/aa8599}

\bibitem[{C.-G. {Kim} {et~al.}(2020){Kim}, {Ostriker}, {Somerville}, {Bryan}, {Fielding}, {Forbes}, {Hayward}, {Hernquist}, \& {Pandya}}]{Kim2020}
{Kim}, C.-G., {Ostriker}, E.~C., {Somerville}, R.~S., {et~al.} 2020, \bibinfo{title}{{First Results from SMAUG: Characterization of Multiphase Galactic Outflows from a Suite of Local Star-forming Galactic Disk Simulations},} \apj, 900, 61, \dodoi{10.3847/1538-4357/aba962}

\bibitem[{J. {Kim} \& D. {Ryu}(2005){Kim} \& {Ryu}}]{Ryu2005}
{Kim}, J., \& {Ryu}, D. 2005, \bibinfo{title}{{Density Power Spectrum of Compressible Hydrodynamic Turbulent Flows},} \apjl, 630, L45, \dodoi{10.1086/491600}

\bibitem[{J.-G. {Kim} {et~al.}(2023){Kim}, {Gong}, {Kim}, \& {Ostriker}}]{Kim_JG2023}
{Kim}, J.-G., {Gong}, M., {Kim}, C.-G., \& {Ostriker}, E.~C. 2023, \bibinfo{title}{{Photochemistry and Heating/Cooling of the Multiphase Interstellar Medium with UV Radiative Transfer for Magnetohydrodynamic Simulations},} \apjs, 264, 10, \dodoi{10.3847/1538-4365/ac9b1d}

\bibitem[{J.-G. {Kim} {et~al.}(2018){Kim}, {Kim}, \& {Ostriker}}]{KimKimOstriker2018}
{Kim}, J.-G., {Kim}, W.-T., \& {Ostriker}, E.~C. 2018, \bibinfo{title}{{Modeling UV Radiation Feedback from Massive Stars. II. Dispersal of Star-forming Giant Molecular Clouds by Photoionization and Radiation Pressure},} \apj, 859, 68, \dodoi{10.3847/1538-4357/aabe27}

\bibitem[{J.-G. {Kim} {et~al.}(2021){Kim}, {Ostriker}, \& {Filippova}}]{KimOstrikerFilippova2021}
{Kim}, J.-G., {Ostriker}, E.~C., \& {Filippova}, N. 2021, \bibinfo{title}{{Star Formation Efficiency and Dispersal of Giant Molecular Clouds with UV Radiation Feedback: Dependence on Gravitational Boundedness and Magnetic Fields},} \apj, 911, 128, \dodoi{10.3847/1538-4357/abe934}

\bibitem[{W.-T. {Kim} {et~al.}(2020){Kim}, {Kim}, \& {Ostriker}}]{KimKimOstriker2020}
{Kim}, W.-T., {Kim}, C.-G., \& {Ostriker}, E.~C. 2020, \bibinfo{title}{{Local Simulations of Spiral Galaxies with the TIGRESS Framework. I. Star Formation and Arm Spurs/Feathers},} \apj, 898, 35, \dodoi{10.3847/1538-4357/ab9b87}

\bibitem[{E.~W. {Koch} {et~al.}(2020){Koch}, {Chiang}, {Utomo}, {Chastenet}, {Leroy}, {Rosolowsky}, \& {Sandstrom}}]{Koch2020}
{Koch}, E.~W., {Chiang}, I.-D., {Utomo}, D., {et~al.} 2020, \bibinfo{title}{{Spatial power spectra of dust across the Local Group: No constraint on disc scale height},} \mnras, 492, 2663, \dodoi{10.1093/mnras/stz3582}

\bibitem[{E.~W. {Koch} \& E.~W. {Rosolowsky}(2015){Koch} \& {Rosolowsky}}]{KochRosolowsky2015}
{Koch}, E.~W., \& {Rosolowsky}, E.~W. 2015, \bibinfo{title}{{Filament identification through mathematical morphology},} \mnras, 452, 3435, \dodoi{10.1093/mnras/stv1521}

\bibitem[{A. {Kolmogorov}(1941){Kolmogorov}}]{Kolmogorov1941}
{Kolmogorov}, A. 1941, \bibinfo{title}{{The Local Structure of Turbulence in Incompressible Viscous Fluid for Very Large Reynolds' Numbers},} Akademiia Nauk SSSR Doklady, 30, 301

\bibitem[{L. {Konstandin} {et~al.}(2016){Konstandin}, {Schmidt}, {Girichidis}, {Peters}, {Shetty}, \& {Klessen}}]{Konstandin2016}
{Konstandin}, L., {Schmidt}, W., {Girichidis}, P., {et~al.} 2016, \bibinfo{title}{{Mach number study of supersonic turbulence: the properties of the density field},} \mnras, 460, 4483, \dodoi{10.1093/mnras/stw1313}

\bibitem[{R.~H. {Kraichnan}(1965){Kraichnan}}]{Kraichnan1965}
{Kraichnan}, R.~H. 1965, \bibinfo{title}{{Inertial-Range Spectrum of Hydromagnetic Turbulence},} Physics of Fluids, 8, 1385, \dodoi{10.1063/1.1761412}

\bibitem[{A.~G. {Kritsuk} {et~al.}(2007){Kritsuk}, {Norman}, {Padoan}, \& {Wagner}}]{Kritsuk2007}
{Kritsuk}, A.~G., {Norman}, M.~L., {Padoan}, P., \& {Wagner}, R. 2007, \bibinfo{title}{{The Statistics of Supersonic Isothermal Turbulence},} \apj, 665, 416, \dodoi{10.1086/519443}

\bibitem[{M.~R. {Krumholz} {et~al.}(2019){Krumholz}, {McKee}, \& {Bland-Hawthorn}}]{Krumholz2019}
{Krumholz}, M.~R., {McKee}, C.~F., \& {Bland-Hawthorn}, J. 2019, \bibinfo{title}{{Star Clusters Across Cosmic Time},} \araa, 57, 227, \dodoi{10.1146/annurev-astro-091918-104430}

\bibitem[{C.~G. {Lacey}(1984){Lacey}}]{Lacey1984}
{Lacey}, C.~G. 1984, \bibinfo{title}{{The influence of massive gas clouds on stellar velocity dispersions in galactic discs},} \mnras, 208, 687, \dodoi{10.1093/mnras/208.4.687}

\bibitem[{C.~J. {Lada} {et~al.}(1978){Lada}, {Elmegreen}, {Cong}, \& {Thaddeus}}]{Lada1978}
{Lada}, C.~J., {Elmegreen}, B.~G., {Cong}, H.~I., \& {Thaddeus}, P. 1978, \bibinfo{title}{{Molecular clouds in the vicinity of W3, W4, and W5.},} \apjl, 226, L39, \dodoi{10.1086/182826}

\bibitem[{J.~C. {Lee} {et~al.}(2023){Lee}, {Sandstrom}, {Leroy}, {Thilker}, {Schinnerer}, {Rosolowsky}, {Larson}, {Egorov}, {Williams}, {Schmidt}, {Emsellem}, {Anand}, {Barnes}, {Belfiore}, {Be{\v{s}}li{\'c}}, {Bigiel}, {Blanc}, {Bolatto}, {Boquien}, {den Brok}, {Cao}, {Chandar}, {Chastenet}, {Chevance}, {Chiang}, {Congiu}, {Dale}, {Deger}, {Eibensteiner}, {Faesi}, {Glover}, {Grasha}, {Groves}, {Hassani}, {Henny}, {Henshaw}, {Hoyer}, {Hughes}, {Jeffreson}, {Jim{\'e}nez-Donaire}, {Kim}, {Kim}, {Klessen}, {Koch}, {Kreckel}, {Kruijssen}, {Li}, {Liu}, {Lopez}, {Maschmann}, {Chen}, {Meidt}, {Murphy}, {Neumann}, {Neumayer}, {Pan}, {Pessa}, {Pety}, {Querejeta}, {Pinna}, {Rodr{\'\i}guez}, {Saito}, {S{\'a}nchez-Bl{\'a}zquez}, {Santoro}, {Sardone}, {Smith}, {Sormani}, {Scheuermann}, {Stuber}, {Sutter}, {Sun}, {Teng}, {Tre{\ss}}, {Usero}, {Watkins}, {Whitmore}, \& {Razza}}]{Lee2023}
{Lee}, J.~C., {Sandstrom}, K.~M., {Leroy}, A.~K., {et~al.} 2023, \bibinfo{title}{{The PHANGS-JWST Treasury Survey: Star Formation, Feedback, and Dust Physics at High Angular Resolution in Nearby GalaxieS},} \apjl, 944, L17, \dodoi{10.3847/2041-8213/acaaae}

\bibitem[{A.~K. {Leroy} {et~al.}(2021){Leroy}, {Schinnerer}, {Hughes}, {Rosolowsky}, {Pety}, {Schruba}, {Usero}, {Blanc}, {Chevance}, {Emsellem}, {Faesi}, {Herrera}, {Liu}, {Meidt}, {Querejeta}, {Saito}, {Sandstrom}, {Sun}, {Williams}, {Anand}, {Barnes}, {Behrens}, {Belfiore}, {Benincasa}, {Be{\v{s}}li{\'c}}, {Bigiel}, {Bolatto}, {den Brok}, {Cao}, {Chandar}, {Chastenet}, {Chiang}, {Congiu}, {Dale}, {Deger}, {Eibensteiner}, {Egorov}, {Garc{\'\i}a-Rodr{\'\i}guez}, {Glover}, {Grasha}, {Henshaw}, {Ho}, {Kepley}, {Kim}, {Klessen}, {Kreckel}, {Koch}, {Kruijssen}, {Larson}, {Lee}, {Lopez}, {Machado}, {Mayker}, {McElroy}, {Murphy}, {Ostriker}, {Pan}, {Pessa}, {Puschnig}, {Razza}, {S{\'a}nchez-Bl{\'a}zquez}, {Santoro}, {Sardone}, {Scheuermann}, {Sliwa}, {Sormani}, {Stuber}, {Thilker}, {Turner}, {Utomo}, {Watkins}, \& {Whitmore}}]{Leroy2021}
{Leroy}, A.~K., {Schinnerer}, E., {Hughes}, A., {et~al.} 2021, \bibinfo{title}{{PHANGS-ALMA: Arcsecond CO(2-1) Imaging of Nearby Star-forming Galaxies},} \apjs, 257, 43, \dodoi{10.3847/1538-4365/ac17f3}

\bibitem[{C. {Lind-Thomsen} {et~al.}(2025){Lind-Thomsen}, {Sneppen}, \& {Steinhardt}}]{LindThomsenSneppenSteinhardt2025}
{Lind-Thomsen}, C., {Sneppen}, A., \& {Steinhardt}, C.~L. 2025, \bibinfo{title}{{A Power Spectral Study of PHANGS Galaxies with JWST MIRI: On the Spatial Scales of Dust and Polycyclic Aromatic Hydrocarbons},} \apj, 985, 144, \dodoi{10.3847/1538-4357/adc808}

\bibitem[{N.~B. {Linzer} {et~al.}(2024){Linzer}, {Kim}, {Kim}, \& {Ostriker}}]{Linzer2024}
{Linzer}, N.~B., {Kim}, J.-G., {Kim}, C.-G., \& {Ostriker}, E.~C. 2024, \bibinfo{title}{{Ultraviolet Radiation Fields in Star-forming Disk Galaxies: Numerical Simulations with TIGRESS-NCR},} \apj, 975, 173, \dodoi{10.3847/1538-4357/ad7733}

\bibitem[{S.~A. {Mao} {et~al.}(2020){Mao}, {Ostriker}, \& {Kim}}]{Mao2020}
{Mao}, S.~A., {Ostriker}, E.~C., \& {Kim}, C.-G. 2020, \bibinfo{title}{{Cloud Properties and Correlations with Star Formation in Self-consistent Simulations of the Multiphase ISM},} \apj, 898, 52, \dodoi{10.3847/1538-4357/ab989c}

\bibitem[{N.~M. {McClure-Griffiths} {et~al.}(2023){McClure-Griffiths}, {Stanimirovi{\'c}}, \& {Rybarczyk}}]{McClure2023}
{McClure-Griffiths}, N.~M., {Stanimirovi{\'c}}, S., \& {Rybarczyk}, D.~R. 2023, \bibinfo{title}{{Atomic Hydrogen in the Milky Way: A Stepping Stone in the Evolution of Galaxies},} \araa, 61, 19, \dodoi{10.1146/annurev-astro-052920-104851}

\bibitem[{C.~F. {McKee} \& E.~C. {Ostriker}(2007){McKee} \& {Ostriker}}]{mckee_ostriker2007}
{McKee}, C.~F., \& {Ostriker}, E.~C. 2007, \bibinfo{title}{{Theory of Star Formation},} \araa, 45, 565, \dodoi{10.1146/annurev.astro.45.051806.110602}

\bibitem[{T. {Naab} \& J.~P. {Ostriker}(2017){Naab} \& {Ostriker}}]{Naab2017}
{Naab}, T., \& {Ostriker}, J.~P. 2017, \bibinfo{title}{{Theoretical Challenges in Galaxy Formation},} \araa, 55, 59, \dodoi{10.1146/annurev-astro-081913-040019}

\bibitem[{D. {Obreschkow} {et~al.}(2013){Obreschkow}, {Power}, {Bruderer}, \& {Bonvin}}]{Obreschkow2013}
{Obreschkow}, D., {Power}, C., {Bruderer}, M., \& {Bonvin}, C. 2013, \bibinfo{title}{{A Robust Measure of Cosmic Structure beyond the Power Spectrum: Cosmic Filaments and the Temperature of Dark Matter},} \apj, 762, 115, \dodoi{10.1088/0004-637X/762/2/115}

\bibitem[{E.~C. {Ostriker} \& C.-G. {Kim}(2022){Ostriker} \& {Kim}}]{ostriker2022}
{Ostriker}, E.~C., \& {Kim}, C.-G. 2022, \bibinfo{title}{{Pressure-regulated, Feedback-modulated Star Formation in Disk Galaxies},} \apj, 936, 137, \dodoi{10.3847/1538-4357/ac7de2}

\bibitem[{E.~C. {Ostriker} {et~al.}(2001){Ostriker}, {Stone}, \& {Gammie}}]{OstrikerStoneGammie2001}
{Ostriker}, E.~C., {Stone}, J.~M., \& {Gammie}, C.~F. 2001, \bibinfo{title}{{Density, Velocity, and Magnetic Field Structure in Turbulent Molecular Cloud Models},} \apj, 546, 980, \dodoi{10.1086/318290}

\bibitem[{P. {Padoan} {et~al.}(2016){Padoan}, {Pan}, {Haugb{\o}lle}, \& {Nordlund}}]{Padoan2016}
{Padoan}, P., {Pan}, L., {Haugb{\o}lle}, T., \& {Nordlund}, {\r{A}}. 2016, \bibinfo{title}{{Supernova Driving. I. The Origin of Molecular Cloud Turbulence},} \apj, 822, 11, \dodoi{10.3847/0004-637X/822/1/11}

\bibitem[{L. {Pan} {et~al.}(2016){Pan}, {Padoan}, {Haugb{\o}lle}, \& {Nordlund}}]{Pan2016}
{Pan}, L., {Padoan}, P., {Haugb{\o}lle}, T., \& {Nordlund}, {\r{A}}. 2016, \bibinfo{title}{{Supernova Driving. II. Compressive Ratio in Molecular-cloud Turbulence},} \apj, 825, 30, \dodoi{10.3847/0004-637X/825/1/30}

\bibitem[{D. {Pathak} {et~al.}(2024){Pathak}, {Leroy}, {Thompson}, {Lopez}, {Belfiore}, {Boquien}, {Dale}, {Glover}, {Klessen}, {Koch}, {Rosolowsky}, {Sandstrom}, {Schinnerer}, {Smith}, {Sun}, {Sutter}, {Williams}, {Bigiel}, {Cao}, {Chastenet}, {Chevance}, {Chown}, {Emsellem}, {Faesi}, {Larson}, {Lee}, {Meidt}, {Ostriker}, {Ramambason}, {Sarbadhicary}, \& {Thilker}}]{Pathak2024}
{Pathak}, D., {Leroy}, A.~K., {Thompson}, T.~A., {et~al.} 2024, \bibinfo{title}{{A Two-Component Probability Distribution Function Describes the Mid-IR Emission from the Disks of Star-forming Galaxies},} \aj, 167, 39, \dodoi{10.3847/1538-3881/ad110d}

\bibitem[{J.~A. {Peacock}(1999){Peacock}}]{Peacock1999}
{Peacock}, J.~A. 1999, {Cosmological Physics}

\bibitem[{T. {Peters} {et~al.}(2017){Peters}, {Naab}, {Walch}, {Glover}, {Girichidis}, {Pellegrini}, {Klessen}, {W{\"u}nsch}, {Gatto}, \& {Baczynski}}]{Peters2017}
{Peters}, T., {Naab}, T., {Walch}, S., {et~al.} 2017, \bibinfo{title}{{The SILCC project - IV. Impact of dissociating and ionizing radiation on the interstellar medium and H{\ensuremath{\alpha}} emission as a tracer of the star formation rate},} \mnras, 466, 3293, \dodoi{10.1093/mnras/stw3216}

\bibitem[{ {Planck Collaboration} {et~al.}(2015){Planck Collaboration}, {Ade}, {Aghanim}, {Alina}, {Alves}, {Armitage-Caplan}, {Arnaud}, {Arzoumanian}, {Ashdown}, {Atrio-Barandela}, {Aumont}, {Baccigalupi}, {Banday}, {Barreiro}, {Battaner}, {Benabed}, {Benoit-L{\'e}vy}, {Bernard}, {Bersanelli}, {Bielewicz}, {Bock}, {Bond}, {Borrill}, {Bouchet}, {Boulanger}, {Bracco}, {Burigana}, {Butler}, {Cardoso}, {Catalano}, {Chamballu}, {Chary}, {Chiang}, {Christensen}, {Colombi}, {Colombo}, {Combet}, {Couchot}, {Coulais}, {Crill}, {Curto}, {Cuttaia}, {Danese}, {Davies}, {Davis}, {de Bernardis}, {de Gouveia Dal Pino}, {de Rosa}, {de Zotti}, {Delabrouille}, {D{\'e}sert}, {Dickinson}, {Diego}, {Donzelli}, {Dor{\'e}}, {Douspis}, {Dunkley}, {Dupac}, {Efstathiou}, {En{\ss}lin}, {Eriksen}, {Falgarone}, {Ferri{\`e}re}, {Finelli}, {Forni}, {Frailis}, {Fraisse}, {Franceschi}, {Galeotta}, {Ganga}, {Ghosh}, {Giard}, {Giraud-H{\'e}raud}, {Gonz{\'a}lez-Nuevo}, {G{\'o}rski}, {Gregorio}, {Gruppuso}, {Guillet}, {Hansen}, {Harrison},
  {Helou}, {Hern{\'a}ndez-Monteagudo}, {Hildebrandt}, {Hivon}, {Hobson}, {Holmes}, {Hornstrup}, {Huffenberger}, {Jaffe}, {Jaffe}, {Jones}, {Juvela}, {Keih{\"a}nen}, {Keskitalo}, {Kisner}, {Kneissl}, {Knoche}, {Kunz}, {Kurki-Suonio}, {Lagache}, {L{\"a}hteenm{\"a}ki}, {Lamarre}, {Lasenby}, {Lawrence}, {Leahy}, {Leonardi}, {Levrier}, {Liguori}, {Lilje}, {Linden-V{\o}rnle}, {L{\'o}pez-Caniego}, {Lubin}, {Mac{\'\i}as-P{\'e}rez}, {Maffei}, {Magalh{\~a}es}, {Maino}, {Mandolesi}, {Maris}, {Marshall}, {Martin}, {Mart{\'\i}nez-Gonz{\'a}lez}, {Masi}, {Matarrese}, {Mazzotta}, {Melchiorri}, {Mendes}, {Mennella}, {Migliaccio}, {Miville-Desch{\^e}nes}, {Moneti}, {Montier}, {Morgante}, {Mortlock}, {Munshi}, {Murphy}, {Naselsky}, {Nati}, {Natoli}, {Netterfield}, {Noviello}, {Novikov}, {Novikov}, {Oxborrow}, {Pagano}, {Pajot}, {Paladini}, {Paoletti}, {Pasian}, {Pearson}, {Perdereau}, {Perotto}, {Perrotta}, {Piacentini}, {Piat}, {Pietrobon}, {Plaszczynski}, {Poidevin}, {Pointecouteau}, {Polenta}, {Popa}, {Pratt}, {Prunet},
  {Puget}, {Rachen}, {Reach}, {Rebolo}, {Reinecke}, {Remazeilles}, {Renault}, {Ricciardi}, {Riller}, {Ristorcelli}, {Rocha}, {Rosset}, {Roudier}, {Rubi{\~n}o-Mart{\'\i}n}, {Rusholme}, {Sandri}, {Savini}, {Scott}, {Spencer}, {Stolyarov}, {Stompor}, {Sudiwala}, {Sutton}, {Suur-Uski}, {Sygnet}, {Tauber}, {Terenzi}, {Toffolatti}, {Tomasi}, {Tristram}, {Tucci}, {Umana}, {Valenziano}, {Valiviita}, {Van Tent}, {Vielva}, {Villa}, \& {Wade}}]{Planck2015}
{Planck Collaboration}, {Ade}, P.~A.~R., {Aghanim}, N., {et~al.} 2015, \bibinfo{title}{{Planck intermediate results. XIX. An overview of the polarized thermal emission from Galactic dust},} \aap, 576, A104, \dodoi{10.1051/0004-6361/201424082}

\bibitem[{S.~K.~N. {Portillo} {et~al.}(2018){Portillo}, {Slepian}, {Burkhart}, {Kahraman}, \& {Finkbeiner}}]{Portillo2018}
{Portillo}, S. K.~N., {Slepian}, Z., {Burkhart}, B., {Kahraman}, S., \& {Finkbeiner}, D.~P. 2018, \bibinfo{title}{{Developing the 3-point Correlation Function for the Turbulent Interstellar Medium},} \apj, 862, 119, \dodoi{10.3847/1538-4357/aacb80}

\bibitem[{W.~H. {Press} \& P. {Schechter}(1974){Press} \& {Schechter}}]{PressSchechter1974}
{Press}, W.~H., \& {Schechter}, P. 1974, \bibinfo{title}{{Formation of Galaxies and Clusters of Galaxies by Self-Similar Gravitational Condensation},} \apj, 187, 425, \dodoi{10.1086/152650}

\bibitem[{T.-E. {Rathjen} {et~al.}(2025){Rathjen}, {Walch}, {Naab}, {N{\"u}rnberger}, {W{\"u}nsch}, {Seifried}, \& {Glover}}]{Rathjen2025}
{Rathjen}, T.-E., {Walch}, S., {Naab}, T., {et~al.} 2025, \bibinfo{title}{{SILCC - VIII: The impact of far-ultraviolet radiation on star formation and the interstellar medium},} \mnras, \dodoi{10.1093/mnras/staf792}

\bibitem[{E. {Rosolowsky}(2005){Rosolowsky}}]{Rosolowsky2005}
{Rosolowsky}, E. 2005, \bibinfo{title}{{The Mass Spectra of Giant Molecular Clouds in the Local Group},} \pasp, 117, 1403, \dodoi{10.1086/497582}

\bibitem[{E. {Rosolowsky} {et~al.}(2021){Rosolowsky}, {Hughes}, {Leroy}, {Sun}, {Querejeta}, {Schruba}, {Usero}, {Herrera}, {Liu}, {Pety}, {Saito}, {Be{\v{s}}li{\'c}}, {Bigiel}, {Blanc}, {Chevance}, {Dale}, {Deger}, {Faesi}, {Glover}, {Henshaw}, {Klessen}, {Kruijssen}, {Larson}, {Lee}, {Meidt}, {Mok}, {Schinnerer}, {Thilker}, \& {Williams}}]{Rosolowsky2021}
{Rosolowsky}, E., {Hughes}, A., {Leroy}, A.~K., {et~al.} 2021, \bibinfo{title}{{Giant molecular cloud catalogues for PHANGS-ALMA: methods and initial results},} \mnras, 502, 1218, \dodoi{10.1093/mnras/stab085}

\bibitem[{D.~B. {Sanders} {et~al.}(1985){Sanders}, {Scoville}, \& {Solomon}}]{Sanders1985}
{Sanders}, D.~B., {Scoville}, N.~Z., \& {Solomon}, P.~M. 1985, \bibinfo{title}{{Giant molecular clouds in the galaxy. II. Characteristics of discretefeatures.},} \apj, 289, 373, \dodoi{10.1086/162897}

\bibitem[{A.~K. {Saydjari} {et~al.}(2021){Saydjari}, {Portillo}, {Slepian}, {Kahraman}, {Burkhart}, \& {Finkbeiner}}]{Saydjari2021}
{Saydjari}, A.~K., {Portillo}, S. K.~N., {Slepian}, Z., {et~al.} 2021, \bibinfo{title}{{Classification of Magnetohydrodynamic Simulations Using Wavelet Scattering Transforms},} \apj, 910, 122, \dodoi{10.3847/1538-4357/abe46d}

\bibitem[{E. {Schinnerer} \& A.~K. {Leroy}(2024){Schinnerer} \& {Leroy}}]{SchinnererLeroy2024}
{Schinnerer}, E., \& {Leroy}, A.~K. 2024, \bibinfo{title}{{Molecular Gas and the Star-Formation Process on Cloud Scales in Nearby Galaxies},} \araa, 62, 369, \dodoi{10.1146/annurev-astro-071221-052651}

\bibitem[{K. {Shiidsuka} \& S. {Ida}(1999){Shiidsuka} \& {Ida}}]{ShiidsukaIda1999}
{Shiidsuka}, K., \& {Ida}, S. 1999, \bibinfo{title}{{Evolution of the velocity dispersion of self-gravitating particles in disc potentials},} \mnras, 307, 737, \dodoi{10.1046/j.1365-8711.1999.02598.x}

\bibitem[{J.~D. {Soler} {et~al.}(2022){Soler}, {Miville-Desch{\^e}nes}, {Molinari}, {Klessen}, {Hennebelle}, {Testi}, {McClure-Griffiths}, {Beuther}, {Elia}, {Schisano}, {Traficante}, {Girichidis}, {Glover}, {Smith}, {Sormani}, \& {Tre{\ss}}}]{Soler2022}
{Soler}, J.~D., {Miville-Desch{\^e}nes}, M.~A., {Molinari}, S., {et~al.} 2022, \bibinfo{title}{{The Galactic dynamics revealed by the filamentary structure in atomic hydrogen emission},} \aap, 662, A96, \dodoi{10.1051/0004-6361/202243334}

\bibitem[{P.~M. {Solomon} {et~al.}(1987){Solomon}, {Rivolo}, {Barrett}, \& {Yahil}}]{Solomon1987}
{Solomon}, P.~M., {Rivolo}, A.~R., {Barrett}, J., \& {Yahil}, A. 1987, \bibinfo{title}{{Mass, Luminosity, and Line Width Relations of Galactic Molecular Clouds},} \apj, 319, 730, \dodoi{10.1086/165493}

\bibitem[{R.~S. {Somerville} \& R. {Dav{\'e}}(2015){Somerville} \& {Dav{\'e}}}]{Somerville2015}
{Somerville}, R.~S., \& {Dav{\'e}}, R. 2015, \bibinfo{title}{{Physical Models of Galaxy Formation in a Cosmological Framework},} \araa, 53, 51, \dodoi{10.1146/annurev-astro-082812-140951}

\bibitem[{L. {Spitzer}(1942){Spitzer}}]{Spitzer1942}
{Spitzer}, Jr., L. 1942, \bibinfo{title}{{The Dynamics of the Interstellar Medium. III. Galactic Distribution.},} \apj, 95, 329, \dodoi{10.1086/144407}

\bibitem[{L. {Spitzer} \& M. {Schwarzschild}(1951){Spitzer} \& {Schwarzschild}}]{SpitzerSchwarzschild1951}
{Spitzer}, Jr., L., \& {Schwarzschild}, M. 1951, \bibinfo{title}{{The Possible Influence of Interstellar Clouds on Stellar Velocities.},} \apj, 114, 385, \dodoi{10.1086/145478}

\bibitem[{L. {Spitzer} \& M. {Schwarzschild}(1953){Spitzer} \& {Schwarzschild}}]{SpitzerSchwarzschild1953}
{Spitzer}, Jr., L., \& {Schwarzschild}, M. 1953, \bibinfo{title}{{The Possible Influence of Interstellar Clouds on Stellar Velocities. II.},} \apj, 118, 106, \dodoi{10.1086/145730}

\bibitem[{S. Stanimirović {et~al.}(1999)Stanimirović, Staveley-Smith, Dickey, Sault, \& Snowden}]{Stanimirovic1999}
Stanimirović, S., Staveley-Smith, L., Dickey, J.~M., Sault, R.~J., \& Snowden, S.~L. 1999, \bibinfo{title}{The large-scale Hi structure of the Small Magellanic Cloud,} Monthly Notices of the Royal Astronomical Society, 302, 417, \dodoi{10.1046/j.1365-8711.1999.02013.x}

\bibitem[{J.~M. {Stone} {et~al.}(2008){Stone}, {Gardiner}, {Teuben}, {Hawley}, \& {Simon}}]{Stone2008}
{Stone}, J.~M., {Gardiner}, T.~A., {Teuben}, P., {Hawley}, J.~F., \& {Simon}, J.~B. 2008, \bibinfo{title}{{Athena: A New Code for Astrophysical MHD},} \apjs, 178, 137, \dodoi{10.1086/588755}

\bibitem[{J. {Sun} {et~al.}(2022){Sun}, {Leroy}, {Rosolowsky}, {Hughes}, {Schinnerer}, {Schruba}, {Koch}, {Blanc}, {Chiang}, {Groves}, {Liu}, {Meidt}, {Pan}, {Pety}, {Querejeta}, {Saito}, {Sandstrom}, {Sardone}, {Usero}, {Utomo}, {Williams}, {Barnes}, {Benincasa}, {Bigiel}, {Bolatto}, {Boquien}, {Chevance}, {Dale}, {Deger}, {Emsellem}, {Glover}, {Grasha}, {Henshaw}, {Klessen}, {Kreckel}, {Kruijssen}, {Ostriker}, \& {Thilker}}]{Sun2022}
{Sun}, J., {Leroy}, A.~K., {Rosolowsky}, E., {et~al.} 2022, \bibinfo{title}{{Molecular Cloud Populations in the Context of Their Host Galaxy Environments: A Multiwavelength Perspective},} \aj, 164, 43, \dodoi{10.3847/1538-3881/ac74bd}

\bibitem[{T. {Tepper-Garc{\'\i}a} {et~al.}(2024){Tepper-Garc{\'\i}a}, {Bland-Hawthorn}, {Vasiliev}, {Agertz}, {Teyssier}, \& {Federrath}}]{TepperGarcia2024}
{Tepper-Garc{\'\i}a}, T., {Bland-Hawthorn}, J., {Vasiliev}, E., {et~al.} 2024, \bibinfo{title}{{NEXUS: a framework for controlled simulations of idealized galaxies},} \mnras, 535, 187, \dodoi{10.1093/mnras/stae2372}

\bibitem[{D. {Utomo} {et~al.}(2019){Utomo}, {Chiang}, {Leroy}, {Sandstrom}, \& {Chastenet}}]{Utomo2019}
{Utomo}, D., {Chiang}, I.-D., {Leroy}, A.~K., {Sandstrom}, K.~M., \& {Chastenet}, J. 2019, \bibinfo{title}{{The Resolved Distributions of Dust Mass and Temperature in Local Group Galaxies},} \apj, 874, 141, \dodoi{10.3847/1538-4357/ab05d3}

\bibitem[{E. {Vazquez-Semadeni}(1994){Vazquez-Semadeni}}]{Vazquez-Semadeni1994}
{Vazquez-Semadeni}, E. 1994, \bibinfo{title}{{Hierarchical Structure in Nearly Pressureless Flows as a Consequence of Self-similar Statistics},} \apj, 423, 681, \dodoi{10.1086/173847}

\bibitem[{M.~D. {Weinberg} {et~al.}(1987){Weinberg}, {Shapiro}, \& {Wasserman}}]{WeinbergShapiroWasserman1987}
{Weinberg}, M.~D., {Shapiro}, S.~L., \& {Wasserman}, I. 1987, \bibinfo{title}{{The Dynamical Fate of Wide Binaries in the Solar Neighborhood},} \apj, 312, 367, \dodoi{10.1086/164883}

\bibitem[{J.~P. {Williams} \& C.~F. {McKee}(1997){Williams} \& {McKee}}]{Williams1997}
{Williams}, J.~P., \& {McKee}, C.~F. 1997, \bibinfo{title}{{The Galactic Distribution of OB Associations in Molecular Clouds},} \apj, 476, 166, \dodoi{10.1086/303588}

\bibitem[{R. {Wolstenhulme} {et~al.}(2015){Wolstenhulme}, {Bonvin}, \& {Obreschkow}}]{WolstenhulmeBonvinObreschkow2015}
{Wolstenhulme}, R., {Bonvin}, C., \& {Obreschkow}, D. 2015, \bibinfo{title}{{Three-point Phase Correlations: A New Measure of Non-linear Large-scale Structure},} \apj, 804, 132, \dodoi{10.1088/0004-637X/804/2/132}

\bibitem[{K.~H. {Yuen} {et~al.}(2023){Yuen}, {Li}, \& {Yan}}]{Yuen2023}
{Yuen}, K.~H., {Li}, H., \& {Yan}, H. 2023, \bibinfo{title}{{Temporal Properties of the Compressible Magnetohydrodynamic Turbulence},} arXiv e-prints, arXiv:2310.03806, \dodoi{10.48550/arXiv.2310.03806}

\bibitem[{H.-X. {Zhang} {et~al.}(2012){Zhang}, {Hunter}, \& {Elmegreen}}]{Zhang2012}
{Zhang}, H.-X., {Hunter}, D.~A., \& {Elmegreen}, B.~G. 2012, \bibinfo{title}{{H I Power Spectra and the Turbulent Interstellar Medium of Dwarf Irregular Galaxies},} \apj, 754, 29, \dodoi{10.1088/0004-637X/754/1/29}

\end{thebibliography}
\bibliographystyle{aasjournalv7}

\end{document}